\documentclass[11pt]{article}
\usepackage{amsfonts}
\usepackage{dsfont}
\usepackage{amssymb}
\usepackage{amsmath}
\usepackage{amsxtra}
\usepackage{graphicx}
\usepackage{psfrag}
\usepackage{mathrsfs}
\usepackage{multirow}
\usepackage[sort, numbers, compress]{natbib}
\usepackage{stmaryrd}
\usepackage{subfigure}
\usepackage{tikz}
\usepackage{empheq}
\usepackage[normalem]{ulem}
\usepackage{pdfpages}
\usepackage{color}
\usepackage{longtable}
\usepackage{footmisc}
\usepackage{hyperref}

\usepackage{lineno}

\begin{document}
\def\BGamma{\mbox{\boldmath$\Gamma$}}
\def\BDelta{\mbox{\boldmath$\Delta$}}
\def\BTheta{\mbox{\boldmath$\Theta$}}
\def\BLambda{\mbox{\boldmath$\Lambda$}}
\def\BXi{\mbox{\boldmath$\Xi$}}
\def\BPi{\mbox{\boldmath$\Pi$}}
\def\BSigma{\mbox{\boldmath$\Sigma$}}
\def\BUpsilon{\mbox{\boldmath$\Upsilon$}}
\def\BPhi{\mbox{\boldmath$\Phi$}}
\def\BPsi{\mbox{\boldmath$\Psi$}}
\def\BOmega{\mbox{\boldmath$\theta$}}
\def\Balpha{\mbox{\boldmath$\alpha$}}
\def\Bbeta{\mbox{\boldmath$\beta$}}
\def\Bgamma{\mbox{\boldmath$\gamma$}}
\def\Bdelta{\mbox{\boldmath$\delta$}}          
\def\Bepsilon{\mbox{\boldmath$\epsilon$}}
\def\Bzeta{\mbox{\boldmath$\zeta$}}
\def\Beta{\mbox{\boldmath$\eta$}}
\def\Btheta{\mbox{\boldmath$\theta$}}
\def\Biota{\mbox{\boldmath$\iota$}}
\def\Bkappa{\mbox{\boldmath$\kappa$}}
\def\Blambda{\mbox{\boldmath$\lambda$}}
\def\Bmu{\mbox{\boldmath$\mu$}}
\def\Bnu{\mbox{\boldmath$\nu$}}
\def\Bxi{\mbox{\boldmath$\xi$}}
\def\Bpi{\mbox{\boldmath$\pi$}}
\def\Brho{\mbox{\boldmath$\rho$}}
\def\Bsigma{\mbox{\boldmath$\sigma$}}
\def\Btau{\mbox{\boldmath$\tau$}}
\def\Bupsilon{\mbox{\boldmath$\upsilon$}}
\def\Bphi{\mbox{\boldmath$\phi$}}
\def\Bchi{\mbox{\boldmath$\chi$}}
\def\Bpsi{\mbox{\boldmath$\psi$}}
\def\Bomega{\mbox{\boldmath$\omega$}}
\def\Bvarepsilon{\mbox{\boldmath$\varepsilon$}}
\def\Bvartheta{\mbox{\boldmath$\vartheta$}}
\def\Bvarpi{\mbox{\boldmath$\varpi$}}
\def\Bvarrho{\mbox{\boldmath$\varrho$}}
\def\Bvarsigma{\mbox{\boldmath$\varsigma$}}
\def\Bvarphi{\mbox{\boldmath$\varphi$}}
\def\bone{\mbox{\boldmath$1$}}
\def\bzero{\mbox{\boldmath$0$}}
\def\bnabla{\mbox{\boldmath$\nabla$}}
\def\bvarepsilon{\mbox{\boldmath$\varepsilon$}}
\def\bA{\mbox{\boldmath$ A$}}
\def\bB{\mbox{\boldmath$ B$}}
\def\bC{\mbox{\boldmath$ C$}}
\def\bD{\mbox{\boldmath$ D$}}
\def\bE{\mbox{\boldmath$ E$}}
\def\bF{\mbox{\boldmath$ F$}}
\def\bG{\mbox{\boldmath$ G$}}
\def\bH{\mbox{\boldmath$ H$}}
\def\bI{\mbox{\boldmath$ I$}}
\def\bJ{\mbox{\boldmath$ J$}}
\def\bK{\mbox{\boldmath$ K$}}
\def\bL{\mbox{\boldmath$ L$}}
\def\bM{\mbox{\boldmath$ M$}}
\def\bN{\mbox{\boldmath$ N$}}
\def\bO{\mbox{\boldmath$ O$}}
\def\bP{\mbox{\boldmath$ P$}}
\def\bQ{\mbox{\boldmath$ Q$}}
\def\bR{\mbox{\boldmath$ R$}}
\def\bS{\mbox{\boldmath$ S$}}
\def\bT{\mbox{\boldmath$ T$}}
\def\bU{\mbox{\boldmath$ U$}}
\def\bV{\mbox{\boldmath$ V$}}
\def\bW{\mbox{\boldmath$ W$}}
\def\bX{\mbox{\boldmath$ X$}}
\def\bY{\mbox{\boldmath$ Y$}}
\def\bZ{\mbox{\boldmath$ Z$}}
\def\ba{\mbox{\boldmath$ a$}}
\def\bb{\mbox{\boldmath$ b$}}
\def\bc{\mbox{\boldmath$ c$}}
\def\bd{\mbox{\boldmath$ d$}}
\def\be{\mbox{\boldmath$ e$}}
\def\bff{\mbox{\boldmath$ f$}}
\def\bg{\mbox{\boldmath$ g$}}
\def\bh{\mbox{\boldmath$ h$}}
\def\bi{\mbox{\boldmath$ i$}}
\def\bj{\mbox{\boldmath$ j$}}
\def\bk{\mbox{\boldmath$ k$}}
\def\bl{\mbox{\boldmath$ l$}}
\def\bm{\mbox{\boldmath$ m$}}
\def\bn{\mbox{\boldmath$ n$}}
\def\bo{\mbox{\boldmath$ o$}}
\def\bp{\mbox{\boldmath$ p$}}
\def\bq{\mbox{\boldmath$ q$}}
\def\br{\mbox{\boldmath$ r$}}
\def\bs{\mbox{\boldmath$ s$}}
\def\bt{\mbox{\boldmath$ t$}}
\def\bu{\mbox{\boldmath$ u$}}
\def\bv{\mbox{\boldmath$ v$}}
\def\bw{\mbox{\boldmath$ w$}}
\def\bx{\mbox{\boldmath$ x$}}
\def\by{\mbox{\boldmath$ y$}}
\def\bz{\mbox{\boldmath$ z$}}
\newcommand{\kg}[1]{\textcolor{blue}{\textbf{(kg)} #1}}
\newcommand{\gt}[1]{\textcolor{red}{\textbf{(gt)} #1}}
\newcommand{\zw}[1]{\textcolor{purple}{\textbf{(zw)} #1}}
\newcommand{\xz}[1]{\textcolor{brown}{\textbf{(xz)} #1}}
\newcommand{\mct}[1]{\textcolor{green}{\textbf{(mct)} #1}}
\title{ System inference for the spatio-temporal evolution of infectious diseases: Michigan in the time of COVID-19}
\author{Z.~Wang, X.~Zhang, G.H.~Teichert, M.~Carrasco-Teja \& K.~Garikipati\thanks{Corresponding author, \texttt{krishna@umich.edu}}\\Mechanical Engineering, Mathematics and the Michigan Institute for Computational \\Discovery \& Engineering, University of Michigan}

\maketitle
\begin{abstract}
    We extend the classical SIR model of infectious disease spread to account for time dependence in the parameters, which also include diffusivities. The temporal dependence accounts for the changing characteristics of testing, quarantine and treatment protocols, while diffusivity incorporates a mobile population. This model has been applied to data on the evolution of the COVID-19 pandemic in the US state of Michigan. For system inference, we use recent advances; specifically our framework for Variational System Identification (Wang et al., \emph{Comp. Meth. App. Mech. Eng.}, \textbf{356}, 44-74, 2019; arXiv:2001.04816 [cs.CE]) as well as Bayesian machine learning methods.
\end{abstract}
\section{Background}
\label{sec:background}
Starting from their origins in the the work of Kermack \& McKendrick\cite{Kermack1927}, the use of differential equation models of the course of infectious diseases has grown to become one of the more accessible instances of the reach of mathematics. The current COVID-19 Pandemic has brought them into the common parlance. Even before this, however, the baseline Susceptible-Infected-Recovered (SIR) model had been extended to include Exposed (E) and Deceased (D) compartments and applied with considerable success to influenza, ebola, malaria, cholera, tuberculosis and several other infectious diseases\cite{Eisenberg2015,Eisenberg2013-SIWRCholera,Wesoloski2012-Malaria,Colizza2007-InfluenzaMetapopulations}. (Some of this literature also includes agent-based models, which we do not consider here.) During the COVID-19 Pandemic, the widespread availability of data in the public domain \cite{1point3acres,yang2020covidnet,jhumap,MI-covid19-data,NYT-data,IHME-data} has served to attract methods of mathematics, computation and data science to analyzing this information, inferring the disease's dynamics and making projections. The present communication is in this spirit, and brings our  recent work in large scale computations of partial differential equations (PDEs), system inference and machine learning to this problem \cite{WangCMAME2019,Wang2020,Teichert2018,Teichert2019,Teichert2020,Zhang2020}.

Of particular interest to us are two lines of enquiry: The first is that for a rapidly evolving disease such as COVID-19, with its public health, population-based, political, travel and economic manifestations, the classical SIR model of ordinary differential equations (ODEs) with constant coefficients seems inadequate. Driven by data that extends the compartments to the deceased (D), we have adopted the SIRD model. The first extension that we have undertaken is to allow the ODE coefficients to vary in time to reflect the evolving contours of testing, quarantine and treatment protocols. This is not necessarily novel, and has been addressed in other work \cite{Eisenberg2015,Jo2020}, although perhaps not with the inference approach of Variational System Identification (VSI) and ODE-constrained optimization that we have adopted. 

The second is the fact of a mobile population. Population mobility has been addressed through metapopulation models that characterize how diseases move between population hubs, across countries, or even internationally. The most widely known are gravity models (e.g. \cite{Truscott2012-GravityModels}), and network and agent based models \cite{Hunter2017-ABMReview}. Given the prominence that quarantine protocols--adorned with the current-day euphemism of ``social distancing''--have played in the COVID-19 Pandemic, it appears natural to seek an extension of the SIRD model to a spatio-temporal PDE model. As the world went into lockdown, but at different rates and degrees of rigor, and then began to emerge from it, the detection of patterns of mobility in space and time presents a compelling avenue for investigation. Such an extension also has been considered--chiefly in the setting of the mathematical analysis of reaction-diffusion systems \cite{Chinviriyasit2010,Gai2020,Angulo2013-spatialBMESIR}. Our contribution to this aspect of the mathematical treatment is to also allow the diffusivity of the S, I and R sub-populations to vary with time.

To these tasks we have brought the abundance of high-quality, public domain, data on the evolution of the various compartment pertaining to the SIRD model in the US state of Michigan. The temporal resolution by days and spatial resolution by the 85 counties of Michigan has allowed us to apply our methods of Variational System Identification \cite{WangCMAME2019,Wang2020}, PDE-constrained optimization and machine learning \cite{Teichert2018,Teichert2019,Teichert2020,Zhang2020} to these data. 

In Section \ref{sec:SIRDmodel} we review the foundational SIRD ODE model. Section \ref{sec:data} is on data preparation. The application of system identification and machine learning to the ODE system are, respectively, in Sections \ref{sec:sys_id} and \ref{sec:NN}. The results for inferred parameters and forward simulation prediction are presented in Section \ref{sec:results}. The extension to inferring mobility via reaction-diffusion systems is in Section \ref{sec:PDE}. Our conclusions appear in Section \ref{sec:concl}.

\section{The compartmental model of infectious disease dynamics}\label{sec:SIRDmodel}
We use the SIRD version of compartmental epidemiology models. The population, taken to remain constant at $N$, is divided into four disjoint compartments with time-dependent sub-populations: $S(t)$ for susceptible, $I(t)$ for infected, $R(t)$ for recovered and  $D(t)$ for deceased individuals. The governing ODEs are:
\begin{align}
    \frac{\text{d} S}{\text{d} t} &=-\frac{\beta}{N} SI+\gamma R\label{eq:S}\\
    \frac{\text{d}I}{\text{d}t} &=\frac{\beta}{N}SI-\mu I-\alpha I\label{eq:I}\\
    \frac{\text{d}R}{\text{d}t} &=\mu I-\gamma R\label{eq:R}\\
    \frac{\text{d}D}{\text{d}t} &=\alpha I\label{eq:D}\\
    N &= S(t) + I(t) + R(t) + D(t).\label{eq:N}
\end{align}
This is the canonical form of the model where the sub-populations are assumed to be well-mixed so that spatial variations can be ignored over the domain of interest. Here $\beta(t)$ is the infection rate, $\mu(t)$ is the recovery rate, $\gamma(t)$ is the rate of immunity loss, and $\alpha(t)$ is the death rate--all allowed to vary with time. Using the natural temporal unit of one day, we note that $1/\mu(t)$ is also the number of days an  individual remains infectious. It follows that $\beta(t)/\mu(t)$ is the effective reproduction rate: the total number of the susceptible population that an infectious individual passes the disease to. This quantity is commonly denoted by $R_0$, but we use $r_0(t) = \beta(t)/\mu(t)$, to distinguish it from the recovered population, and emphasizing that it, too, varies with time.

We reiterate what we have outlined in the Background (Section \ref{sec:background}). Given the rapidly varying nature of testing, reporting, treatment protocols and quarantine conditions over the course of an epidemic, it is natural to allow the coefficients in the SIRD model, Equations (\ref{eq:S}-\ref{eq:D}) to vary with time. Such variation is evident in epidemiological data. The reader may be familiar with the time varying nature of such factors over the course of the COVID-19 Pandemic. It is a central feature of data preparation in the following section.

\begin{figure}[h]
    \centering
    \includegraphics[width=0.7\textwidth]{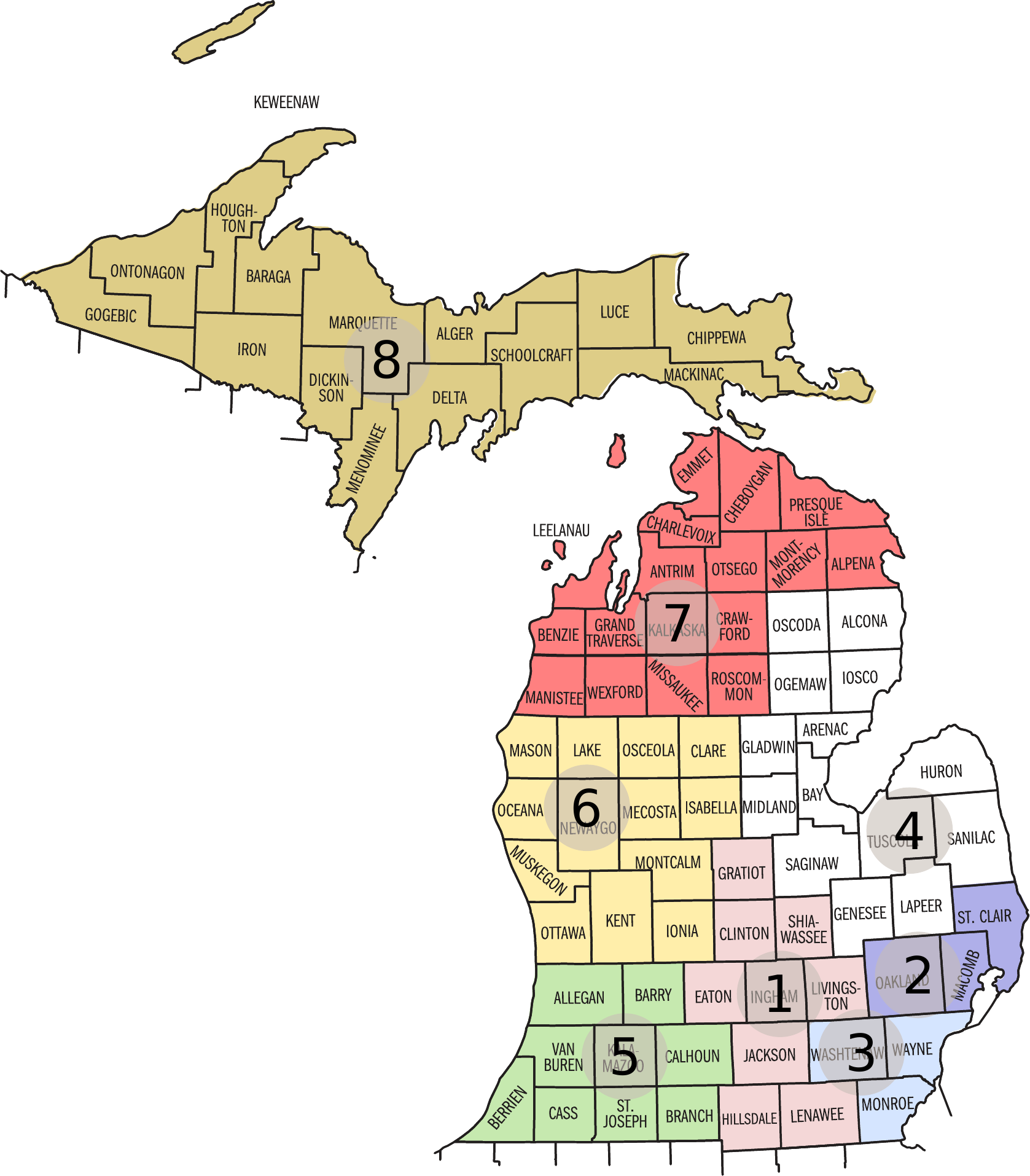}
    \caption{The map of Michigan delineating the counties and regions (modified from \cite{MI_map}).}
    \label{fig:Michmap}
\end{figure}

\section{Data preparation} \label{sec:data}
\begin{figure}[h]
    \centering
    \includegraphics[width=0.45\textwidth]{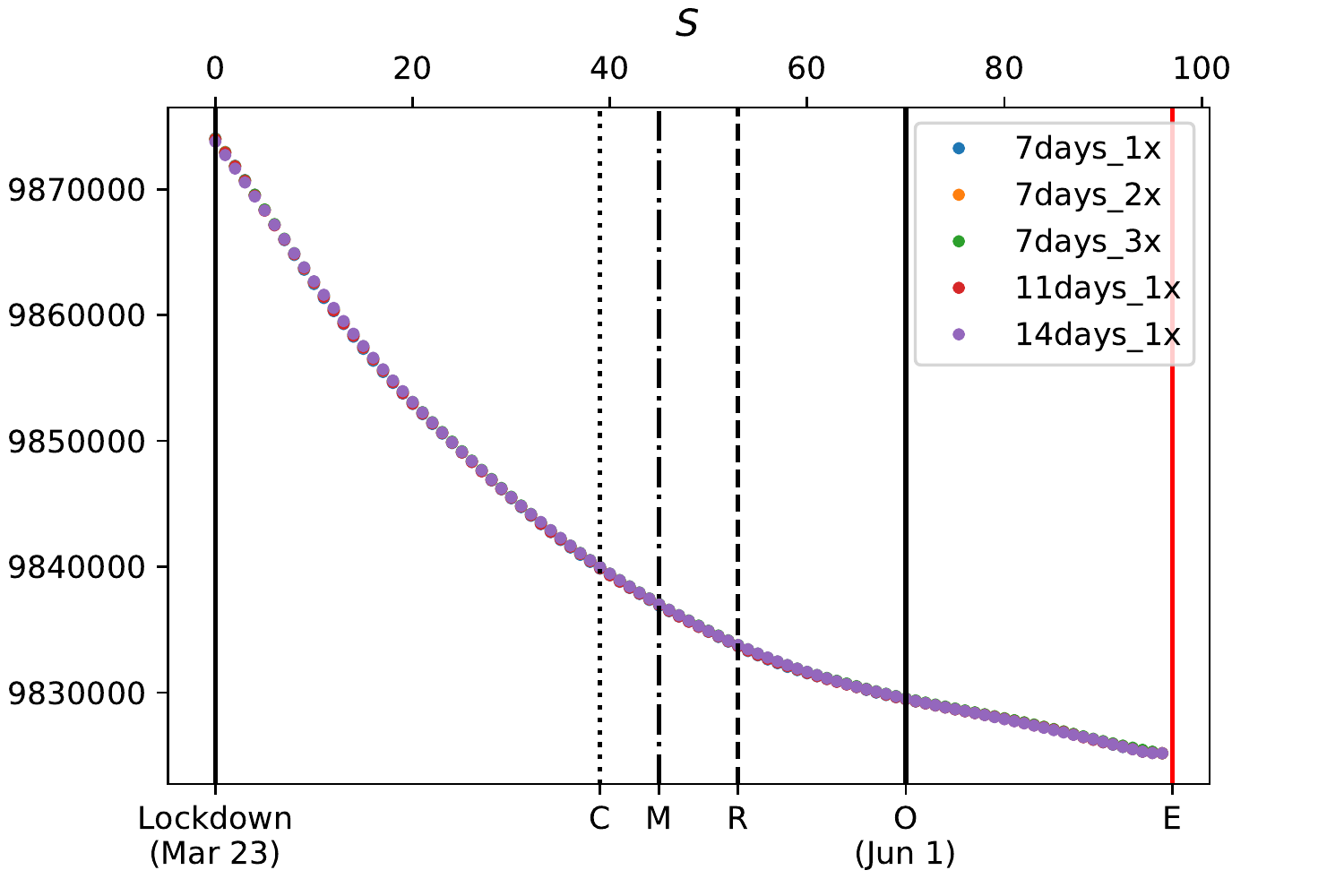}
    \includegraphics[width=0.45\textwidth]{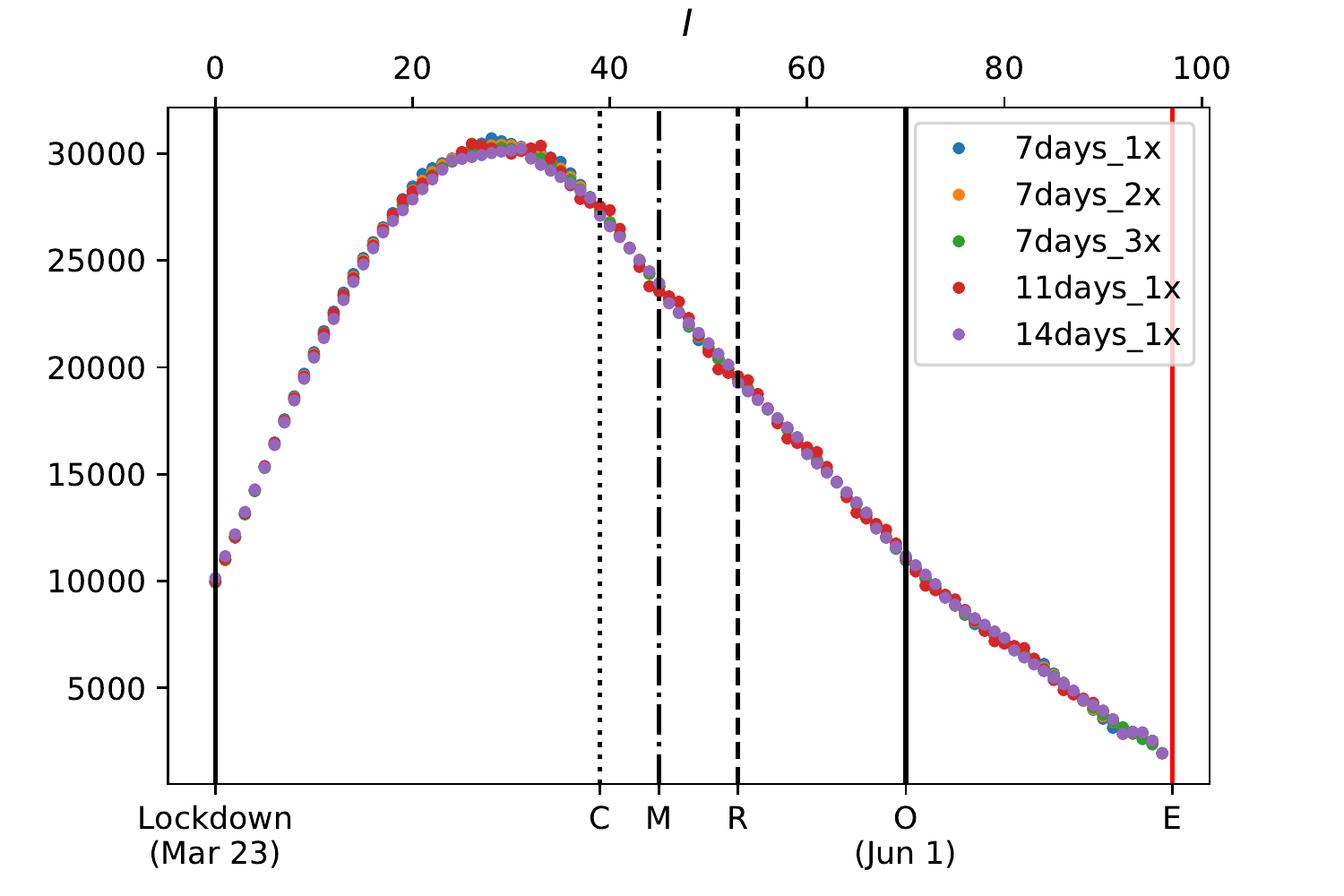}
    \includegraphics[width=0.45\textwidth]{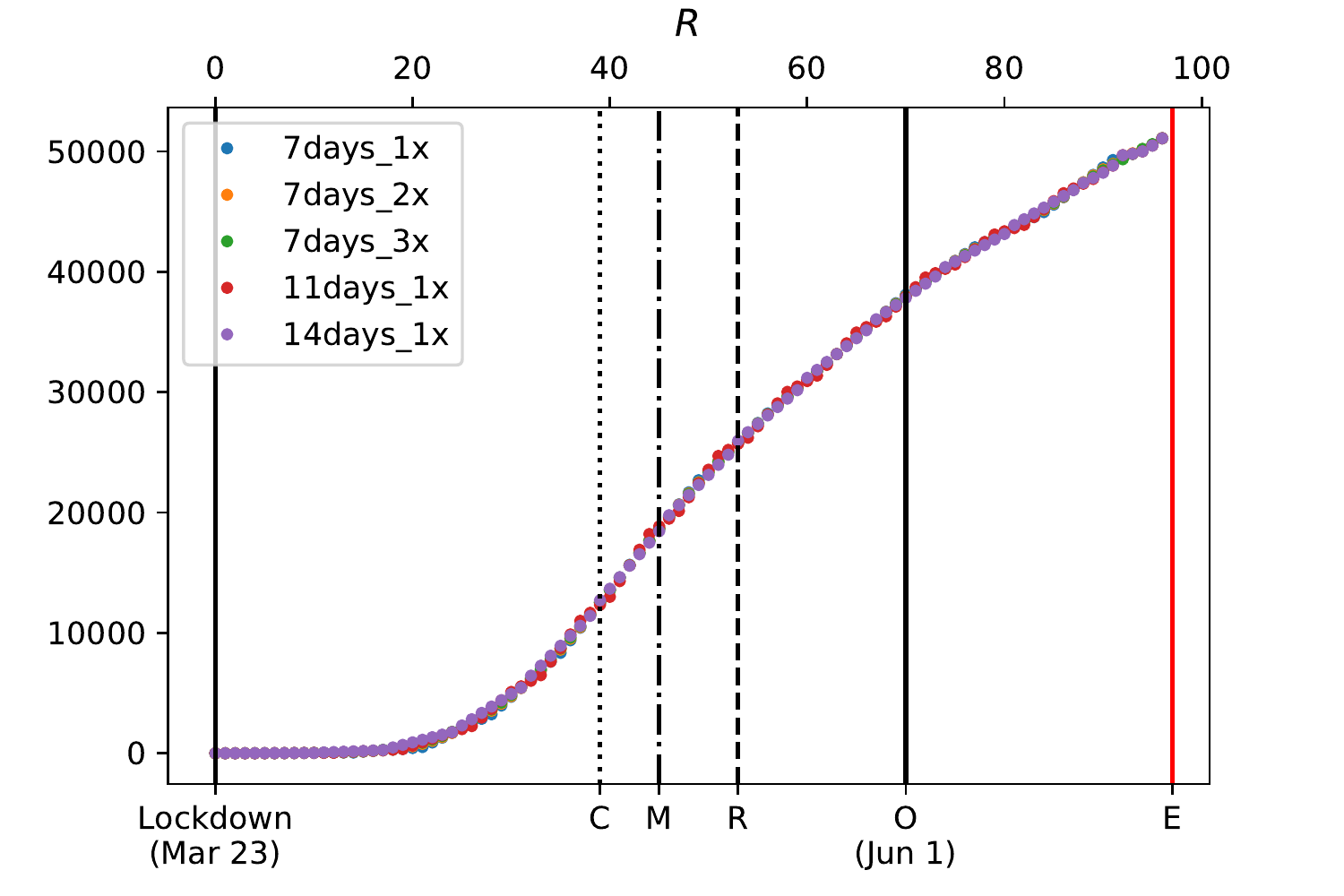}
    \includegraphics[width=0.45\textwidth]{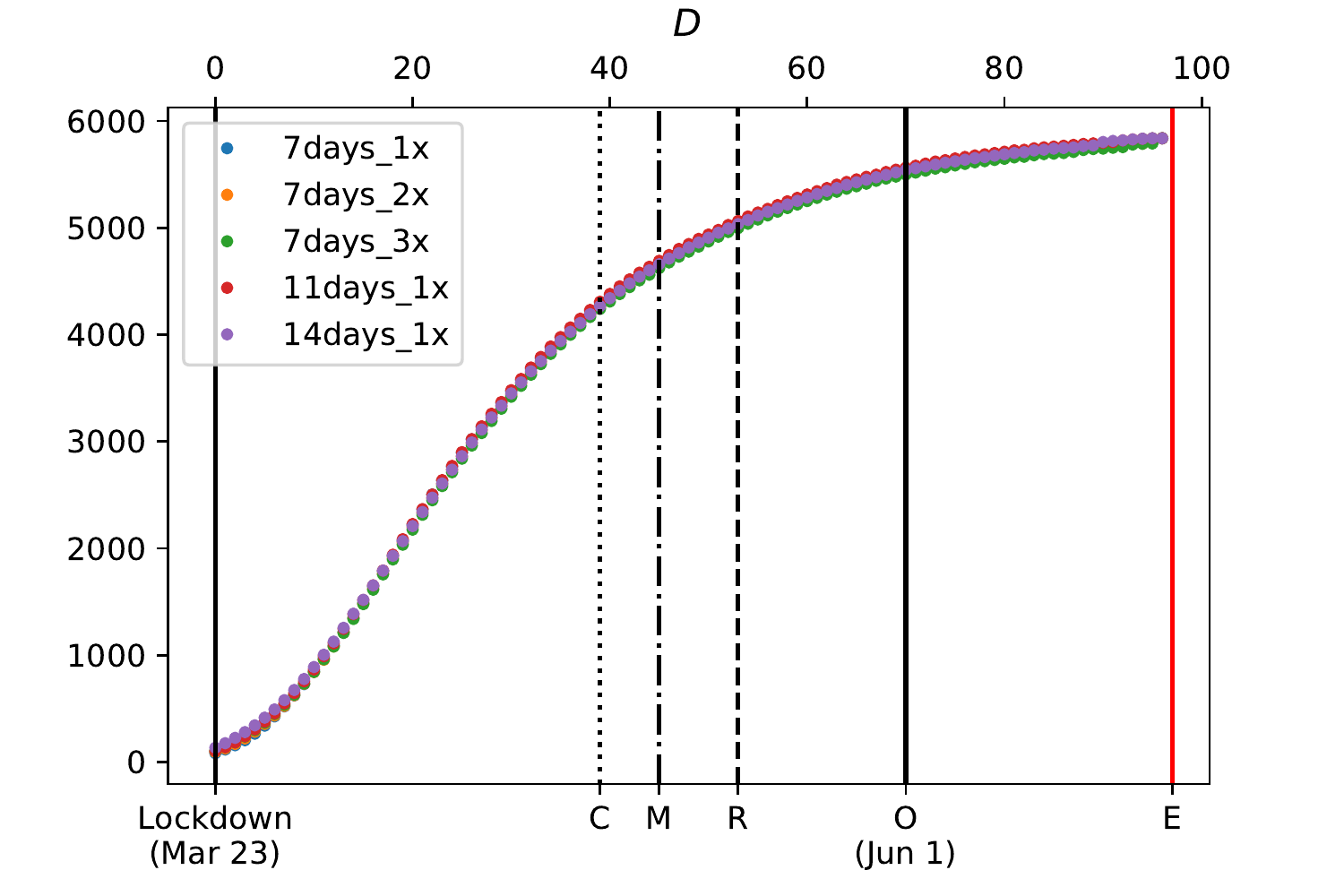}
    \caption{Cumulative data with different kernel widths and multiples of application of the smoothing filter: $7\text{days}\_1\times$ represents a 7-day filter applied once. Important dates are marked with the lockdown on March 23, reopening of construction and real estate sites (C) on May 1, reopening of manufacturing sites (M) on May 7, permission. to restart laboratory research (R) on May 15, lifting of stay-at-home order (O) on June 1, and the end of our data collection (E) on June 28. }
    \label{fig:smoothing1}
\end{figure}
\begin{figure}[h]
    \centering
    \includegraphics[width=0.45\textwidth]{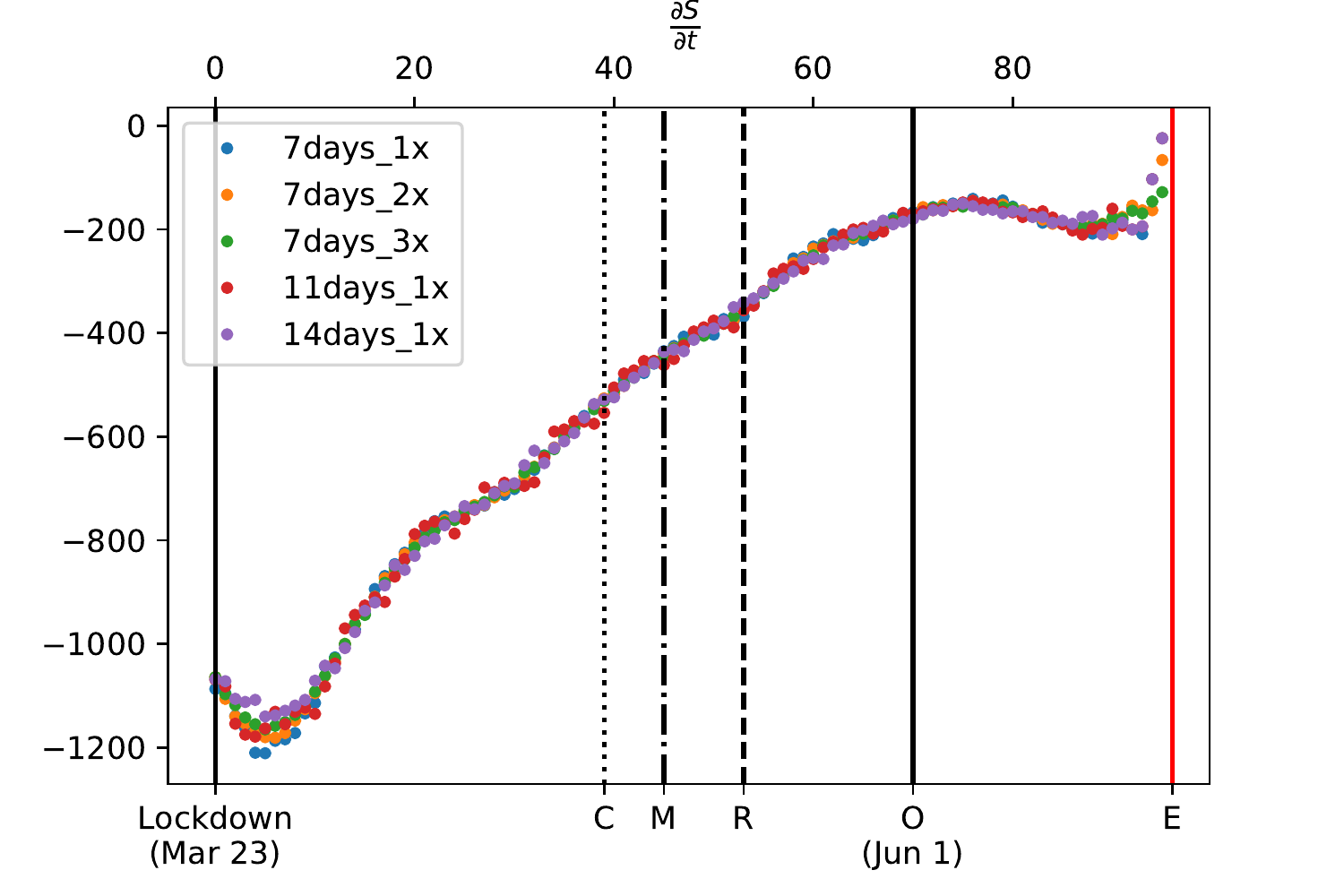}
    \includegraphics[width=0.45\textwidth]{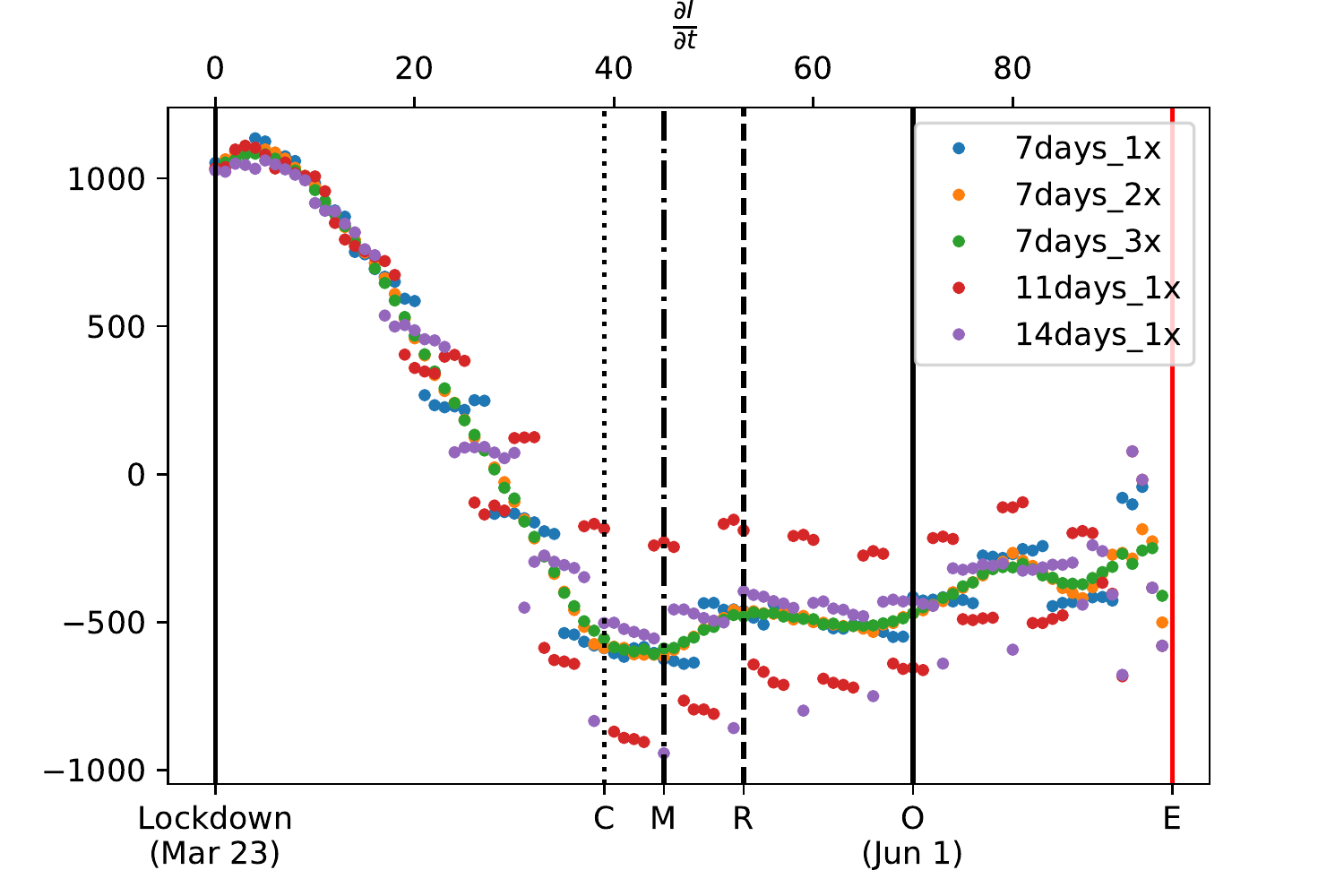}
    \includegraphics[width=0.45\textwidth]{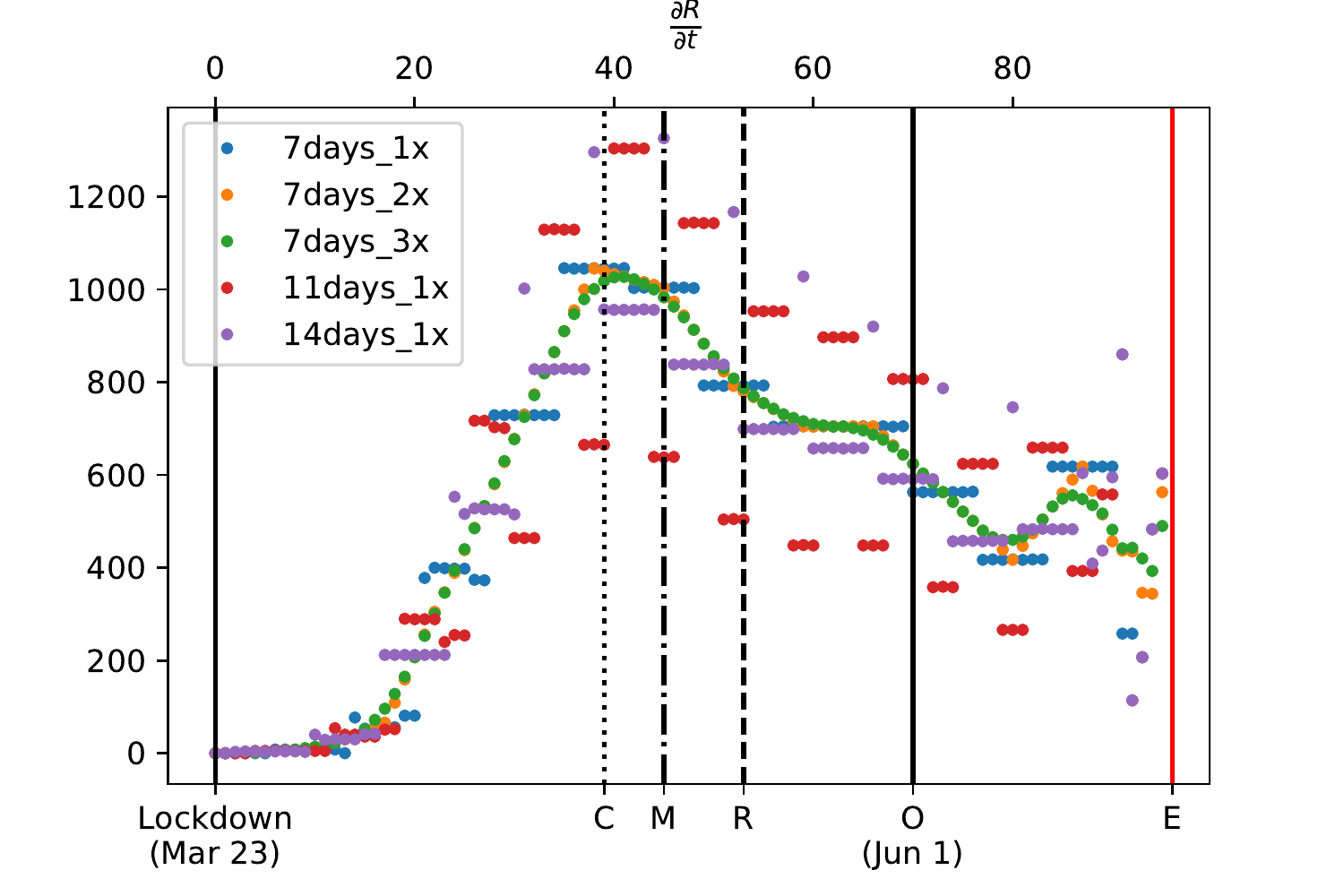}
    \includegraphics[width=0.45\textwidth]{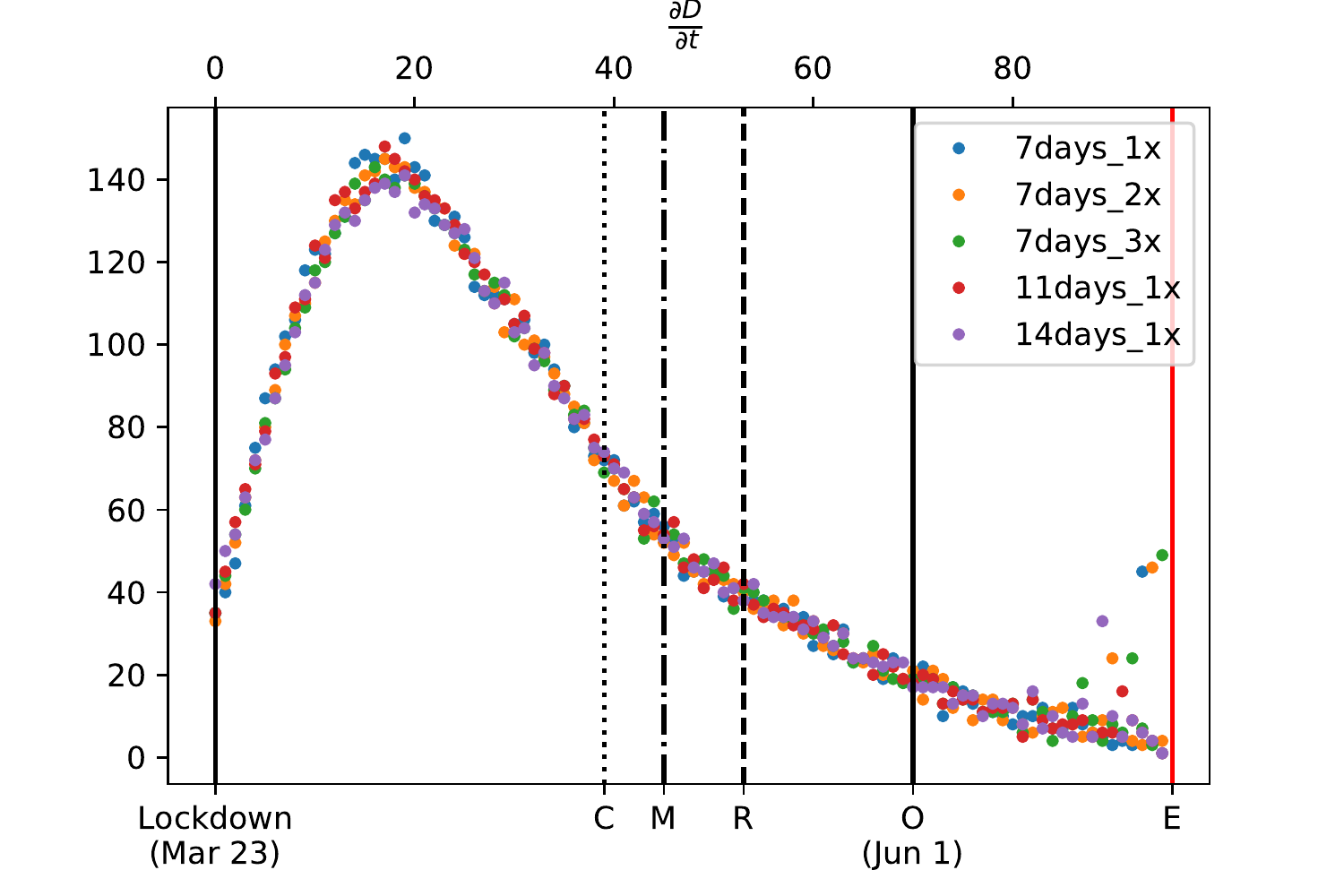}
    \caption{Time derivatives (daily change) of sub-population data with different kernel widths and multiples of application of the smoothing filter. Important dates are marked with the lockdown on March 23, reopening of construction and real estate sites (C) on May 1, reopening of manufacturing sites (M) on May 7, permission. to restart laboratory research (R) on May 15, lifting of stay-at-home order (O) on June 1, and the end of our data collection (E) on June 28.}
    \label{fig:smoothing2}
\end{figure}
Counts of new confirmed infected cases $I(t)$ and deaths $D(t)$ were reported in the public domain on a daily basis by the state of Michigan for each county \cite{CasesDeathsMI}, while total recovered cases in the state $R(t)$ were reported weekly \cite{RecoveredMI}. See Figure \ref{fig:Michmap} for the counties and regions that Michigan is partitioned into. Since county specific recovery data was not reported, the distribution of recovered cases across counties was approximated to be the same as the distribution of cumulative infected cases, $\int_0^t I(\tau)\text{d}\tau$ across counties. Estimates for the populations of Michigan's counties \cite{census2019} were used to determine the susceptible population, $S(t)$, from Equation (\ref{eq:N}).

Some amount of data smoothing was necessary, particularly to account for the weekly instead of daily reporting of the number of recovered cases. To compare the effect of the smoothing method on the data, a moving average filter was applied using 7, 11, and 15-day windows, guided by the week-long period of oscillation in the raw data for daily new infections $I(t) - I(t-1)$. The 7-day window was applied one, two, and three times. As seen in Figure \ref{fig:smoothing1}, the method of smoothing has little effect on the trends of the data. However, and as expected, there is a strong effect on the numerical time derivatives (see Figure \ref{fig:smoothing2}). It is clear that multiple passes of the filter are required to remove jumps in $\text{d}R/\text{d}t$ and $\text{d}I/\text{d}t$. Since the additional smoothing is helpful for system inference in Section \ref{sec:sys_id} and does not negatively affect the data, the 7-day moving average filter applied three times was used for data smoothing.

The lockdown in Michigan began on March 23, 2020. For brevity, we use C for the date when the outdoor construction industry was allowed to resume on May 1, 2020, M for the restart of some manufacturing on May 7, 2020, R for reopening of research laboratories on May 15, 2020, O for broader opening of most other activities and lifting of the stay at home order on June 1, 2020 (albeit with distancing guidelines in place), and E for the end of the data period that we considered (June 28, 2020). This notation is used for the rest of this communication. 

\section{System Identification and ODE-constrained optimization}
\label{sec:sys_id}

The SIRD model, Equations (\ref{eq:S}-\ref{eq:D}) was time-discretized using the Backward Euler method and written as:
\begin{align}
    \frac{S^\text{d}_{m} - S^\text{d}_{m-1}}{\Delta t} +\frac{\beta}{N} S^\text{d}_{m}I^\text{d}_{m}-\gamma R^\text{d}_{m}\label{eq:Sdisc} &=0\\
    \frac{I^\text{d}_{m} - I^\text{d}_{m-1}}{\Delta t} -\frac{\beta}{N}S^\text{d}_{m}I^\text{d}_{m}+\mu I^\text{d}_{m}+\alpha I^\text{d}_{m} &= 0\label{eq:Idisc}\\
    \frac{R^\text{d}_{m} - R^\text{d}_{m-1}}{\Delta t} -\mu I^\text{d}_{m}+\gamma R^\text{d}_{m} &= 0\label{eq:Rdisc}\\
    \frac{D^\text{d}_{m} - D^\text{d}_{m-1}}{\Delta t} - \alpha I^\text{d}_m &= 0\label{eq:Ddisc}
\end{align}
where $S^\text{d}_m, I^\text{d}_m, R^\text{d}_m, D^\text{d}_m$ are the corresponding  data (smoothed as in Section \ref{sec:data}) at time $t_m$ (the end of the $m^\text{th}$ day), $\Delta t = 1$ day and Equation (\ref{eq:N}) holds: $S^\text{d}_m = N - I^\text{d}_m - R^\text{d}_m - D^\text{d}_m$. 

The system identification problem is to infer the time-dependent coefficients $\beta(t), \gamma(t), \mu(t), \alpha(t)$, which we choose to expand in a polynomial basis (other choices of bases are admissible).
\begin{align}
    \beta(t)&=\theta_0+\theta_1t+\theta_2t^2+\theta_3t^3\label{eq:beta}\\
    \gamma(t)&=\theta_4+\theta_5t+\theta_6t^2+\theta_7t^3\label{eq:gamma}\\
    \mu(t)&=\theta_8+\theta_9t+\theta_{10}t^2+\theta_{11}t^3\label{eq:mu}\\
    \alpha(t)&=\theta_{12}+\theta_{13}t+\theta_{14}t^2+\theta_{15}t^3\label{eq:alpha}
\end{align}
The parameters to be inferred are collected into a vector $\boldsymbol{\theta} = \langle \theta_0,\dots,\theta_{15}\rangle^\text{T}$. Since the data are known the label vector can be constructed as:
\begin{equation}
    \boldsymbol{y}_{m} = \left\{\begin{array}{c}
        \frac{S^\text{d}_{m} - S^\text{d}_{m-1}}{\Delta t}   \\
        \frac{I^\text{d}_{m} - I^\text{d}_{m-1}}{\Delta t}\\
        \frac{R^\text{d}_{m} - R^\text{d}_{m-1}}{\Delta t} \\
        \frac{D^\text{d}_{m} - D^\text{d}_{m-1}}{\Delta t}
    \end{array}\right\}
    \label{eq:y}
\end{equation}
and a matrix can be assembled from the reaction terms in the time-discretized SIRD equations (\ref{eq:Sdisc}-\ref{eq:Ddisc}):
\begin{equation}
    \boldsymbol{\Xi}_{m} = \left[\begin{array}{cccc}
        \frac{S^\text{d}_m I^\text{d}_m}{N}\langle 1\;t_m\;t_m^2\;t_m^3\rangle & -R^\text{d}_m\langle 1\;t_m\;t_m^2\;t_m^3\rangle & \langle 0\;0\;0\;0\rangle & \langle 0\;0\;0\;0\rangle\\
         -\frac{S^\text{d}_m I^\text{d}_m}{N}\langle 1\;t_m\;t_m^2\;t_m^3\rangle & \langle 0\;0\;0\;0\rangle & I^\text{d}_m\langle 1\;t_m\;t_m^2\;t_m^3\rangle & I_m\langle 1\;t_m\;t_m^2\;t_m^3\rangle\\
         \langle 0\;0\;0\;0\rangle & R^\text{d}_m\langle 1\;t_m\;t_m^2\;t_m^3\rangle & - I^\text{d}_m\langle 1\;t_m\;t_m^2\;t_m^3\rangle & \langle 0\;0\;0\;0\rangle\\
         \langle 0\;0\;0\;0\rangle & \langle 0\;0\;0\;0\rangle & \langle 0\;0\;0\;0\rangle & -I^\text{d}_m\langle 1\;t_m\;t_m^2\;t_m^3\rangle
    \end{array}\right]
    \label{eq:Xi}
\end{equation}
The columns of $\boldsymbol{\Xi}_m$ can be regarded as discretized versions of the basis operators that appear as reaction terms on the right hand-side of the SIRD model (\ref{eq:S}-\ref{eq:D}). The label vectors and matrices of basis operators at times $t_0,\dots t_M$ are collected into
\begin{equation}
    \boldsymbol{y} = \underbrace{\left\{\begin{array}{c}
         \boldsymbol{y}_0 \\
         \vdots\\
         \boldsymbol{y}_M
    \end{array}\right\}}_{4(M+1)\times 1},\quad \boldsymbol{\Xi} = \underbrace{\left[\begin{array}{c}
         \boldsymbol{\Xi}_0 \\
         \vdots\\
         \boldsymbol{\Xi}_M
    \end{array}\right]}_{4(M+1)\times 16}
\end{equation}
and the residual vector is defined:
\begin{equation}
     \mathscr{R}(\boldsymbol{\theta}) = \boldsymbol{y} - \boldsymbol{\Xi}\boldsymbol{\theta}
     \label{eq:residual}
\end{equation}

Our approach to inference combines system identification by stepwise regression \cite{WangCMAME2019,Wang2020} and ODE-constrained optimization using adjoints. We define a loss function that incorporates penalization on $\boldsymbol{\theta}$ (leading to ridge regression below):
\begin{equation}
    \ell(\boldsymbol{\theta}) = \vert\mathscr{R}(\boldsymbol{\theta}) \vert^2 + \frac{1}{2}\lambda\vert\boldsymbol{\theta}\vert^2
    \label{eq:loss}
\end{equation}
Our stepwise regression techniques incorporate two algorithms listed next:

\noindent\fbox{%
\parbox{0.9\textwidth}{%
\textbf{Algorithm 1: Model selection by Stepwise regression:}
\\
\\
$j = 0$, $Q = 0$,  \\
\texttt{While} $j \le P-1-Q$ \texttt{do}\\
\texttt{Step 0:}\\
\text{\qquad}\texttt{Establish target vector $\boldsymbol{y}$ and matrix of bases $\boldsymbol{\Xi}$.}
\vspace{0.25cm}

\texttt{Step 1:}\\
\text{\qquad}\texttt{Solve for $\boldsymbol{\theta}_j$ by ridge regression:}\\ 
\begin{align}
    \boldsymbol{\theta}_j &=\text{arg}\;\underset{\widetilde{\boldsymbol{\theta}}}\min \text{ } \ell(\widetilde{\boldsymbol{\theta}})\nonumber\\
         &= \left(\boldsymbol{\Xi}^\text{T}\boldsymbol{\Xi} + \lambda\boldsymbol{1} \right)^{-1}\boldsymbol{\Xi}^\text{T}\boldsymbol{y}
    \label{eq:ridgereg}
\end{align}

\text{\qquad}\texttt{Calculate the loss function at this iteration, $\ell_j$.}
\vspace{0.25cm}

\texttt{Step 2:}\\
\text{\qquad}\texttt{Apply the $F$-test introduced below.}\\
\text{\qquad}\texttt{IF $F$-test eliminates an operator}\\
\text{\qquad}\texttt{THEN Set $Q = Q + 1$}\\
\text{\qquad\quad}\texttt{Set to zero the corresponding component of $\boldsymbol{\theta}$.}\\
\text{\qquad\quad}\texttt{GOTO Step 1. \%Loss function remains small (\text{$\ell_{j}\sim \ell_{j-1}$}); solution may be overfit.}\\
\text{\qquad}\texttt{ENDIF}\\
\vspace{0.25cm}

\texttt{Step 3:}\\
\text{\qquad}\texttt{Stop if the $F$-test does not allow elimination of any more basis operators.}\\
\text{\qquad}\texttt{\% Beyond this, the loss function increases dramatically for any further reduction. }
}
}

 There are several possible criteria for eliminating basis terms. Here, we adopt a widely used statistical criterion called the $F$-test, also used by us previously \cite{WangCMAME2019,Wang2020}. The significance of the change between the model at iterations $j$ and $j-1$ is evaluated by:
\begin{align}
F=\frac{ \frac{\ell_j -\ell_{j-1}}{p_{j-1}-p_{j}}}{\frac{l_{j-1}}{P-p_{j-1}}}
\end{align}
where $p_j$ is the number of bases at iteration $j$ and $P = 16$ is the total number of operator bases. The $F$-test is achieved through the following algorithm:
\\
\\
\noindent\fbox{%
\parbox{0.9\textwidth}{%
\textbf{Algorithm 2: Application of the $\boldsymbol{F}$-test:}
\\


\texttt{Step 1: \%Find the least significant basis}\\
$i =0$\\
\texttt{While $i \le P-1-Q$  }
\text{\qquad}\texttt{Tentatively eliminate each basis corresponding to non-zero coefficients in $\boldsymbol{\theta}$}\\
\text{\qquad}\texttt{Set the corresponding coefficient to zero in $\boldsymbol{\theta}$.}\\
\text{\qquad}\texttt{Evaluate the loss function followed by ridge regression on the reduced basis set.}\\
\text{\qquad}\texttt{Compute the $F$-value on the reduced basis set with smallest loss function. Label this coefficient $\theta_k$}
\vspace{0.25cm}

\texttt{Step 2:}\\
\text{\qquad}\texttt{IF $F<\alpha$}\\
\text{\qquad}\texttt{THEN formally eliminate the term corresponding to $\theta_k$  in matrix $\boldsymbol{\Xi}$, by deleting the corresponding column.} \\
\text{\qquad}\texttt{ENDIF}
}
}
\\    
    
    Model selection thus finds $\boldsymbol{\theta}$ consisting of a minimal set of non-zero components, ensuring that the coefficients $\beta(t),\dots,\alpha(t)$ admit a parsimonious representation as polynomials in $t$. For clarity, we collect this set of \emph{non-zero} coefficients into another vector, $\boldsymbol{\vartheta}_0$. Using $\text{dim}(\bullet)$ to represent the dimension of a Euclidean vector, we have $\text{dim}(\boldsymbol{\vartheta}_0) \le \text{dim}(\boldsymbol{\theta})$.
    
    The next step is to further refine the values of the non-zero polynomial coefficients using ODE-constrained optimization starting from the initial guess $\boldsymbol{\vartheta}_0$, and regarding $S_m(\widetilde{\boldsymbol{\vartheta}}),I_m(\widetilde{\boldsymbol{\vartheta}}),R_m(\widetilde{\boldsymbol{\vartheta}}),I_m(\widetilde{\boldsymbol{\vartheta}})$ as the forward  solution to the discretized SIRD model (\ref{eq:Sdiscm}-\ref{eq:Ddiscm}) with coefficient $\beta(t),\dots,\alpha(t)$ values drawn from $\widetilde{\boldsymbol{\vartheta}}$:
\begin{equation}
    \boldsymbol{\vartheta} = \text{arg}\;\underset{\widetilde{\boldsymbol{\vartheta}}}\min \quad\sum_{m=0}^M \left(\frac{S_m(\widetilde{\boldsymbol{\vartheta}})-S^\text{d}_m}{W_1}\right)^2+\left(\frac{I_m(\widetilde{\boldsymbol{\vartheta}})-I^\text{d}_m}{W_2}\right)^2+\left(\frac{R_m(\widetilde{\boldsymbol{\vartheta}})-R^\text{d}_m}{W_3}\right)^2+\left(\frac{D_m(\widetilde{\boldsymbol{\vartheta}})-D^\text{d}_m}{W_4}\right)^2 \label{eq:optimloss}
\end{equation}    
Subject to the discretized SIRD model:  
\begin{align}
\forall\; m &\in \{0,\dots,M\}\nonumber \\
    \frac{S_{m}(\widetilde{\boldsymbol{\vartheta}}) - S_{m-1}(\widetilde{\boldsymbol{\vartheta}})}{\Delta t} +\frac{\beta}{N} S_{m}(\widetilde{\boldsymbol{\vartheta}})I_{m}(\widetilde{\boldsymbol{\vartheta}})-\gamma R_{m}(\widetilde{\boldsymbol{\vartheta}})\label{eq:Sdiscm} &=0\\
    \frac{I_{m}(\widetilde{\boldsymbol{\vartheta}}) - I_{m-1}(\widetilde{\boldsymbol{\vartheta}})}{\Delta t} -\frac{\beta}{N}S_{m}(\widetilde{\boldsymbol{\vartheta}})I_{m}(\widetilde{\boldsymbol{\vartheta}})+\mu I_{m}(\widetilde{\boldsymbol{\vartheta}})+\alpha I_{m}(\widetilde{\boldsymbol{\vartheta}}) &= 0\label{eq:Idiscm}\\
    \frac{R_{m}(\widetilde{\boldsymbol{\vartheta}}) - R_{m-1}(\widetilde{\boldsymbol{\vartheta}})}{\Delta t} -\mu I_{m}(\widetilde{\boldsymbol{\vartheta}})+\gamma R_{m}(\widetilde{\boldsymbol{\vartheta}}) &= 0\label{eq:Rdiscm}\\
    \frac{D_{m}(\widetilde{\boldsymbol{\vartheta}}) - D_{m-1}(\widetilde{\boldsymbol{\vartheta}})}{\Delta t} - \alpha I_m(\widetilde{\boldsymbol{\vartheta}}) &= 0\label{eq:Ddiscm}
\end{align}
where 
\begin{align*}
    W_1 &= \underset{m}\max\, S^\text{d}_m - \underset{m}\min\, S^\text{d}_m \\
    W_2 &= \underset{m}\max\, I^\text{d}_m - \underset{m}\min\, I^\text{d}_m \\
    W_3 &= \underset{m}\max\, R^\text{d}_m - \underset{m}\min\, R^\text{d}_m \\
    W_4 &= \underset{m}\max\, D^\text{d}_m - \underset{m}\min\, D^\text{d}_m 
\end{align*}

The ODE-constrained optimization problem is solved iteratively, and requires the gradient of the ODE constraint (\ref{eq:Sdiscm}-\ref{eq:Ddiscm}) with respect to $\widetilde{\boldsymbol{\vartheta}}$. We adopt the classical approach requiring a single solution of the adjoint equation of the original ODE-constraint in each iteration. In this work we use the L-BFGS-B optimization algorithm from \texttt{SciPy}\cite{2020SciPy} and the \texttt{dolfin-adjoint} software library \cite{dolfin-adjoint} to compute the gradient.

\section{Deep and Bayesian neural networks}
\label{sec:NN}
We also explore multilayer feedforward neural networks (NNs), which are universal function approximators \cite{hornik1989multilayer}, to learn the disease's dynamics via the data $S^\text{d}_m, I^\text{d}_m, R^\text{d}_m, D^\text{d}_m$ at discrete times  and to infer the coefficients in Equations (\ref{eq:S}-\ref{eq:D}), as an alternative to the approach presented in Section \ref{sec:sys_id}. Specifically, we construct two NNs to represent the data, with one as a deterministic model and the other being a probabilistic model. 
Both NNs take $\{I^\text{d}_m, R^\text{d}_m, D^\text{d}_m\, \Delta t \}$ as features and $\{I^\text{d}_{m+k}, R^\text{d}_{m+k}, D^\text{d}_{m+k}\}$ as labels.
Thus, the two NNs  make predictions on case numbers at day $m+k$ based on case numbers reported at day $m$. In this work, $k$ is chosen to vary from 1 to $M-m$, where $M = 97$ is the number of days that we used data for. In both types of NNs, $S^\text{d}_{m}$ and $S^\text{d}_{m+k}$ are computed based on the constraint Equation (\ref{eq:N}).

The deterministic model is a deep neural network (DNN) that consists of multiple fully connected layers, whose model parameters (i.e. weights and bias) can be obtained in a straightforward manner by minimizing the loss function
\begin{equation}
\mathcal{L}_\text{DNN} = \text{MSE}
\end{equation}
through an optimization algorithm, such as stochastic gradient descent, via backpropagation. The probabilistic model is a Bayesian neural network (BNN), which also consists of multiple fully connected layers, but with its model parameters (i.e. weights and bias) being sampled from a posterior distribution $P(\mathbf{\boldsymbol{\theta}} | \mathscr{D})$ that is computed based on Bayes' theorem
\begin{equation}
  P(\mathbf{\boldsymbol{\theta}} | \mathscr{D}) = \frac{P(\mathscr{D}|\mathbf{\boldsymbol{\theta}}) P(\mathbf{\boldsymbol{\theta}})}{P(\mathscr{D})},
  \label{eq:bayes}
\end{equation}
where $\mathscr{D}$ denote the i.i.d.~observations (training data) and $P$ represents the probability density function.
In Equation (\ref{eq:bayes}), $P(\mathscr{D}|\mathbf{\boldsymbol{\theta}})$ is the likelihood, $P(\mathbf{\boldsymbol{\theta}})$ is the prior probability, and $P(\mathscr{D})$ is the evidence, respectively. The posterior distribution of $\mathbf{\boldsymbol{\theta}}$ is computed based on  variational inference (VI), which approximates the exact posterior distribution $P(\mathbf{\boldsymbol{\theta}} | \mathscr{D})$ with a more tractable distribution $Q(\mathbf{\boldsymbol{\theta}})$ by minimizing the Kullback-Leibler (KL) divergence \cite{Liu2016Wang-Stein-variational-gradient-descent,Blei2017Variational-inference-review,Graves2011Varational-inference}
\begin{equation}
  Q^* = \text{arg~min~KL}(Q(\mathbf{\boldsymbol{\theta}})||P(\mathbf{\boldsymbol{\theta}} | \mathscr{D})).
  \label{eq:kl-divergence}
\end{equation}
The KL divergence is computed as 
\begin{equation}
  \text{KL}(Q(\mathbf{\boldsymbol{\theta}})||P(\mathbf{\boldsymbol{\theta}} | \mathscr{D})) = \mathbb{E}[\log Q(\mathbf{\boldsymbol{\theta}})] - \mathbb{E}[\log P(\mathbf{\boldsymbol{\theta}}, \mathscr{D})]  + \log P(\mathscr{D}),
  \label{eq:kl-divergence-calculation}
\end{equation}
which requires computing the logarithm of the evidence, $\text{log}P(\mathscr{D})$ in Equation (\ref{eq:bayes}) \cite{Blei2017Variational-inference-review}. Since $P(\mathscr{D})$ is hard to compute, it is challenging to directly evaluate the objective function in Equation (\ref{eq:kl-divergence}). Alternately, we can optimize the evidence lower bound (ELBO) defined as
\begin{equation}
  \text{ELBO}(Q) = \mathbb{E}[\log P(\mathbf{\boldsymbol{\theta}}, \mathscr{D})] - \mathbb{E}[\log Q(\mathbf{\boldsymbol{\theta}})],
  \label{eq:elbo}
\end{equation}
which is equivalent to the KL-divergence up to an additive constant coming from the evidence. Thus, maximizing the ELBO is equivalent to minimizing the KL-divergence. The loss function for the BNN has the following form:
\begin{equation}
  \mathcal{L}_\text{BNN} = \omega_1 \text{MSE} +  \omega_2 \text{ELBO},
  \label{eq:loss-bnn}
\end{equation}
where $\omega_1$ and $\omega_2$ are weighting parameters, with $\omega_1 = 50$ and $\omega_2=1$ being chosen in this work. 
A specific weight perturbation method, known as Flipout \cite{Wen2018GrosseFlipout}, is followed to infer $Q(\mathbf{\boldsymbol{\theta}})$ by minimizing Equation (\ref{eq:loss-bnn}) through mini-batch training via backpropagation with stochastic optimization algorithms. Flipout has been implemented in the \texttt{TensorFlow} Probability Library. The architectures of both NNs are summarized in Table \ref{tab:NNs}. Both NNs were trained by using the Adam optimizer following an exponentially decaying learning rate
\begin{equation}
  \text{lr} = \text{lr}_0 \cdot \text{pow}\left(  v_\text{decay}, \frac{N_\text{total}}{N_\text{decay}}\right)
  \label{eq:lr-step}
\end{equation}
with an initial learning rate $\text{lr}_0 = 0.001$, a decay rate $v_\text{decay} = 0.91$, a decay step $N_\text{decay} = 100$, and a final $N_\text{total} = 10000$ epochs. 

\begin{table}[t!]
  \centering

  \begin{tabular}{l | l }
    \hline
     Layer type & Description \\ \hline \hline
     \multicolumn{2}{c}{DNN}   \\ \hline
    Input layer (features) & $I^\text{d}_m, R^\text{d}_m, D^\text{d}_m\, \Delta t $  \\
    \texttt{Dense} layer & neurons = 40 (Sigmoid) \\
    \texttt{Dense} layer & neurons = 40 (Sigmoid) \\
    Output \texttt{Dense} Layer (labels) & $I^\text{d}_{m+k}, R^\text{d}_{m+k}, D^\text{d}_{m+k}$  (Softplus)  \\ \hline \hline
    \multicolumn{2}{c}{BNN}   \\ \hline
    Input layer (features) & $I^\text{d}_m, R^\text{d}_m, D^\text{d}_m\, \Delta t $  \\
    \texttt{DenseFlipout} layer & neurons = 40 (Sigmoid) \\
    \texttt{DenseFlipout} layer & neurons = 40 (Sigmoid) \\
    Output \texttt{DenseFlipout} Layer (labels) & $I^\text{d}_{m+k}, R^\text{d}_{m+k}, D^\text{d}_{m+k}$ (Softplus)  \\ \hline
  \end{tabular}
  \caption{Model architecture for the DNN and BNN. Note: \texttt{Dense} and \texttt{DenseFlipout} refer to specific NN architectures (layers) used in the \texttt{TensorFlow} Library.}
  \label{tab:NNs}
\end{table}

\section{Results}
\label{sec:results}
Because of the extremely nonuniform distribution of the population of Michigan, we first studied the SIRD model for the entire state consisting of the lower and upper peninsulas (Figure \ref{fig:Michmap}). Following this, the SIRD models were inferred for the eight Regions (also shown in Figure \ref{fig:Michmap}) individually as one direct approach to study the effect of spatial variations in the populations and sub-populations corresponding to the model's compartments.
\label{sec:sysidresults}
\subsection{System identification and ODE-constrained optimization}
\begin{figure}[h!]
    \centering
    \includegraphics[width=0.65\textwidth]{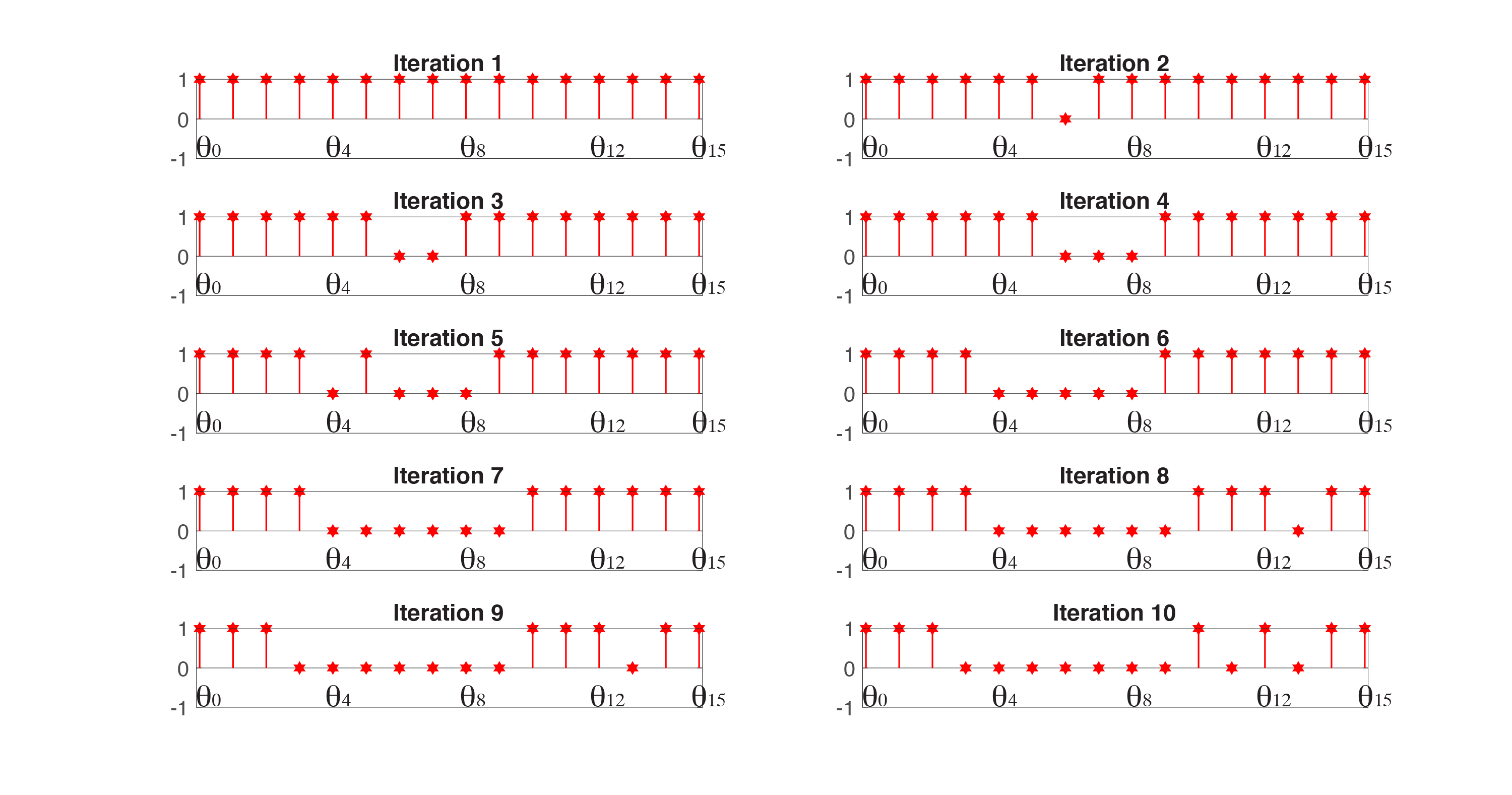}
    \includegraphics[width=0.3\textwidth]{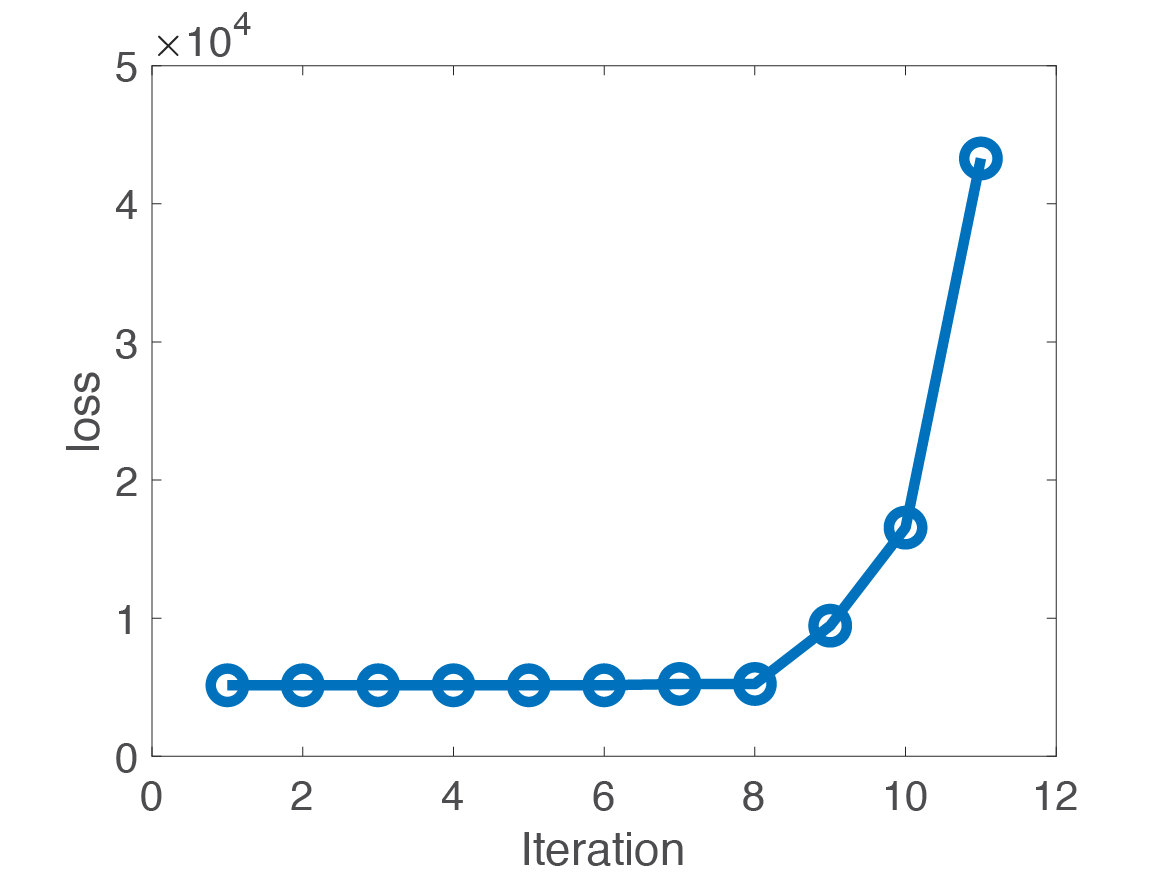}
    \caption{Left: Stem-and-leaf plot illustrating system identification of active time-dependent SIRD parameters using data for the entire state of Michigan. Each stem-and-leaf represents one term of $\theta_0,\dots \theta_{15}t^3$, scaled to 1 (active) or 0 (inactive). Right: The changing loss as terms are eliminated from the set of time-dependent coefficients. System identification converges at Iteration 10 as the loss increases dramatically for further elimination of terms. }
    \label{fig:results-system-id-lp-stem-loss}
\end{figure}

 Figure \ref{fig:results-system-id-lp-stem-loss} shows the progression of stepwise regression to infer the active time-dependent terms in Equations (\ref{eq:beta}-\ref{eq:alpha}) via Algorithms 1 and 2.   The stem-and-leaf plots on the left illustrate the fate of the terms $\theta_0\,-\,\theta_{15}t^3$ over eight iterations of stepwise regression. Each stem-and-leaf represents one term out of $\theta_0\,-\,\theta_{15}t^3$ and the values are scaled to 1 (active) or 0 (inactive) for each iteration. On the right is the loss, which remains low until Iteration 10 and increases dramatically in Iteration 11, if any further terms are eliminated. Following the $F$-test used in Algorithm 2, the large increase in loss after Iteration 10 exceeds the threshold for acceptable model error. Thus system identification converges to the inferred model in ten iterations.
 
\begin{figure}[h!]
    \centering
    \includegraphics[width=0.45\textwidth]{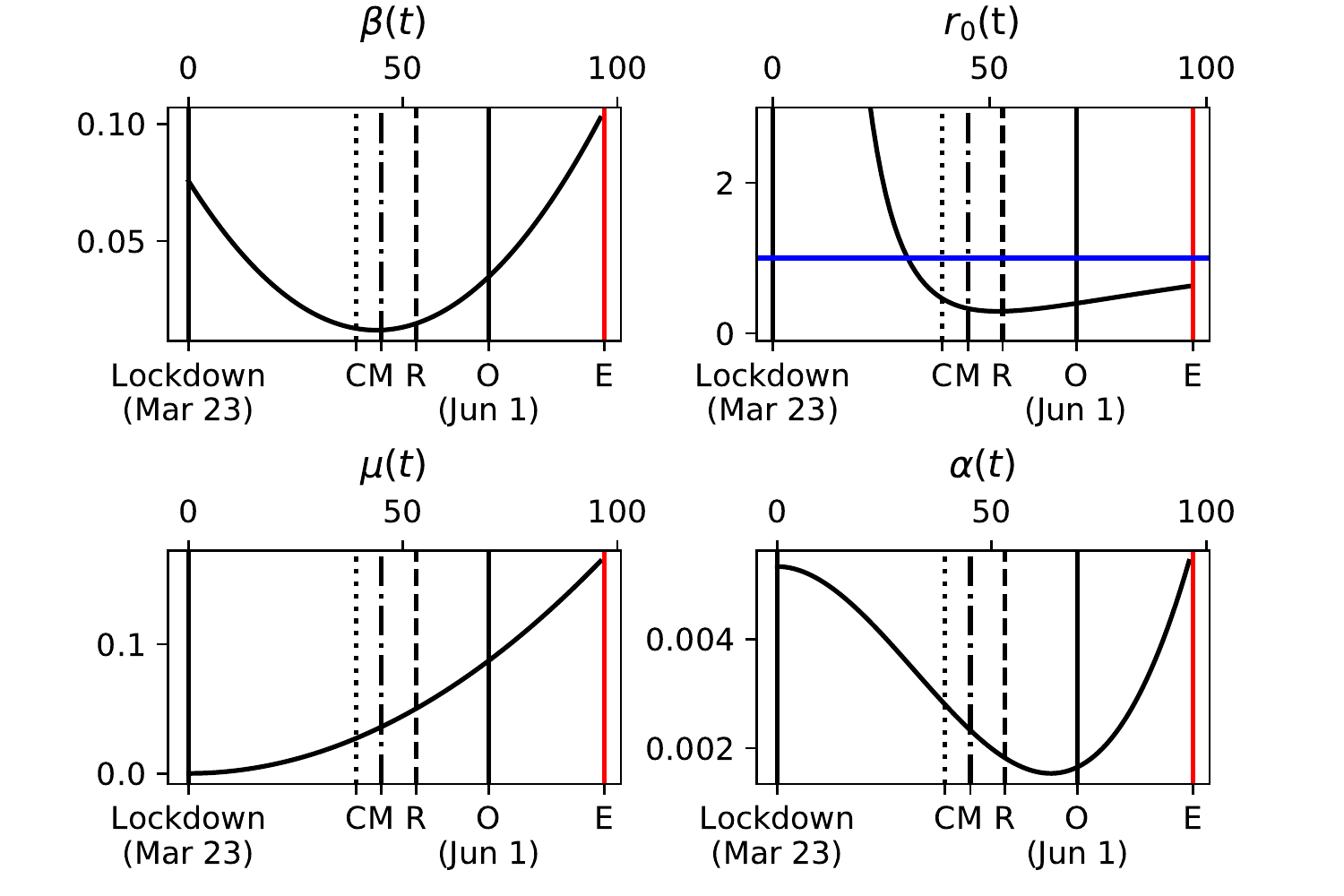}
    \includegraphics[width=0.45\textwidth]{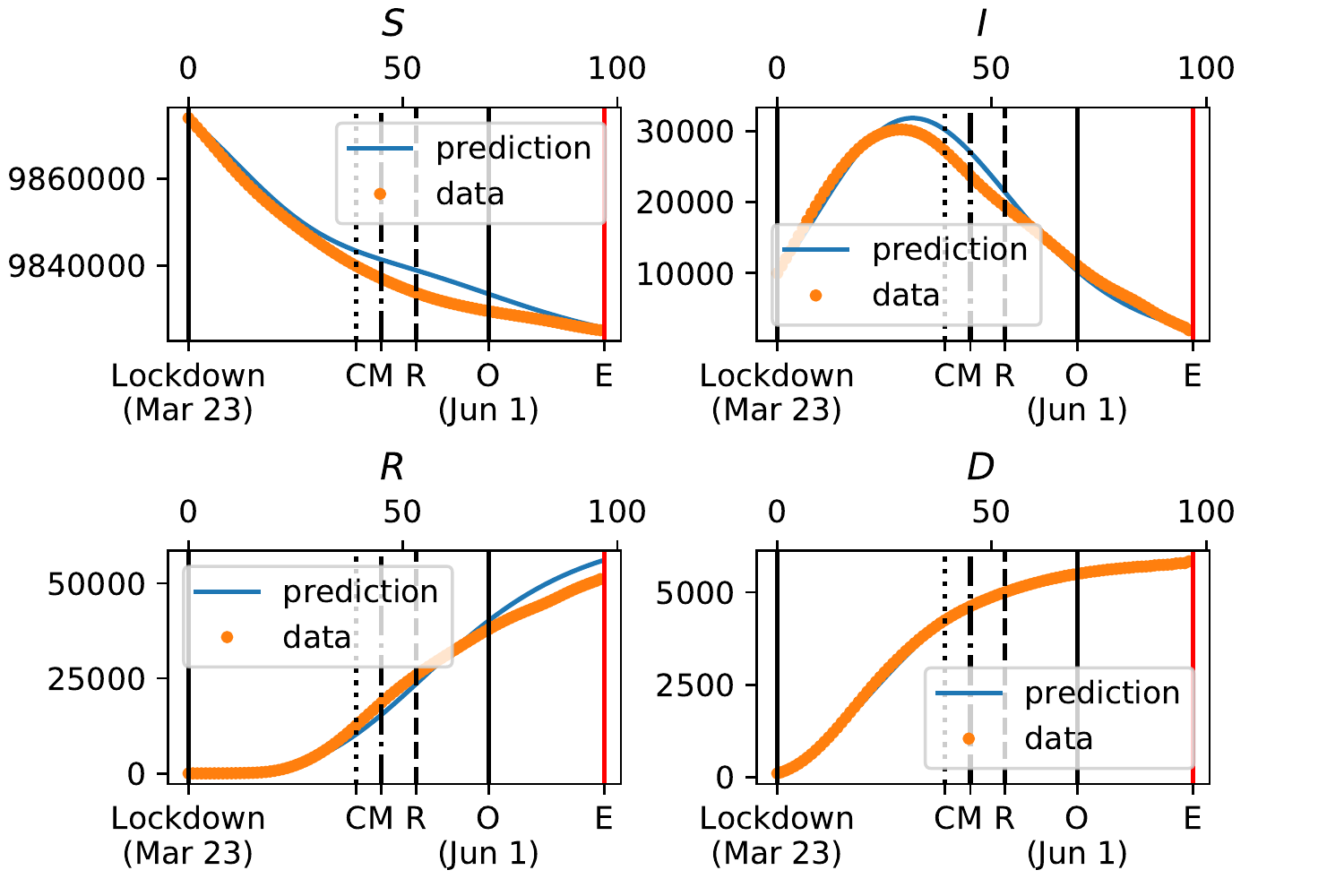}
    \caption{Left: The time-dependent SIRD parameters after tuning by ODE-constrained optiization following. system identification are: $\beta(t)=0.0756-0.0029t+3.33\times10^{-5}t^3,\gamma(t)=0,\mu(t)=1.78\times10^{-5}t^2,\alpha(t)=0.0053-2.8\times10^{-6}t^2+2.93\times10^{-8}t^3 $. Right: Simulation of the four compartments using the inferred ODE SIRD model, in comparison with the data.}
    \label{fig:results-system-id-lp-pred-para}
\end{figure}

Figure \ref{fig:results-system-id-lp-pred-para} shows, on the left, the evolution of SIRD model parameters and, on the right, a comparison of the predictions of the inferred model \emph{versus} the data after ODE-constrained optimization that follows the system identification step. It is important to recall that these results are representative of the population of the entire state of Michigan. 
The SIRD model, having only four compartments, and applied to data that are the outcome of changing characteristics of testing, quarantine and treatment protocols, does not resolve many details of the public health aspects of the epidemic. The immunological characteristics of the disease itself are accounted for only in a very aggregated sense. 

In Figure \ref{fig:results-system-id-lp-pred-para}, the  important dates when the lockdown was imposed, and its gradual lifting are indicated by vertical lines to aid an understanding of the results. We first draw attention to the conclusion that $\gamma(t) = 0$; the inference indicates that recovery from COVID-19 confers permanent immunity--an important conclusion, that remains to be confirmed by immunologists. As may be expected, the population's infection rate, $\beta(t)$, declined as the initially higher rates of positive diagnoses fell with fewer infected individuals. However, it began to rise again upon the opening of construction activities (C), and continued to do so through the lifting of stay at home orders (O). The recovery rate, $\mu(t)$, showed a long initial increase as growing numbers of infected individuals recovered. Our interpretation of the initially high death rate, $\alpha(t)$, is that many of the early cases already had advanced progression of the disease. Its rapid decline can be attributed to the ramp up of the public health campaign, hospitalization and emergency response of the medical system. The success that the state of Michigan gained by mandating an aggressive lockdown of nearly all societal, educational, commercial and industrial activity is best reflected in the rapid decline of the effective reproduction rate, $r_0(t)$. According to the inference presented here, $r_0(t) < 1.0$ for $m > 32$ (April 24,  2020), after which the typical infected individual passed the disease on to less than one other person. The death rate increased over the last few days for which data were obtained, perhaps as some number of individuals who had been infected for a longer time failed to recover. This affected the recovery rate as well, which fell. The close match between the simulations with the inferred ODE SIRD model and data (Figure \ref{fig:results-system-id-lp-pred-para}, right plot) validate the systems inferred. Such validation against the data holds for all the inferred results presented in this communication, although the non-uniqueness of inverse problems does not preclude the existence of multiple sets of inferred coefficients.

\begin{figure}[p!]
    \centering
    \subfigure[Region 1]{\includegraphics[width=0.35\textwidth]{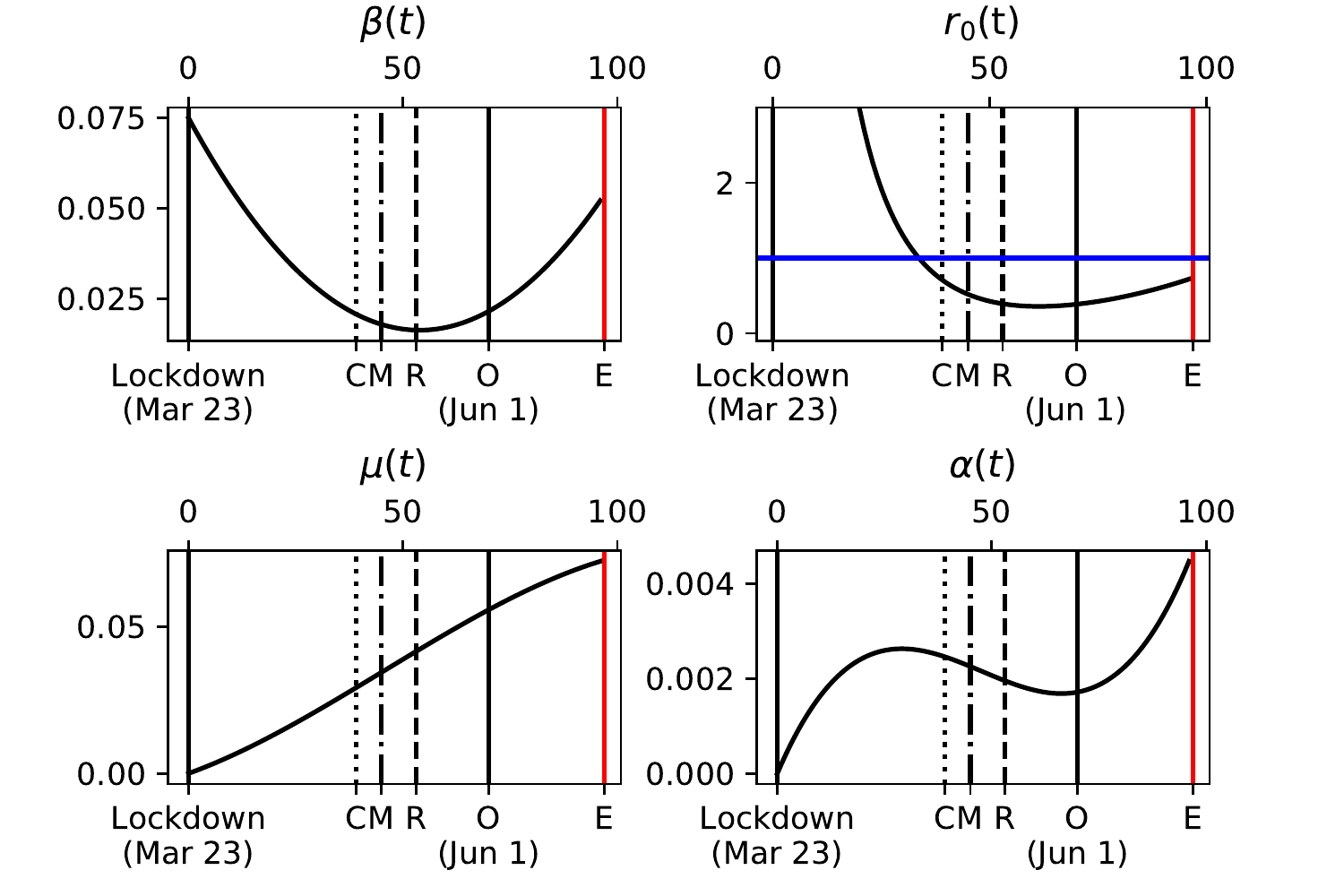}}
    \subfigure[Region 2]{\includegraphics[width=0.35\textwidth]{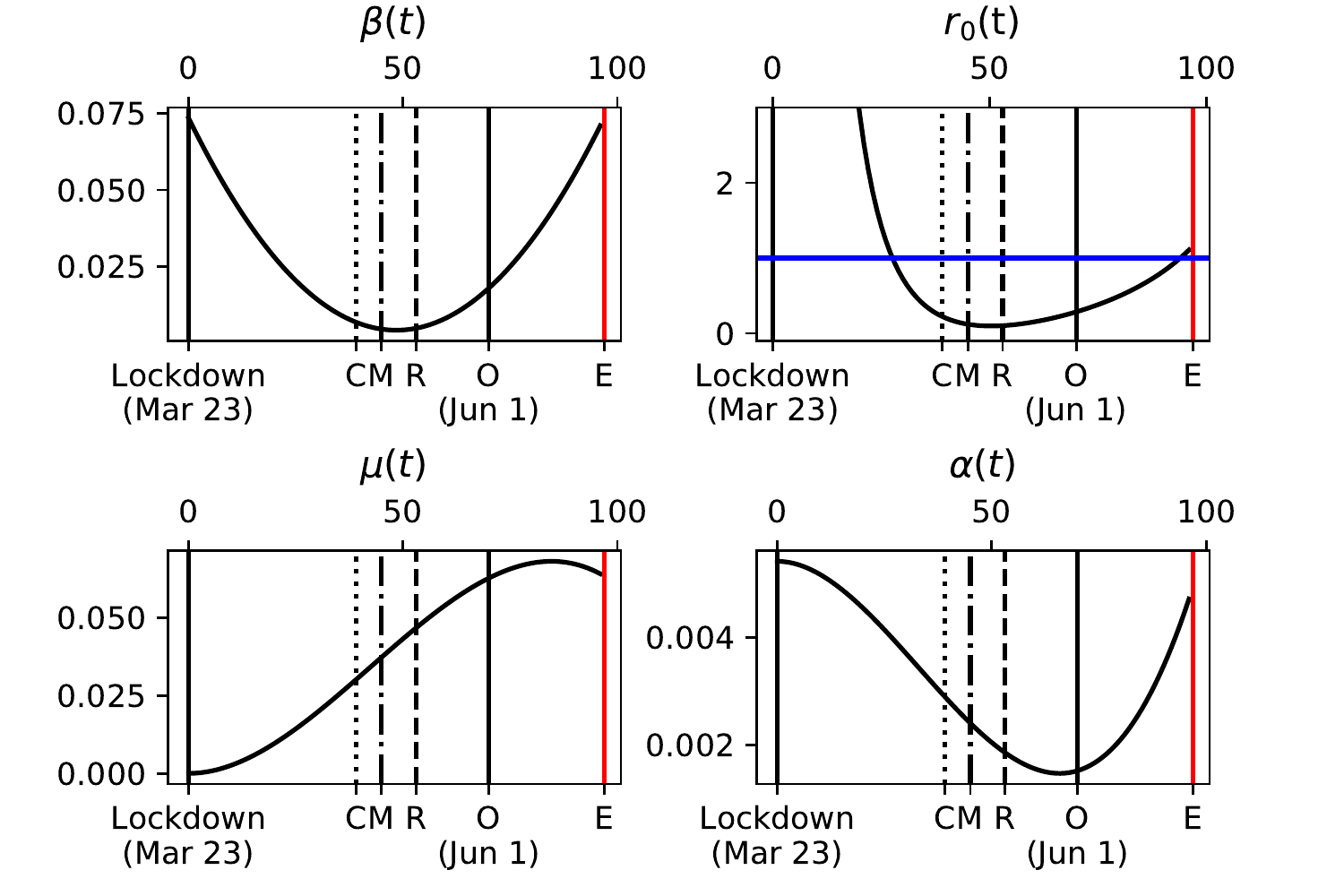}}
    \subfigure[Region 3]{\includegraphics[width=0.35\textwidth]{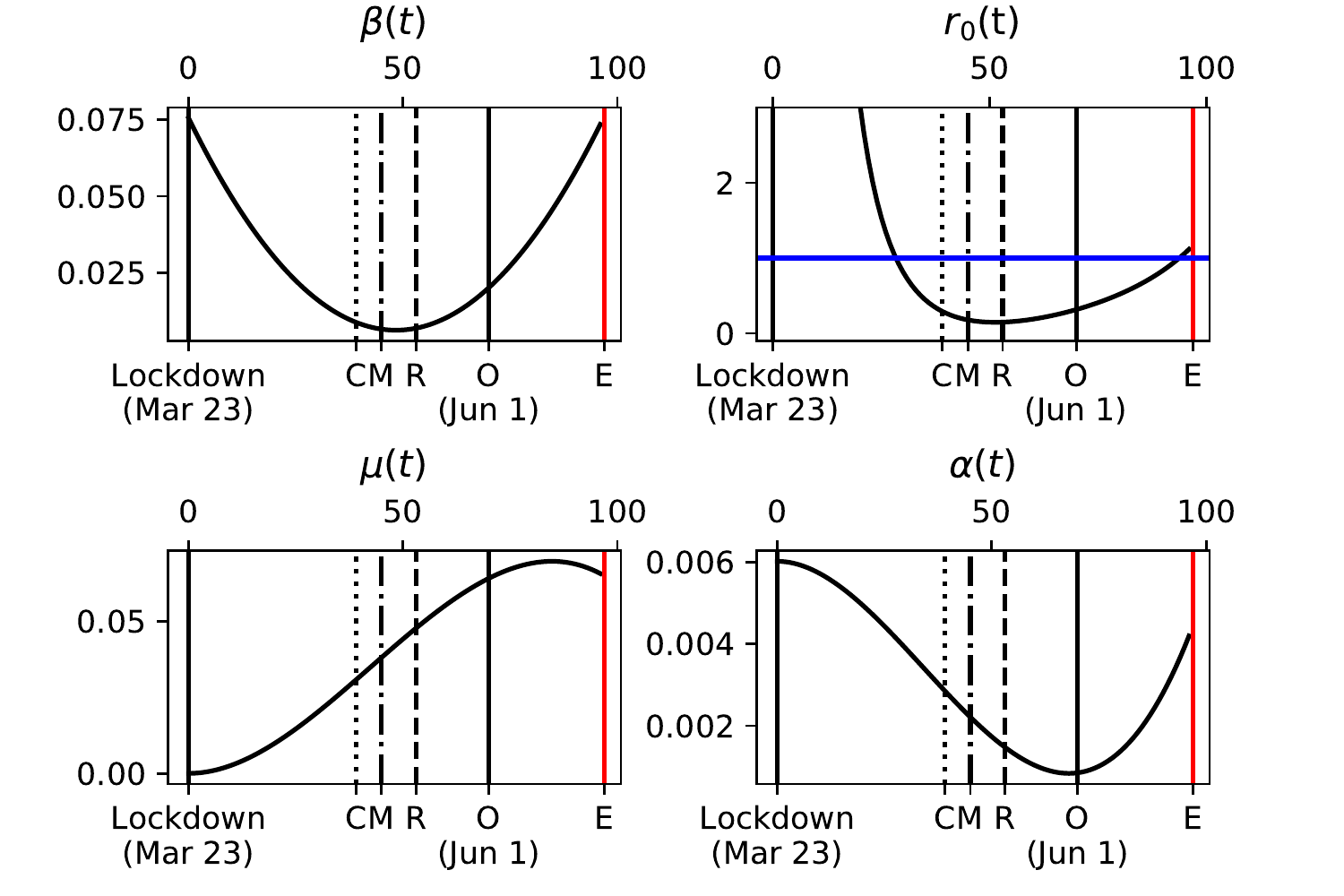}}
    \subfigure[Region 4]{\includegraphics[width=0.35\textwidth]{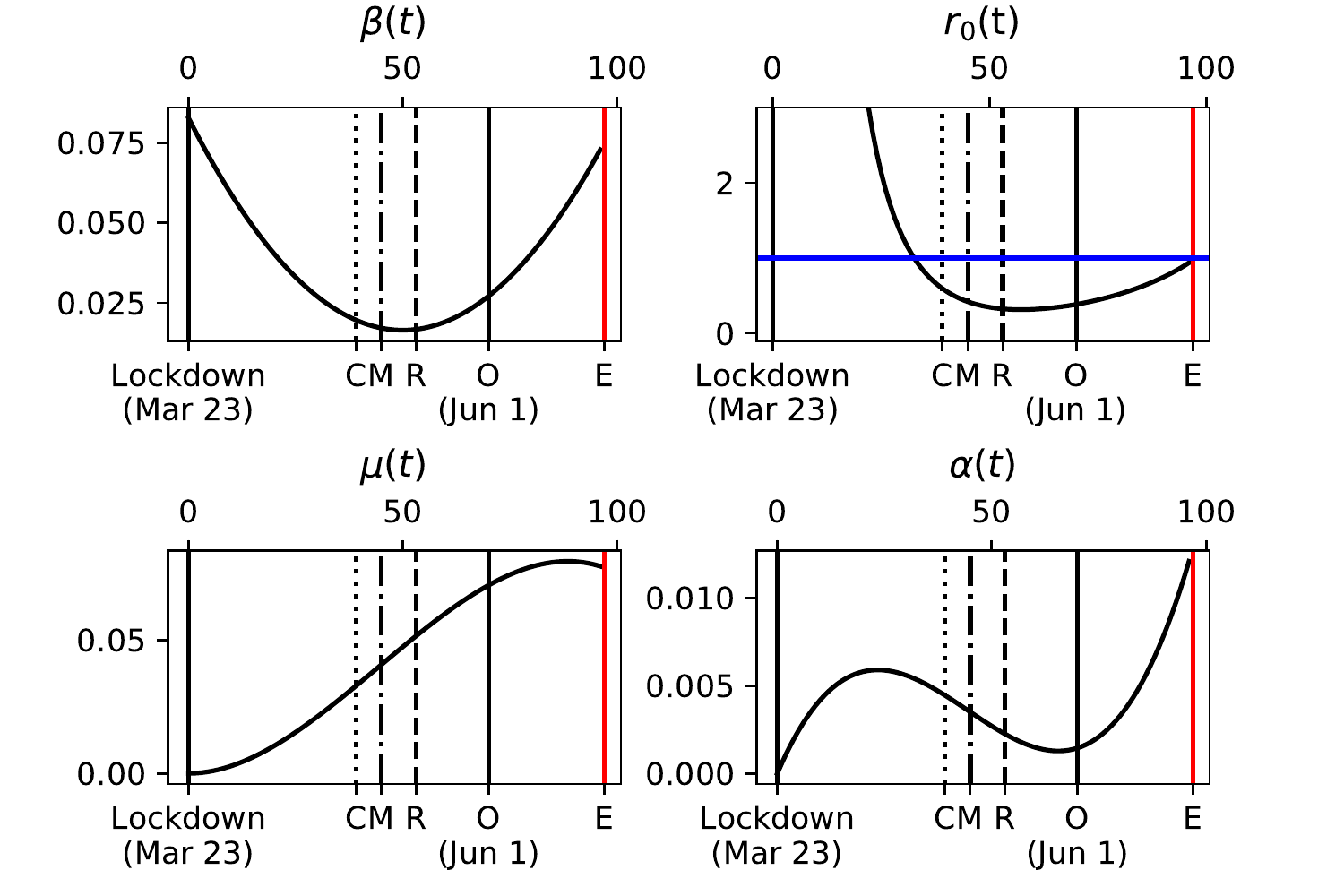}}
    \subfigure[Region 5]{\includegraphics[width=0.35\textwidth]{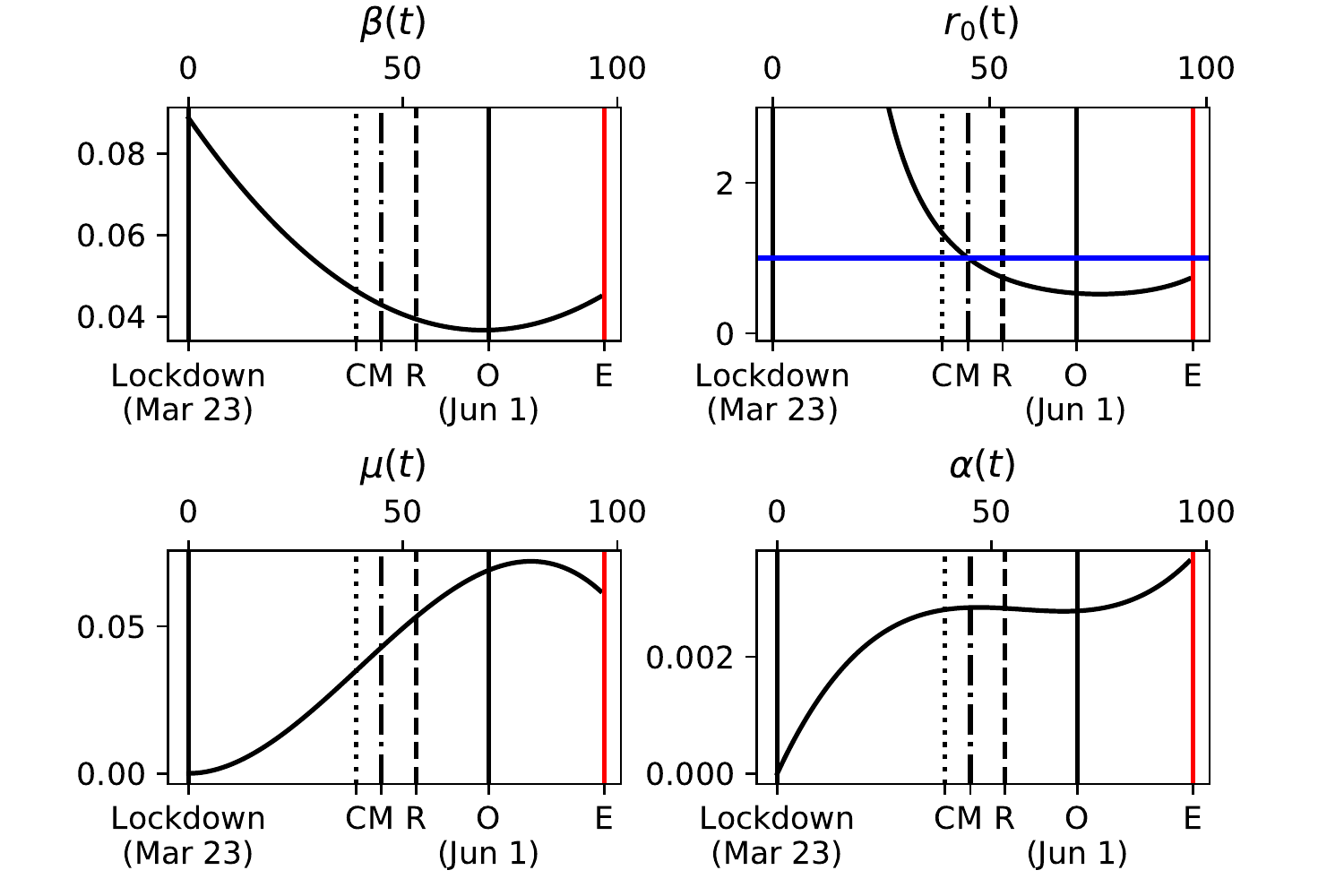}}
    \subfigure[Region 6]{\includegraphics[width=0.35\textwidth]{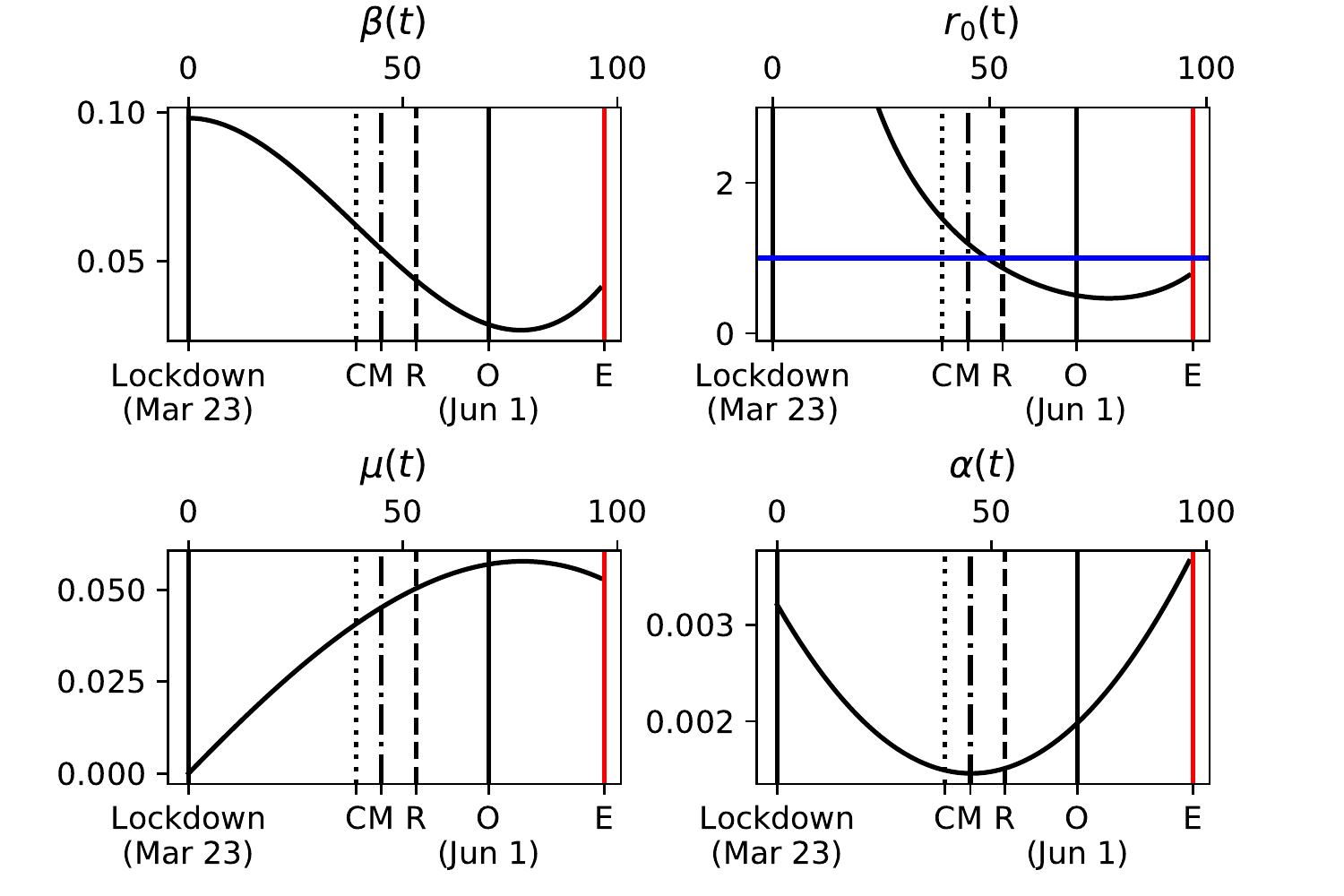}}
    \subfigure[Region 7]{\includegraphics[width=0.35\textwidth]{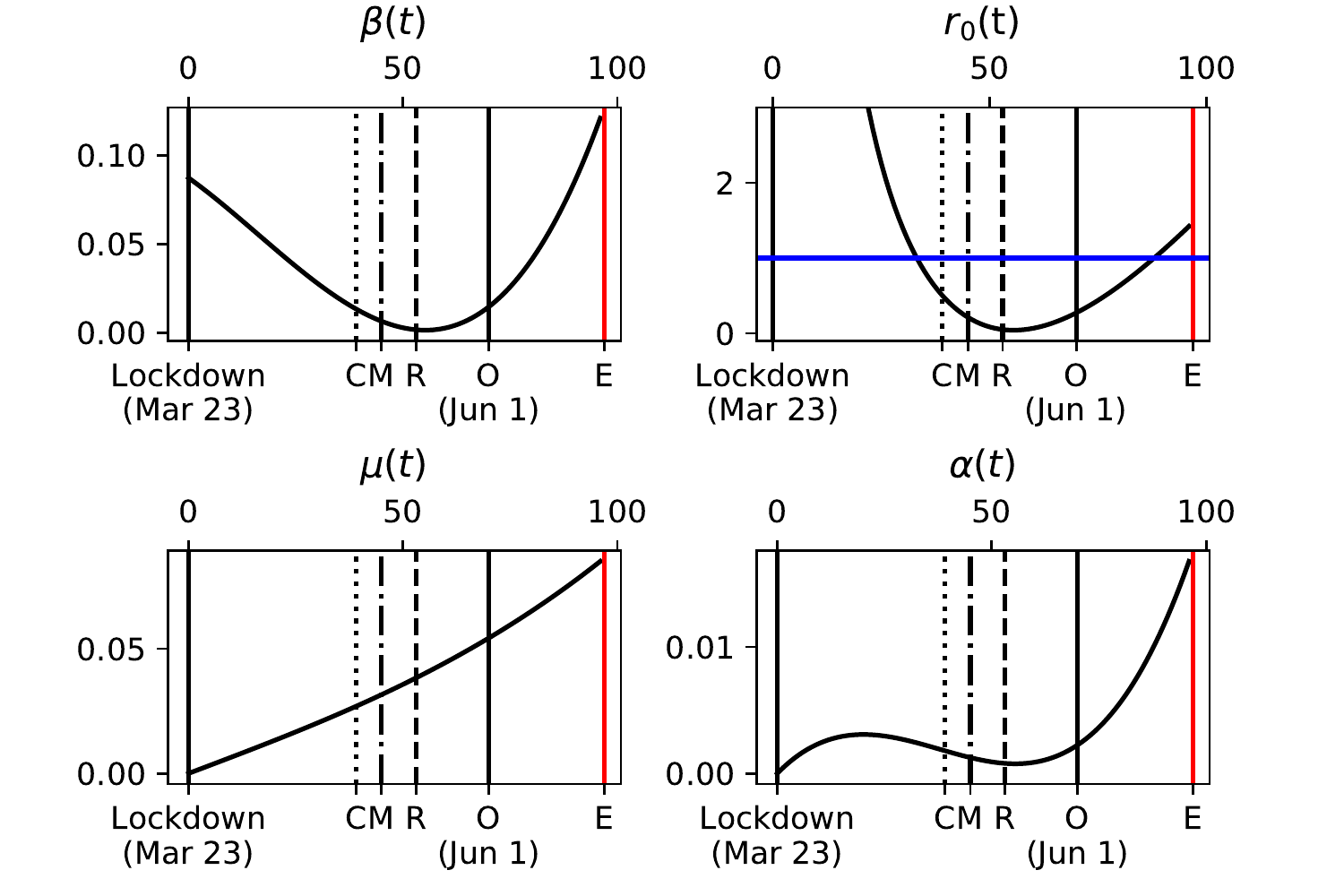}}
    \subfigure[Region 8]{\includegraphics[width=0.35\textwidth]{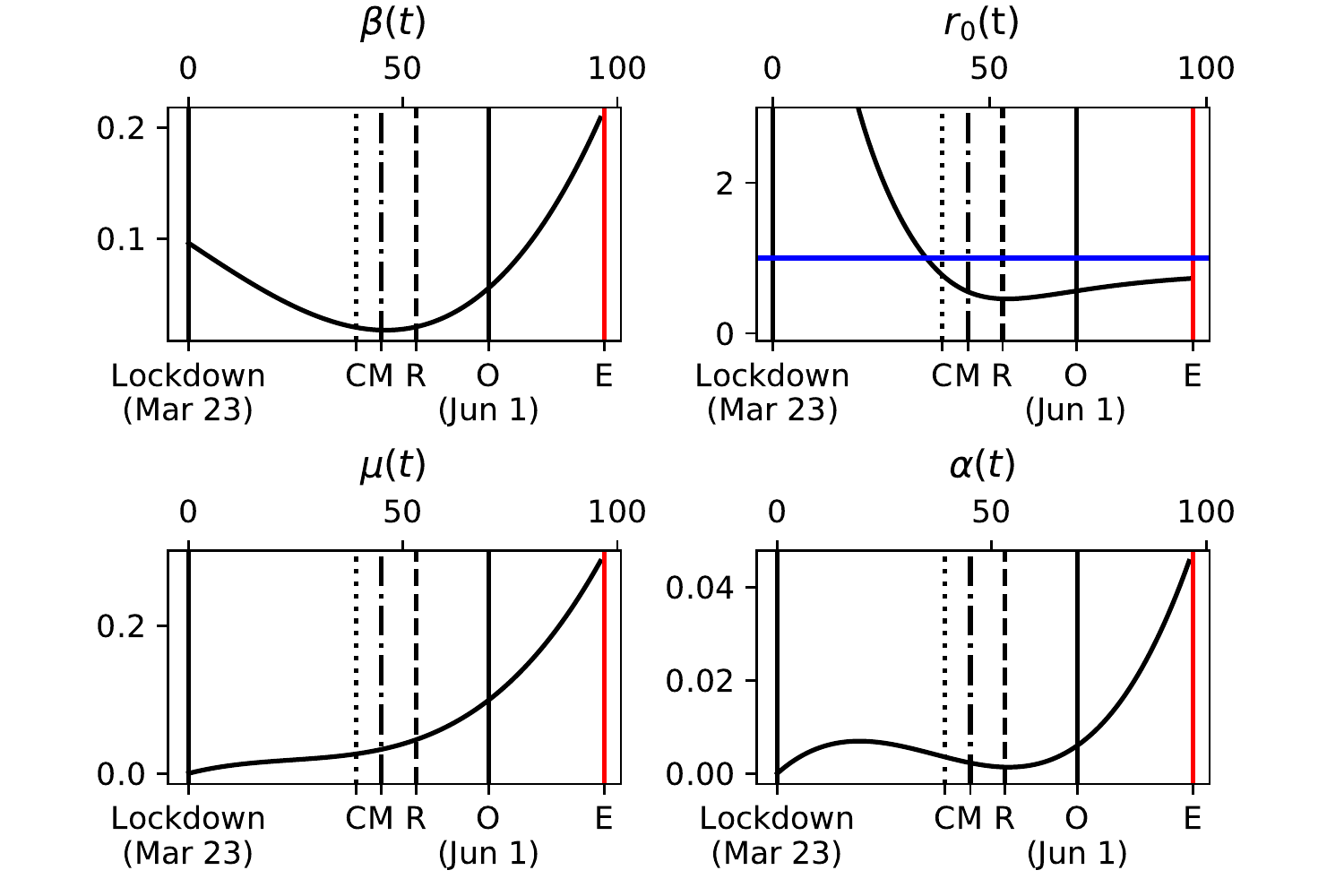}}
    \caption{Parameters of time-dependent SIRD coefficients, $\beta(t), \mu(t), \alpha(t)$, and the effective reproduction rate, $r_0(t)$, for Regions 1-8 (see Figure \ref{fig:Michmap}) of Michigan. }
    \label{fig:paramregions}
\end{figure}

\begin{figure}[p!]
    \centering
    \subfigure[Region 1]{\includegraphics[width=0.35\textwidth]{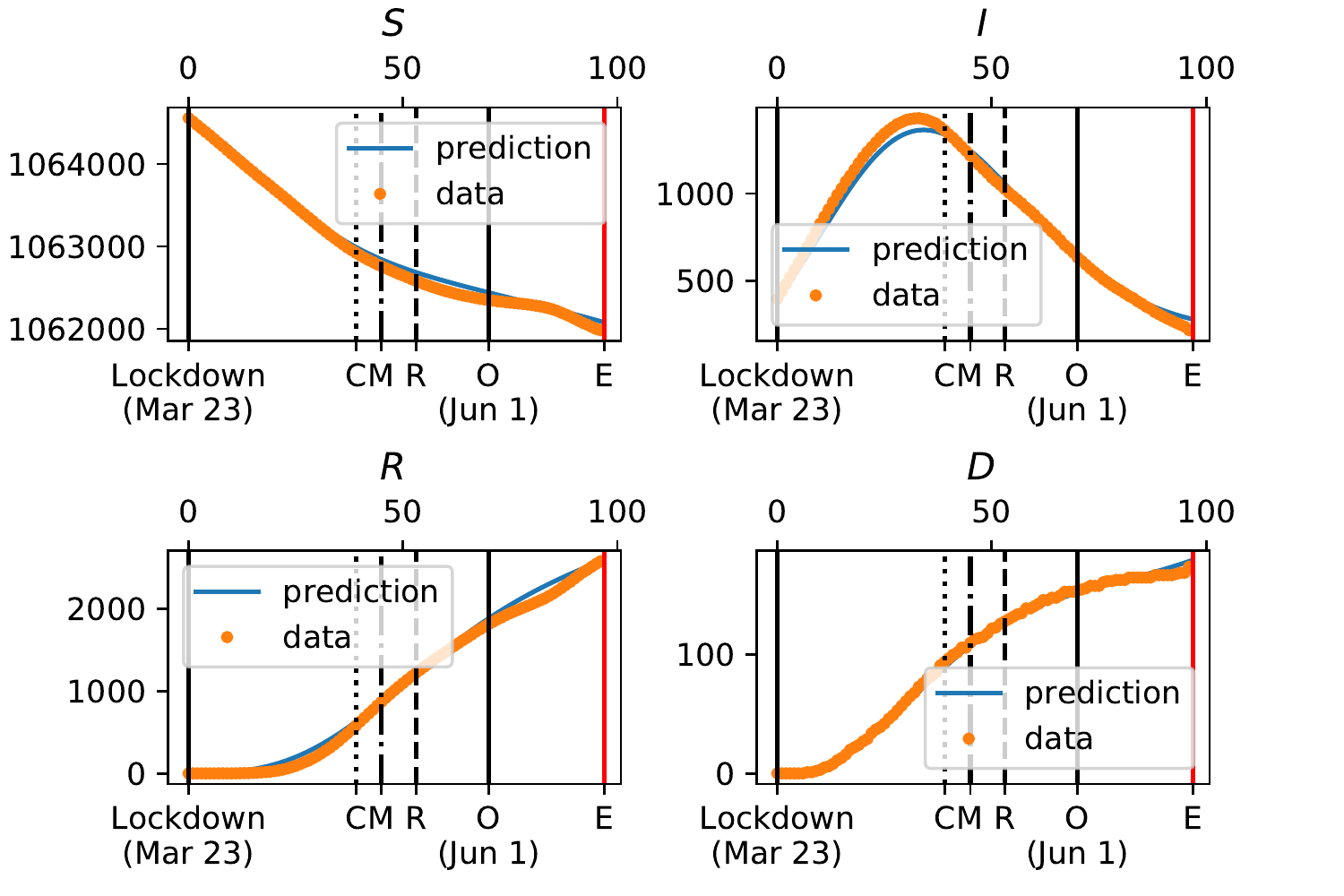}}
    \subfigure[Region 2]{\includegraphics[width=0.35\textwidth]{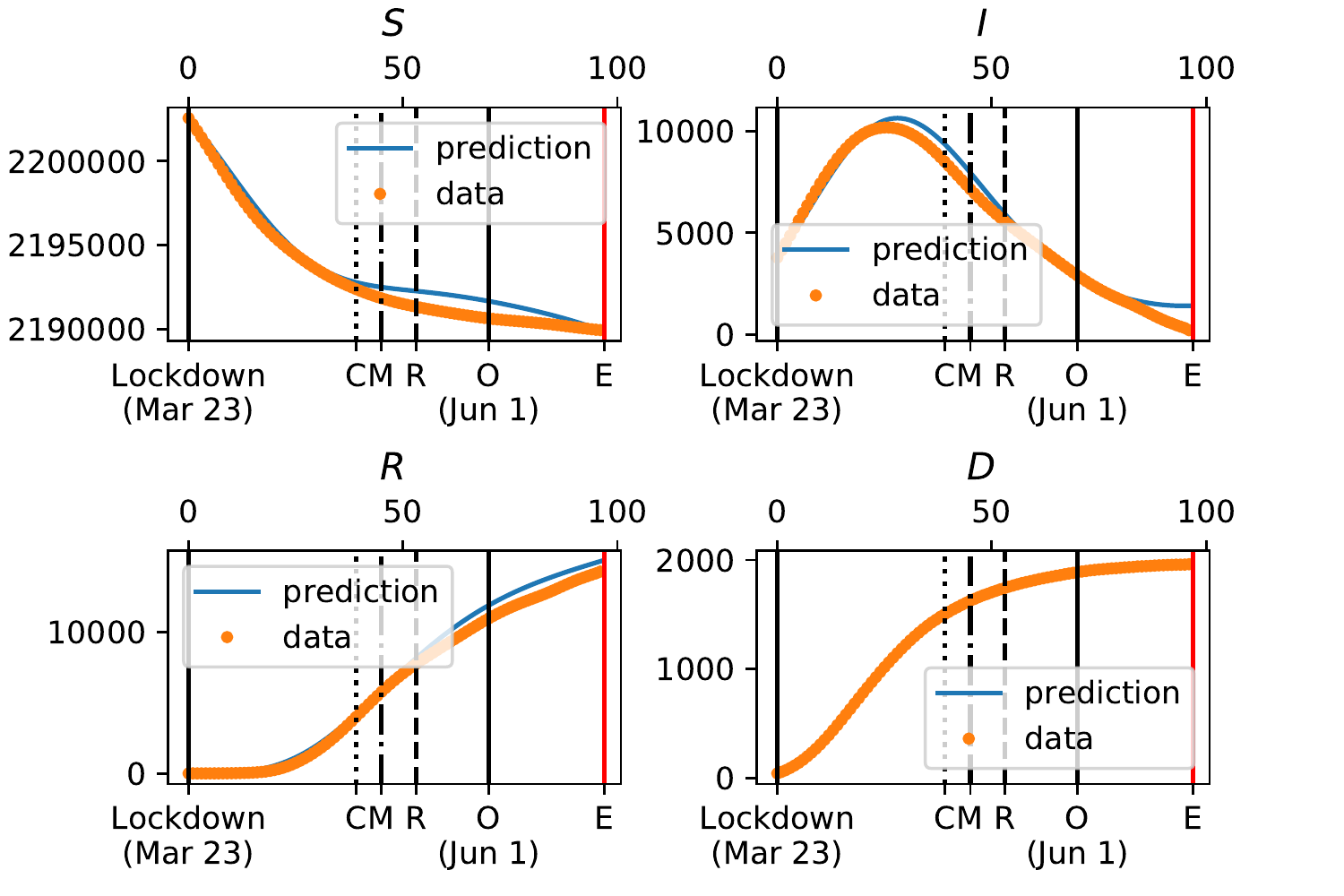}}
    \subfigure[Region 3]{\includegraphics[width=0.35\textwidth]{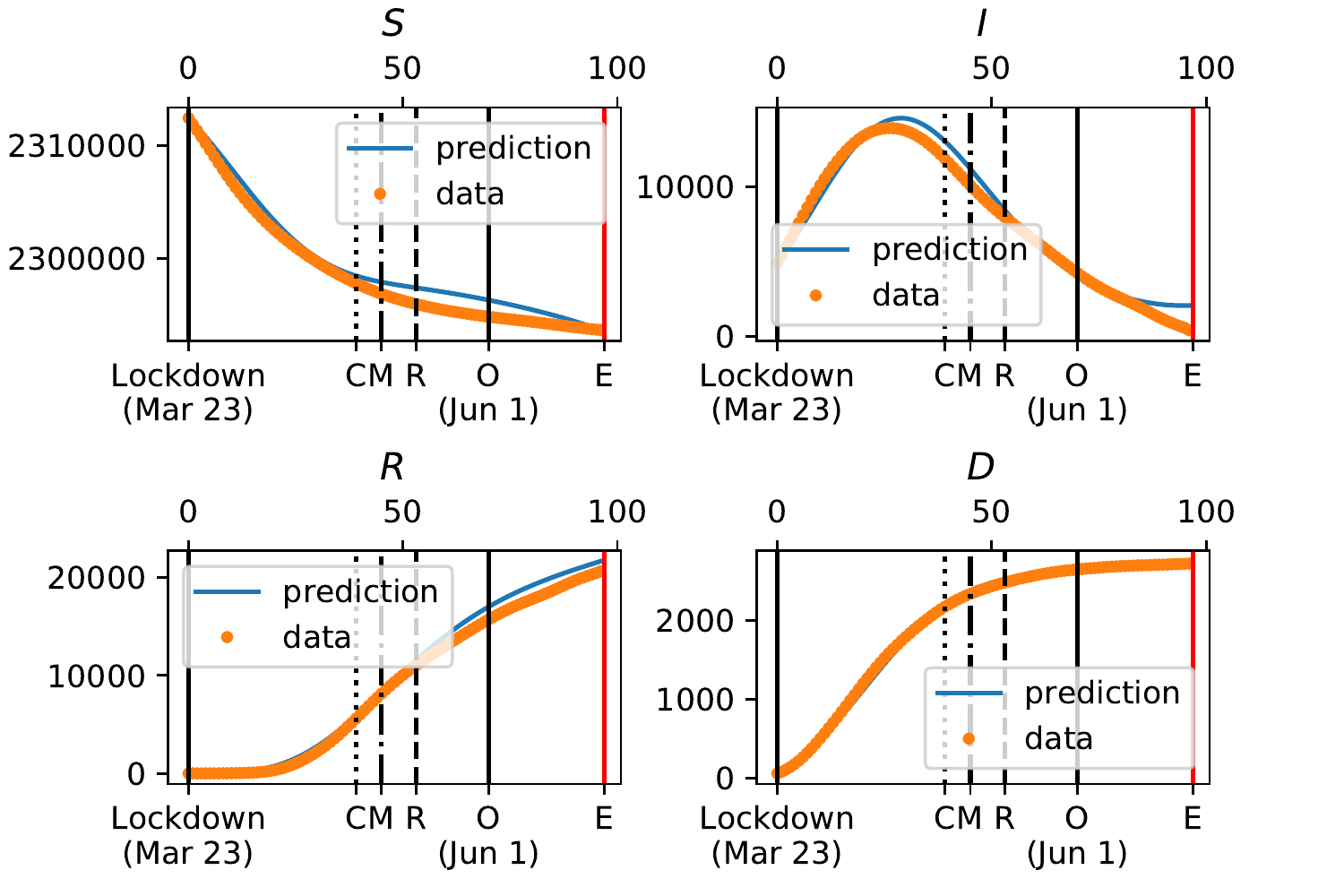}}
    \subfigure[Region 4]{\includegraphics[width=0.35\textwidth]{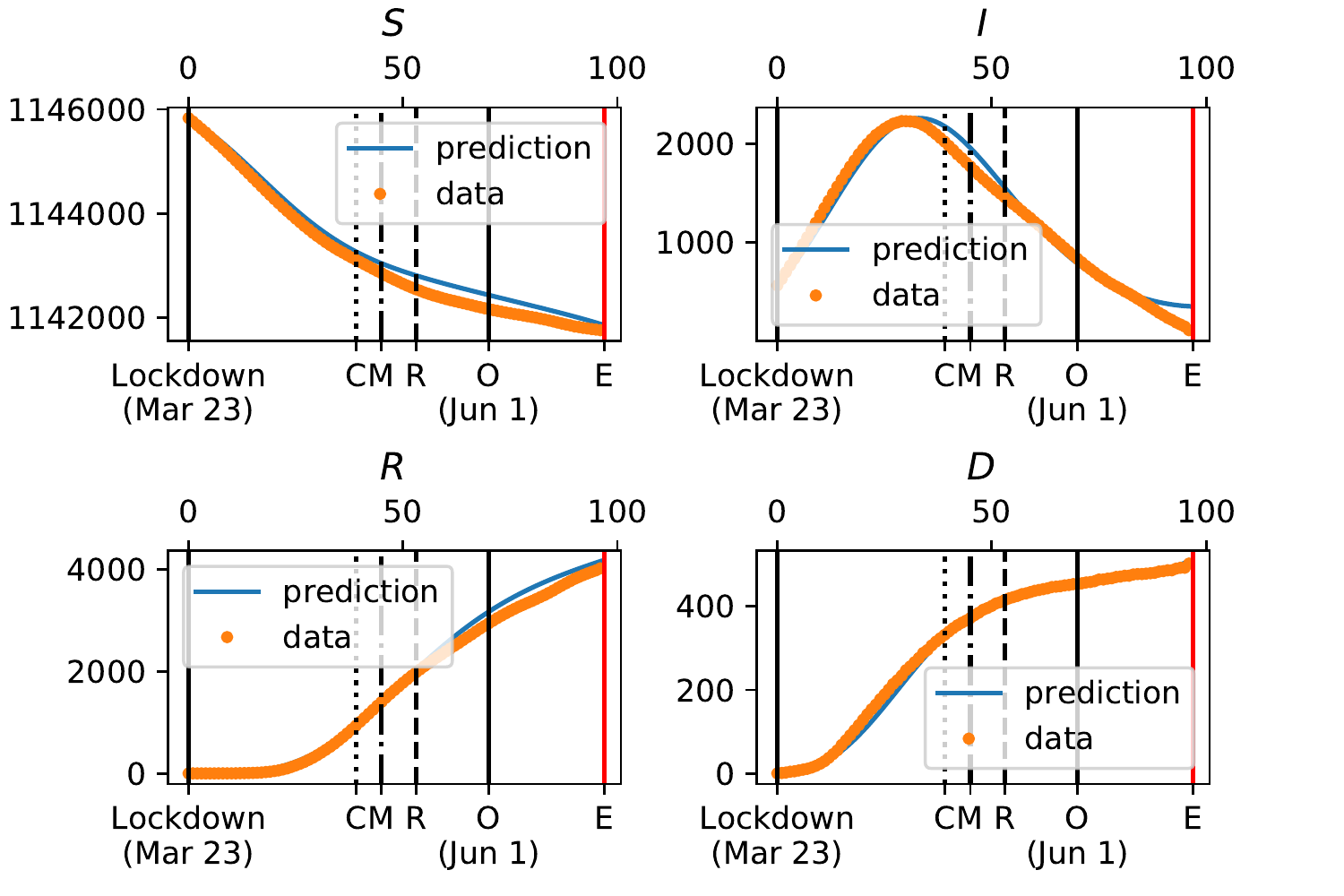}}
    \subfigure[Region 5]{\includegraphics[width=0.35\textwidth]{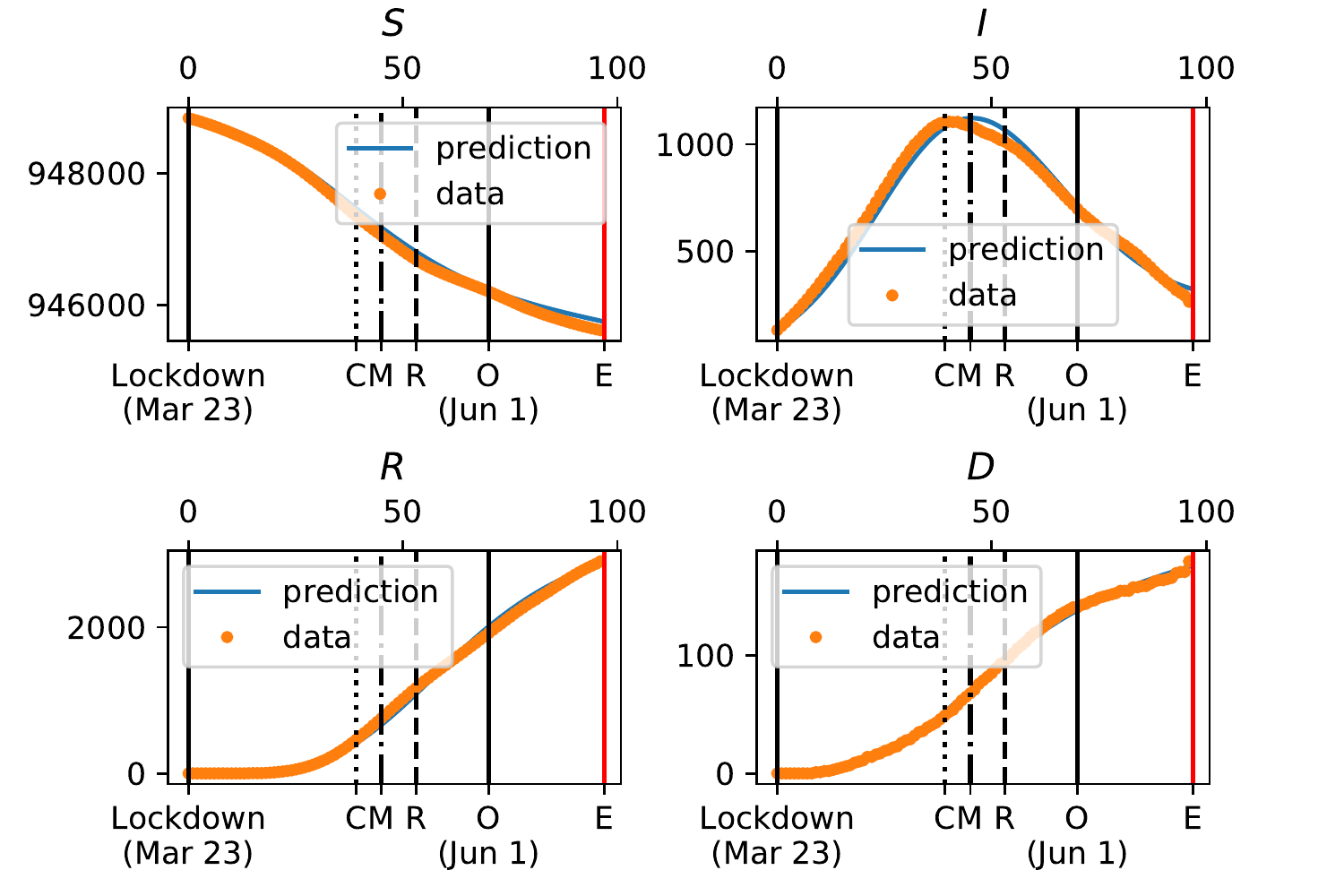}}
    \subfigure[Region 6]{\includegraphics[width=0.35\textwidth]{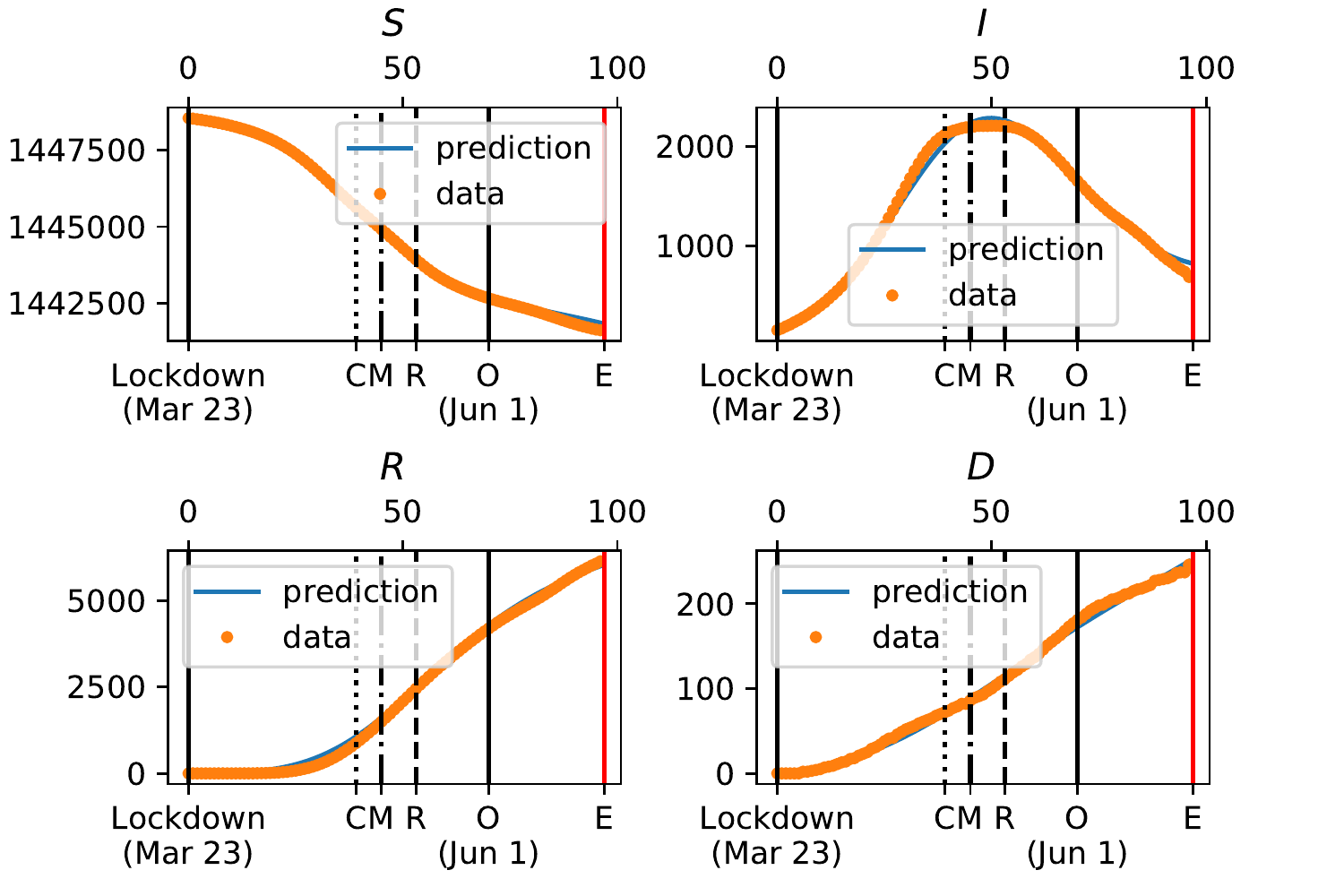}}
    \subfigure[Region 7]{\includegraphics[width=0.35\textwidth]{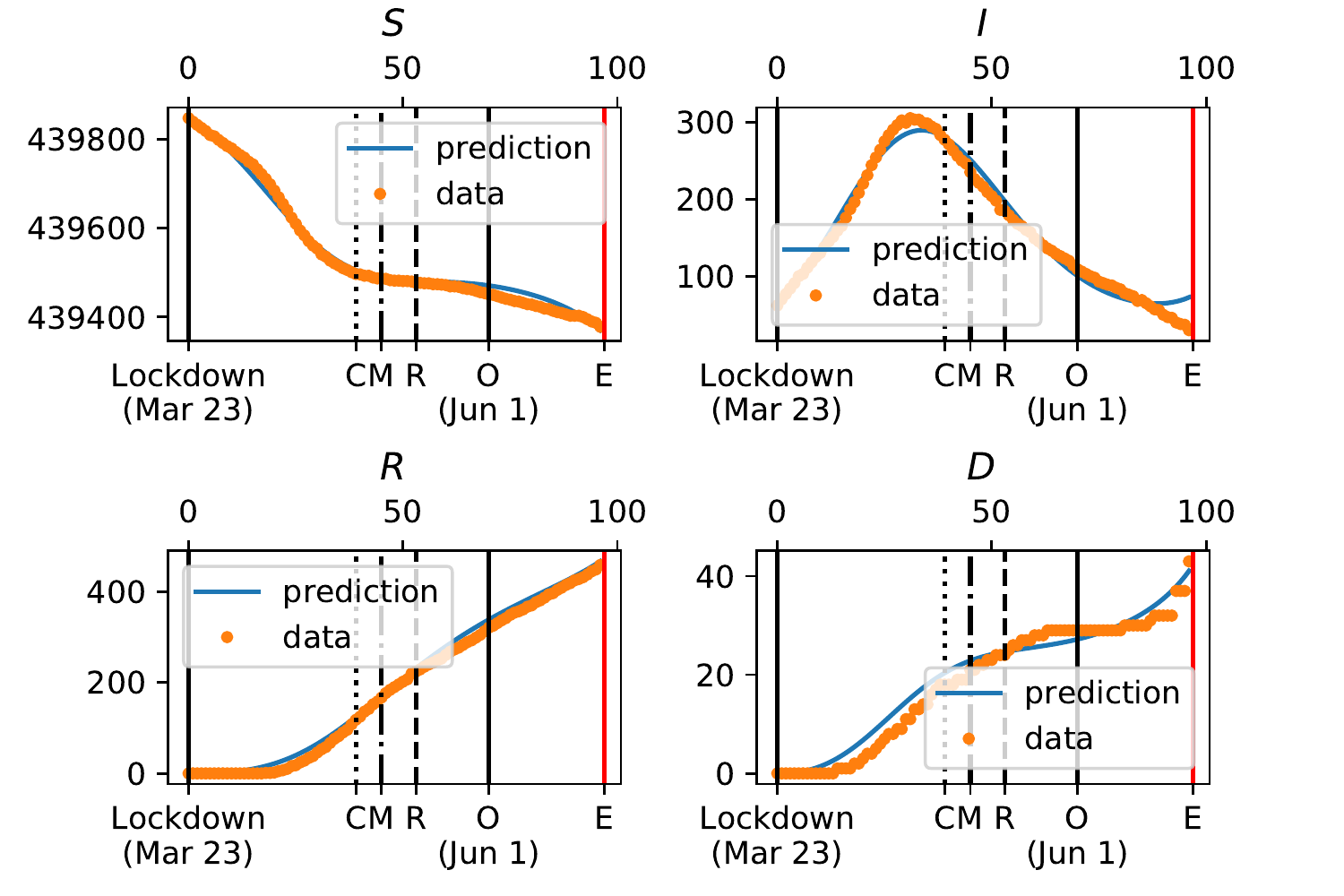}}
    \subfigure[Region 8]{\includegraphics[width=0.35\textwidth]{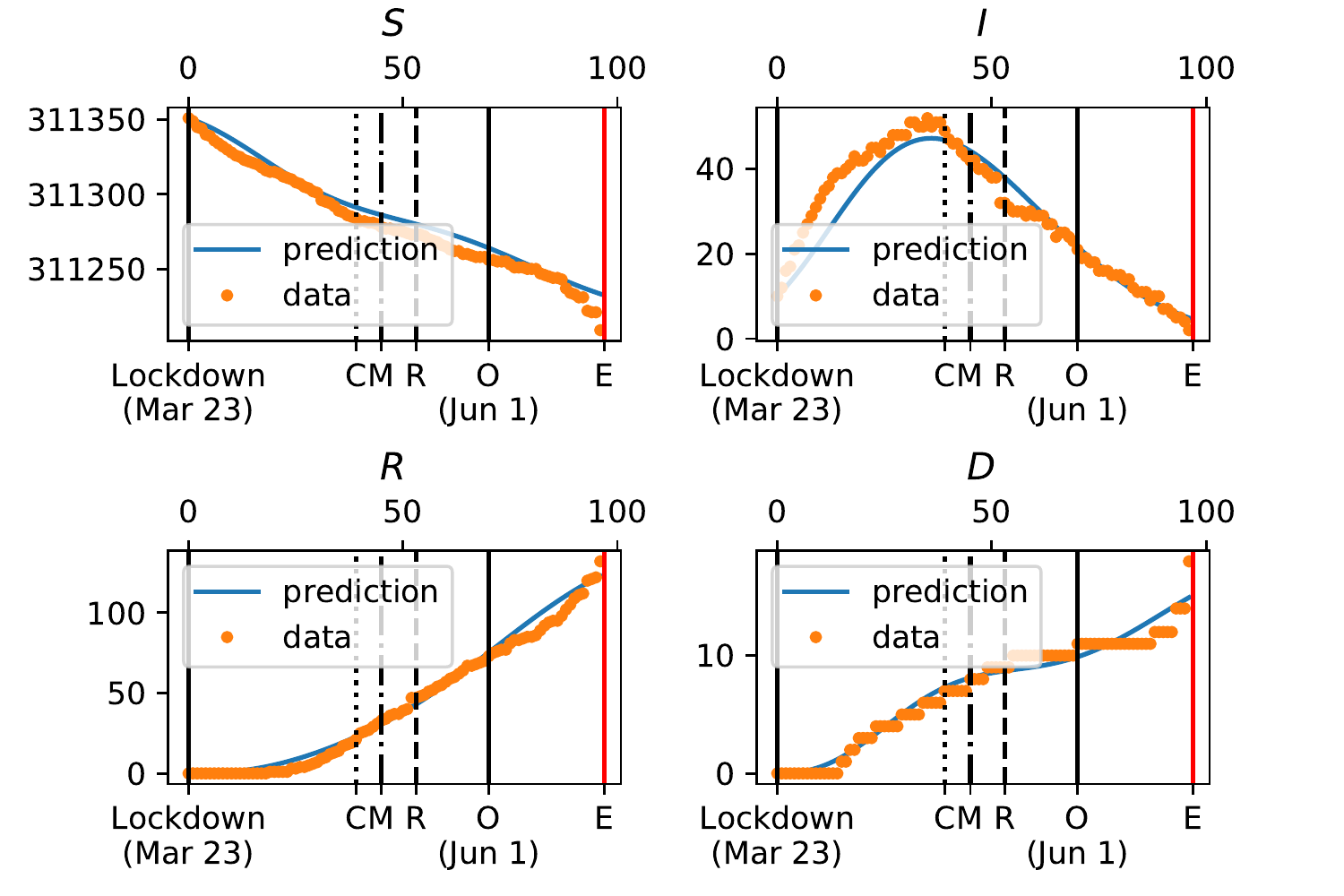}}
    \caption{Comparison of the simulation using inferred SIRD parameters (Figure \ref{fig:paramregions}) for Regions 1-8 of Michigan.}
    \label{fig:simregions}
\end{figure}

Figures \ref{fig:paramregions} and \ref{fig:simregions} illustrate the time-dependent SIRD coefficients and comparison between data and simulated (with the inferred ODE SIRD model) of the disease for Regions 1-8 delineated in Figure \ref{fig:Michmap}. This is an important step toward a more fine-grained understanding of the geographical distribution of the disease in the state. The Southeastern part of the state is more heavily populated, especially Regions 2 and 3, which also bore the greatest burden of the disease. The city of Detroit, at the Western tip of Region 3, was the worst affected, reflecting its well-known socio-economic challenges. By contrast, Washtenaw County, about 50 km to the East, but also in Region 3, bore among the lowest burdens, per capita. At the risk of stating the obvious, we note that Regions 1-4, which account for nearly 80\% of the state's population displayed very similar characteristics in the evolution of the data, as well as in SIRD coefficients and forward simulation results. We do not enter a more detailed analysis of these results here, deferring a different approach to spatial aspects of the spread of the disease to Section \ref{sec:PDE}.

\subsection{Deep and Bayesian neural networks}
\label{sec:results-nn}

\begin{figure}[t!]
    \centering
    \subfigure[time-dependent coefficients]{\includegraphics[height=60mm]{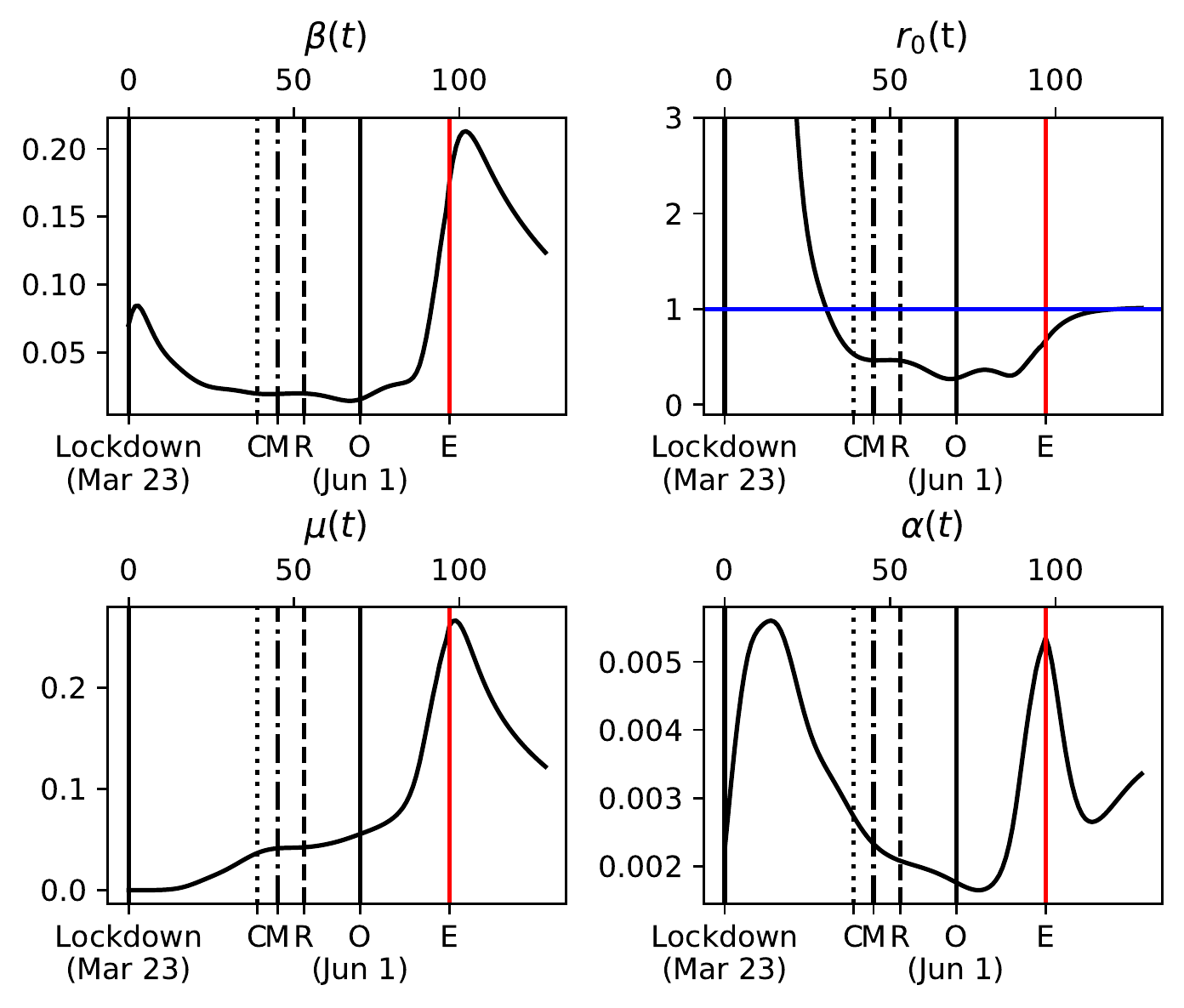}} \hspace{5mm}
    \subfigure[DNN prediction]{\includegraphics[height=60mm]{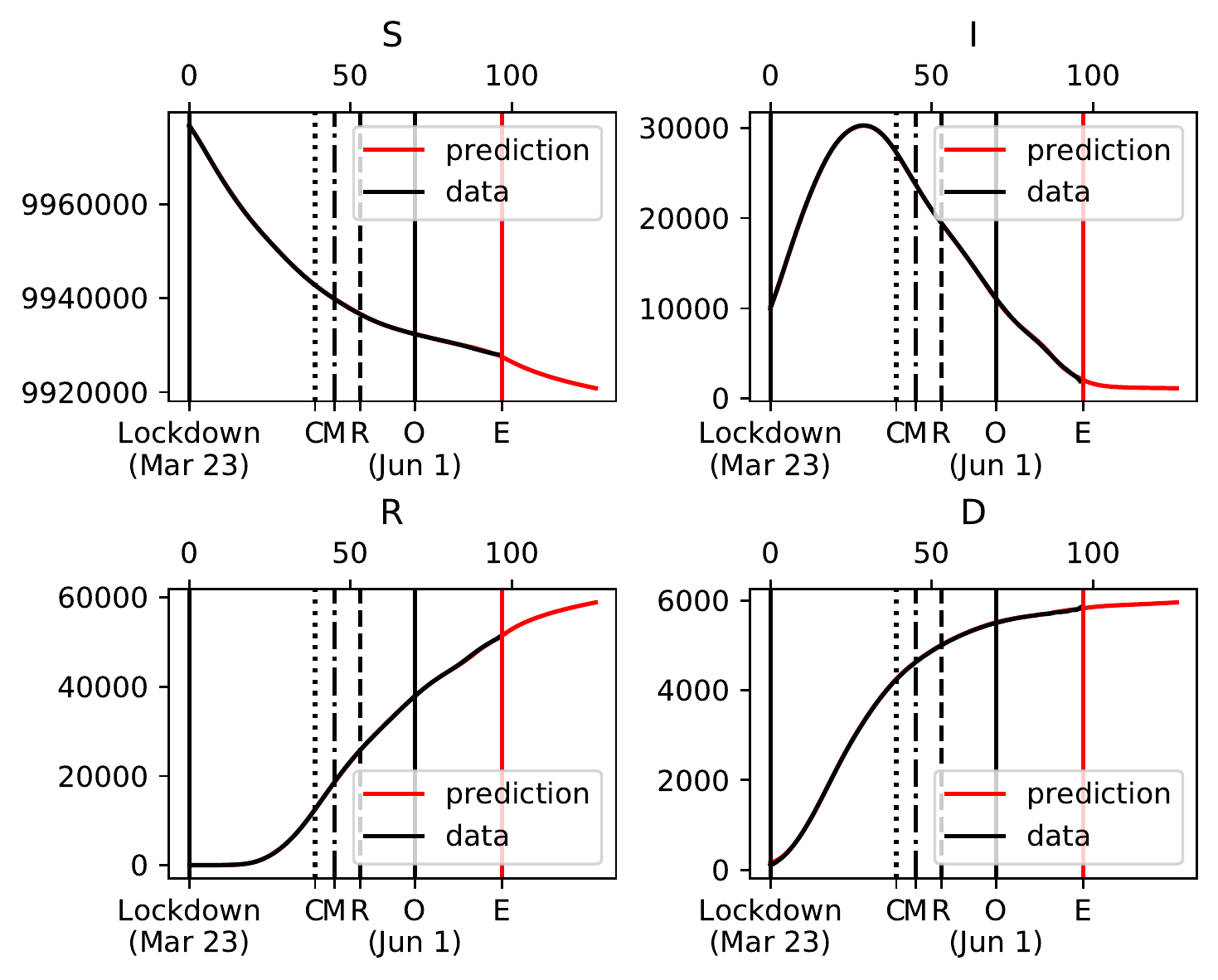}}
    \caption{\label{fig:results-dnn-mi}(a) Time-dependent coefficients identified by DNNs, where an increased infection rate after the open (O) of lockdown on June 1st is observed. (b) DNNs learned $S(t), I(t), R(t), D(t)$ based on the full extent of data point, and made a 30-day prediction.}
\end{figure}

\begin{figure}[t!]
    \centering
    \subfigure[time-dependent coefficients]{\includegraphics[height=60mm]{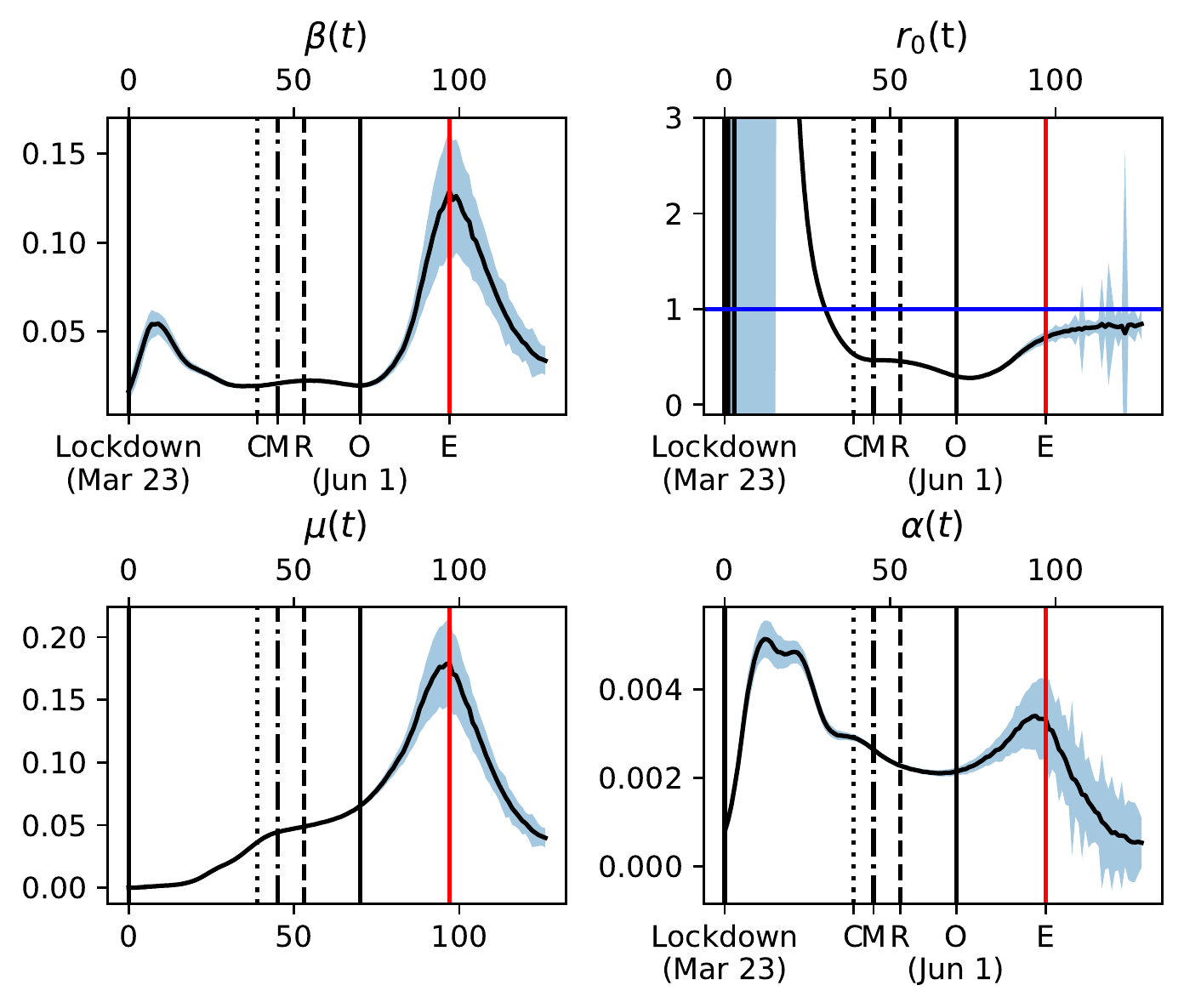}} \hspace{5mm}
    \subfigure[BNN prediction]{\includegraphics[height=60mm]{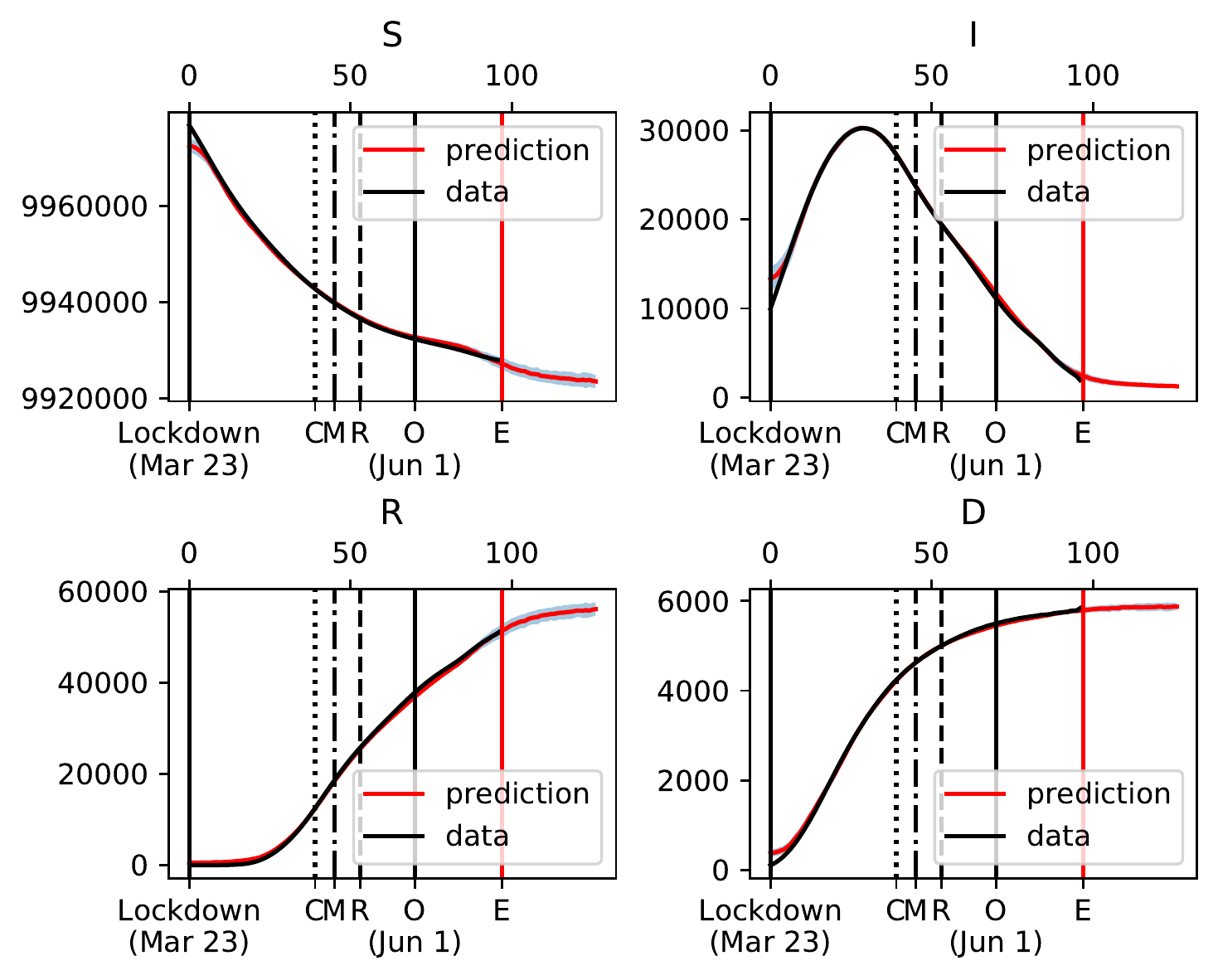}}
 \caption{
 (a) Time-dependent coefficients identified by BNNs, where an increased infection rate after the opening (O) on June 1st is observed. (b) BNNs learned $S(t), I(t), R(t), D(t)$ based on the full extent of data point, and made a 30-day prediction. Bands correspond to $\pm$ standard deviation over the mean.}
 \label{fig:results-bnn-mi}
 \end{figure}

To infer the coefficients $\beta(t)$, $\gamma(t)$, $\mu(t)$, $\alpha(t)$, we first compute the time derivatives of $S(t), I(t), R(t), D(t)$ by using the automatic differentiation API from \texttt{TensorFlow}. The coefficients are then computed by inverting Equations (\ref{eq:S}-\ref{eq:D}) at each time instant. For DNNs, we obtained deterministic results for all the coefficients. With BNNs, a Monte Carlo Sampling is performed to compute the mean and the standard deviation of the coefficients. 

The constraint Equation (\ref{eq:N}) is used to obtain $S^\text{d}_m$ from $I^\text{d}_m, R^\text{d}_m, D^\text{d}_m$. This ensures that the discrete time derivatives in Equations (\ref{eq:Sdiscm}-\ref{eq:Ddiscm}) satisfy

\begin{equation}
    \frac{S^\text{d}_{m} - S^\text{d}_{m-1}}{\Delta t} = -\frac{I^\text{d}_{m} - I^\text{d}_{m-1}}{\Delta t} -\frac{R^\text{d}_{m} - R^\text{d}_{m-1}}{\Delta t} -\frac{D^\text{d}_{m} - D^\text{d}_{m-1}}{\Delta t}
    \label{eq:dataconstraint}
\end{equation}

The constraint Equation (\ref{eq:N}) also has been imposed in the DNN and BNN representations by training networks for $I, R$ and $D$ and then defining the network for $S$ by this conservation of total population. Therefore, in using Equations (\ref{eq:S}-\ref{eq:D}) to invert the DNN/BNN representations for $\beta(t),\gamma(t),\mu(t),\alpha(t)$ at each time instant, a linear dependence is encountered:  The summed left and right hand-sides of (\ref{eq:I}-\ref{eq:D}) exactly equal the left and right hand-side of (\ref{eq:S}), respectively. A unique solution for $\beta(t),\gamma(t),\mu(t),\alpha(t)$ is not possible due to linear dependence introduced by the population constraint. To circumvent this indeterminacy, we endow the system with additional information by requiring that $\gamma(t) = 0$. This represents the conferral of immunity on the recovered population, and importantly, is detected by our inference results using system identification and ODE-constrained minimization, as discussed in Section \ref{sec:sysidresults}.

The inferred values, extended to a 30-day prediction (until July 28, 2020) for $\beta(t), \mu(t), \alpha(t), r_0(t)$ and $S(t), I(t), R(t), D(t)$  obtained from both DNNs and BNNs for Michigan are presented in Figures \ref{fig:results-dnn-mi} and \ref{fig:results-bnn-mi}, while the results for the eight Regions are given in Appendix \ref{sec:DNN-results-regions} and \ref{sec:BNN-results-regions}. 
One can observe that these time-dependent coefficients in Figures \ref{fig:results-dnn-mi}(a) and \ref{fig:results-bnn-mi}(a) have a similar \emph{initial} trend as those inferred by the system inference approach in Figure \ref{fig:results-system-id-lp-pred-para}. The effective reproduction rate $r_0(t) < 1$ for $m > 30$ (April 23), in good agreement with its value obtained via system inference in Figure \ref{fig:results-system-id-lp-pred-para}. As polynomial approximation is used by the system inference approach, the inferred coefficients in Figure \ref{fig:results-system-id-lp-pred-para} are very smooth, whereas inversion using the NN approach captures the detailed fluctuation of these coefficients, particularly, the rising infection rate after the open of the lockdown on June 1, 2020. In Figure \ref{fig:results-bnn-mi}, the band around the inferred coefficients and the NN predictions shows the mean $\pm$ one standard deviation of the  corresponding results. Note the high standard deviation in parameters at early times, due to the noise in the data at small numbers. The regional results in Appendix \ref{sec:DNN-results-regions} and \ref{sec:BNN-results-regions} indicate that an accelerating infection rate for all the regions after the open of the lockdown. In particular, Region 7 and 8 have a predicted $r_0(t)$ value that is greater than 1. In addition, we observed that the BNN inferred coefficients in the regional results have a narrower range compared to those from DNN.

More broadly, we note the difference in trends between the inferred time-dependent coefficients with the DNNs and BNNs in Figures \ref{fig:results-dnn-mi} and \ref{fig:results-bnn-mi} in comparison with those in Figure \ref{fig:results-system-id-lp-pred-para}. This is due to the local inversion at each data point to infer the coefficients with the DNNs and BNNs \emph{versus} the global optimization of losses for system inference in Section \ref{sec:sysidresults}. As was referred to above, inverse problems allow non-unique solutions. It will be instructive to compare the predictions made by the DNN and BNN representations with the data when they become available.

A result that is consistent across all inference methods: system identification with ODE-constrained optimization, DNNs and BNNs, and for the state as a whole as well as its Regions is the following: The infection rate, $\beta(t)$, initially fell with the public health campaign, especially driven by the lockdown orders. However, it began to rise with the first step of opening (C), and even accelerated as more aspects of public, recreational, commercial and industrial activities were relaxed (M, R, O). Yet, every one of the versions of forward simulations with corresponding and consistently inferred systems matched very well with the data, which confirm that the state has largely controlled the pandemic, and continues to do so. As the number of remaining infected individuals, $I(t)$, has fallen steeply, there are fewer conveyors of infection, and even the higher $\beta(t)$ has not yet led to another explosion of infection. This also can be seen by the sharply rising recovery rate, $\mu(t)$, and is verified by the effective reproduction rate $r_0(t)$ falling below $1.0$ after April 20 or 23 (the later date according to the DNN and BNN inference methods). A warning bell, however, must be rung as the results also indicate that $r_0(t) \to 1.0$ from below as we approach the end of our data and the time of writing. Michigan's numbers for $I(t)$ are rising, although not yet exponentially. See Figures \ref{fig:results-system-id-lp-pred-para}-\ref{fig:results-bnn-mi}, and Sections \ref{sec:DNN-results-regions}, \ref{sec:BNN-results-regions}.
\section{Two dimensional SIRD model with diffusion}
\label{sec:PDE}

Classical epidemiological models hold in the well-mixed limit, which is reflected in the compartments and sub-populations, $S, I, R, D$ being being total numbers over some geographical region. Spatial effects have been introduced by simply resolving smaller regions and treating them individually, as demonstrated here with our inference of SIRD coefficients over the regions of Michigan's lower peninsula (Figures \ref{fig:paramregions} and \ref{fig:simregions}). However, while affording a spatially finer-grained treatment, this approach cannot, of course, address the mobility of the population. This is an important consideration, especially in light of the imposition and lifting of quarantines. In the COVID-19 Pandemic, the effects of social distancing, and the possibility of surges with their lifting revolve on the question of the time (and spatially) varying mobility of the population. At the finest resolution, this must be approached  via agent-based models refined to resolve individuals. However, an intriguing question to explore is whether simple reaction-diffusion models can detect the evidence of mobility in these data. With our approach to model inference, we have access to methods of identifying mechanisms from data in which their action, while weak, may hold the key to important insights to the system. In this section, we embark down such a path, while noting that reaction-diffusion models of epidemiology have been considered previously from the perspective of analysis of the corresponding PDEs \cite{Chinviriyasit2010,Gai2020, Angulo2013-spatialBMESIR}.

We now extend the SIRD model to PDEs in two spatial dimensions using the same compartments. However, the population variables are now replaced with spatio-temporally varying densities, $\widehat{S}(\boldsymbol{x},t),\widehat{I}(\boldsymbol{x},t),\widehat{R}(\boldsymbol{x},t),\widehat{D}(\boldsymbol{x},t)$ defined as numbers per unit area.

\begin{align}
    \frac{\partial \widehat{S}}{\partial t} &=\mathcal{D}_\text{S}\nabla^2 \widehat{S}-\frac{\beta}{\widehat{N}} \widehat{S}\widehat{I}+\gamma \widehat{R}\label{eq:S_2D}\\
    \frac{\partial\widehat{I}}{\partial t} &=\mathcal{D}_\text{I}\nabla^2 \widehat{I}+\frac{\beta}{\widehat{N}}\widehat{S}\widehat{I}-\mu \widehat{I}-\alpha \widehat{I}\label{eq:I_2D}\\
    \frac{\partial\widehat{R}}{\partial t} &=\mathcal{D}_\text{R}\nabla^2 \widehat{R}+\mu \widehat{I}-\gamma \widehat{R}\label{eq:R_2D}\\
    \frac{\partial\widehat{D}}{\partial t} &=\alpha \widehat{I}\label{eq:D_2D}
\end{align}
Where $\mathcal{D}_\text{S}, \mathcal{D}_\text{I}, \mathcal{D}_\text{R}$ are diffusivities of the corresponding  compartments, and represent the mobility of the population via random walks. We define $\widehat{(\bullet)}=(\bullet)/\int_{\Omega}\text{d}A$ where $\Omega$ is the domain of the lower peninsula of Michigan, to which we restrict our PDE SIRD studies. Furthermore the population constraint holds: $\int_{\Omega}\widehat{N}\text{d}A = \int_{\Omega}\widehat{S}(t)\text{d}A  + \int_{\Omega}\widehat{I}(t)\text{d}A + \int_{\Omega}\widehat{R}(t)\text{d}A + \int_{\Omega}\widehat{D}(t)\text{d}A$.

\subsection{Inference on the PDE form of the SIRD model}
\label{sec:pdeSIRD}
We adopt the weak form, and specifically, the finite element framework for inference on the above system of PDEs. For a generic, finite-dimensional field $u^h$, the problem is stated as follows: Find $u^h\in \mathscr{S}^h \subset \mathscr{S}$, where $\mathscr{S}^h= \{ u^h \in \mathscr{H}^1(\Omega) ~\vert  ~u^h = ~\bar{u}\; \mathrm{on}\;  \Gamma^u\}$,  such that $\forall ~w^h \in \mathscr{V}^h \subset \mathscr{V}$, where $\mathscr{V}^h= \{ w^h \in\mathscr{H}^1(\Omega)~\vert  ~w^h = ~0 \;\mathrm{on}\;  \Gamma^u\}$, the finite-dimensional (Galerkin) weak form of the problem is satisfied. The variations $w^h$ and trial solutions $u^h$ are defined component-wise using a finite number of basis functions,
\begin{equation}
w^h = \sum_{a=1}^{n_\mathrm{b}} c^a N^a, \quad \qquad u^h = \sum_{a=1}^{n_\mathrm{b}} d^a N^a,
\label{eq:basisdef}
\end{equation}
\noindent where $n_\mathrm{b}$ is the dimensionality of the function spaces $\mathscr{S}^h$ and $\mathscr{V}^h$, and $N^a$ represents the basis functions. To obtain the Galerkin weak forms, we multiply each strong form by the corresponding weighting function, use Backward Euler method for time-discretized, integrate by parts and apply boundary conditions appropriately, leading to:
\begin{align}
    \int_{\Omega}w^h_1 \frac{\widehat{S}^h_{m} - \widehat{S}^h_{m-1}}{\Delta t}  \text{d}s &=-\int_{\Omega} \mathcal{D}_\text{S}\nabla w^h_1\cdot\nabla \widehat{S}^h_m\text{d}s -\int_{\Omega}w^h_1\left(\frac{\beta}{\widehat{N}} \widehat{S}^h_m\widehat{I}^h_m+\gamma \widehat{R}^h_m \right)\text{d}s\label{eq:S_2D_weak}\\
    \int_{\Omega}w^h_2\frac{\widehat{I}^h_{m} - \widehat{I}^h_{m-1}}{\Delta t}\text{d}s &=-\int_{\Omega}\mathcal{D}_\text{I}\nabla w^h_2\cdot\nabla \widehat{I}^h_m\text{d}s+\int_{\Omega}w^h_2\left(\frac{\beta}{\widehat{N}}\widehat{S}^h_m\widehat{I}^h_m-\mu \widehat{I}^h_m-\alpha \widehat{I}^h_m \right)\text{d}s\label{eq:I_2D_weak}\\
    \int_{\Omega}w^h_3\frac{\widehat{R}^h_{m} - \widehat{R}^h_{m-1}}{\Delta t}\text{d}s &=-\int_{\Omega}\mathcal{D}_\text{R}\nabla w^h_3\cdot\nabla \widehat{R}^h_m\text{d}s +\int_{\Omega}w^h_3\left(\mu \widehat{I}^h_m-\gamma \widehat{R}^h_m\right)\text{d}s\label{eq:R_2D_weak}\\
   \int_{\Omega}w^h_4 \frac{\widehat{D}_{m} - \widehat{D}_{m-1}}{\Delta t}\text{d}s &=\int_{\Omega}w^h_4\alpha \widehat{I}^h_m\text{d}s\label{eq:D_2D_weak}
\end{align}
Where, boundary terms disappear because we assume that the populations do not cross the state boundary, or into the upper peninsula. The system identification problem is to infer the time-dependent coefficients $\mathcal{D}_\text{S}(t),\mathcal{D}_\text{I}(t), \mathcal{D}_\text{R}(t)$, and we also choose to expand them in a polynomial basis 
\begin{align}
    D_s(t)=\theta_{16}+\theta_{17}t+\theta_{18}t^2+\theta_{19}t^3\label{eq:Ds}\\
    D_i(t)=\theta_{20}+\theta_{21}t+\theta_{22}t^2+\theta_{23}t^3\label{eq:Di}\\
    D_r(t)=\theta_{24}+\theta_{25}t+\theta_{26}t^2+\theta_{27}t^3\label{eq:Dr}
\end{align}
along with the time-dependent coefficients $\beta(t), \gamma(t), \mu(t), \alpha(t)$ shown in Equation (\ref{eq:beta}-\ref{eq:alpha}). We expect that the effect of mobility on the evolution of population densities is small over the course of the COVID-19 Pandemic. However, our interest is in inferring the presence of this effect in the data following the relaxation of lockdown orders. In order to identify the diffusivities despite the expected dominance of the reaction terms in the data obeying Equations (\ref{eq:D_2D_weak}), we adopt two stage Variational System Identification \cite{Wang2020}.

We define Stage 1 by choosing $w^h_i=1, i=1,..,4$, yielding:
\begin{align}
    \int_{\Omega} \frac{\widehat{S}^h_{m} - \widehat{S}^h_{m-1}}{\Delta t}  \text{d}A &= -\beta\int_{\Omega}\frac{1}{\widehat{N}} \widehat{S}^h\widehat{I}^h \text{d}s-\gamma\int_{\Omega}\widehat{R}^h\text{d}A   \label{eq:S_2D_weak_stage1}\\
    \int_{\Omega}\frac{\widehat{I}^h_{m} - \widehat{I}^h_{m-1}}{\Delta t}\text{d}s &=\beta\int_{\Omega}\frac{1}{\widehat{N}}\widehat{S}^h\widehat{I}^h\text{d}s-\int_{\Omega}\mu \widehat{I}^h\text{d}s-\alpha \int_{\Omega}\widehat{I}^h \text{d}A\label{eq:I_2D_weak_stage1}\\
    \int_{\Omega}\frac{\widehat{R}^h_{m} - \widehat{R}^h_{m-1}}{\Delta t}\text{d}s &= \mu\int_{\Omega}\widehat{I}^h\text{d}s-\gamma \int_{\Omega}\widehat{R}^h\text{d}s\label{eq:R_2D_weak_stage1}\\
   \int_{\Omega} \frac{\widehat{D}^h_{m} - \widehat{D}^h_{m-1}}{\Delta t}\text{d}s &=\alpha \int_{\Omega}\widehat{I}^h\text{d}A\label{eq:D_2D_weak_stage1}
\end{align}

The diffusion operators vanish since, for a constant weighting function, $\nabla w = 0$. In order to avoid a proliferation of superscripts and subscripts, we simply denote the  data interpolated over the finite element mesh at time $m$ by $\widehat{(\bullet)}^\text{d}_m$, dispensing with the superscipt $(\bullet)^h$ for the finite-dimensional fields  The label vector and matrix of bases  can be constructed as:
\begin{equation}
    \boldsymbol{y}_{m} = \left\{\begin{array}{c}
        \int_{\Omega}\frac{\widehat{S}^\text{d}_{m} - \widehat{S}^\text{d}_{m-1}}{\Delta t}\text{d}A   \\
        \int_{\Omega}\frac{\widehat{I}^\text{d}_{m} - \widehat{I}^\text{d}_{m-1}}{\Delta t}\text{d}A\\
        \int_{\Omega}\frac{\widehat{R}^\text{d}_{m} - \widehat{R}^\text{d}_{m-1}}{\Delta t}\text{d}A \\
        \int_{\Omega}\frac{D^\text{d}_{m} - D^\text{d}_{m-1}}{\Delta t}\text{d}A
    \end{array}\right\}
    \label{eq:y_2d}
\end{equation}
\begin{equation}
    \boldsymbol{\Xi}_{m} = \left[\begin{array}{cccc}
        \int_{\Omega}\frac{\widehat{S}^\text{d}_m \widehat{I}^\text{d}_m}{N}\text{d}s\langle 1\;t_m\;t_m^2\;t_m^3\rangle & \int_{\Omega}-\widehat{R}^\text{d}_m\text{d}s\langle 1\;t_m\;t_m^2\;t_m^3\rangle & \langle 0\;0\;0\;0\rangle & \langle 0\;0\;0\;0\rangle\\
         \int_{\Omega}-\frac{\widehat{S}^\text{d}_m \widehat{I}^\text{d}_m}{N}\text{d}s\langle 1\;t_m\;t_m^2\;t_m^3\rangle & \langle 0\;0\;0\;0\rangle & \int_{\Omega}\widehat{I}^\text{d}_m\text{d}s\langle 1\;t_m\;t_m^2\;t_m^3\rangle & \int_{\Omega}I_m\text{d}s\langle 1\;t_m\;t_m^2\;t_m^3\rangle\\
         \langle 0\;0\;0\;0\rangle & \int_{\Omega}\widehat{R}^\text{d}_m\text{d}s\langle 1\;t_m\;t_m^2\;t_m^3\rangle & \int_{\Omega}- \widehat{I}^\text{d}_m\text{d}s\langle 1\;t_m\;t_m^2\;t_m^3\rangle & \langle 0\;0\;0\;0\rangle\\
         \langle 0\;0\;0\;0\rangle & \langle 0\;0\;0\;0\rangle & \langle 0\;0\;0\;0\rangle & \int_{\Omega}-\widehat{I}^\text{d}_m\text{d}s\langle 1\;t_m\;t_m^2\;t_m^3\rangle
    \end{array}\right]
    \label{eq:Xi_2D}
\end{equation}
Once the reaction terms are identified, we return to the original weak forms Equations (\ref{eq:S_2D_weak}-\ref{eq:D_2D_weak}). Accounting for the arbitrariness of $w^h$ in $\mathscr{V}^h$, the finite-dimensionality leads to a system of residual equations for each degree of freedom (DOF):
\begin{align}
\mathscr{R}_i=\mathscr{F}_i\left(S^\text{d}_{m-1}, S^\text{d}_m, \nabla S^\text{d}_m,..., D_\text{s}\langle 1\;t_m\;t_m^2\;t_m^3\rangle,..., N,\nabla  N ...\right),
\label{eq:residual_weak}
\end{align}
where $\mathscr{R}_i$ is the $i^\text{th}$ component of the residual vector. The diffusion terms can then be identified by the two stage approach to Variational System Identification detailed in \cite{WangCMAME2019}.

\subsection{Data preparation on the 2D map of Michigan}

\begin{figure}[p!]
    \centering
    \includegraphics[width=0.7\textwidth]{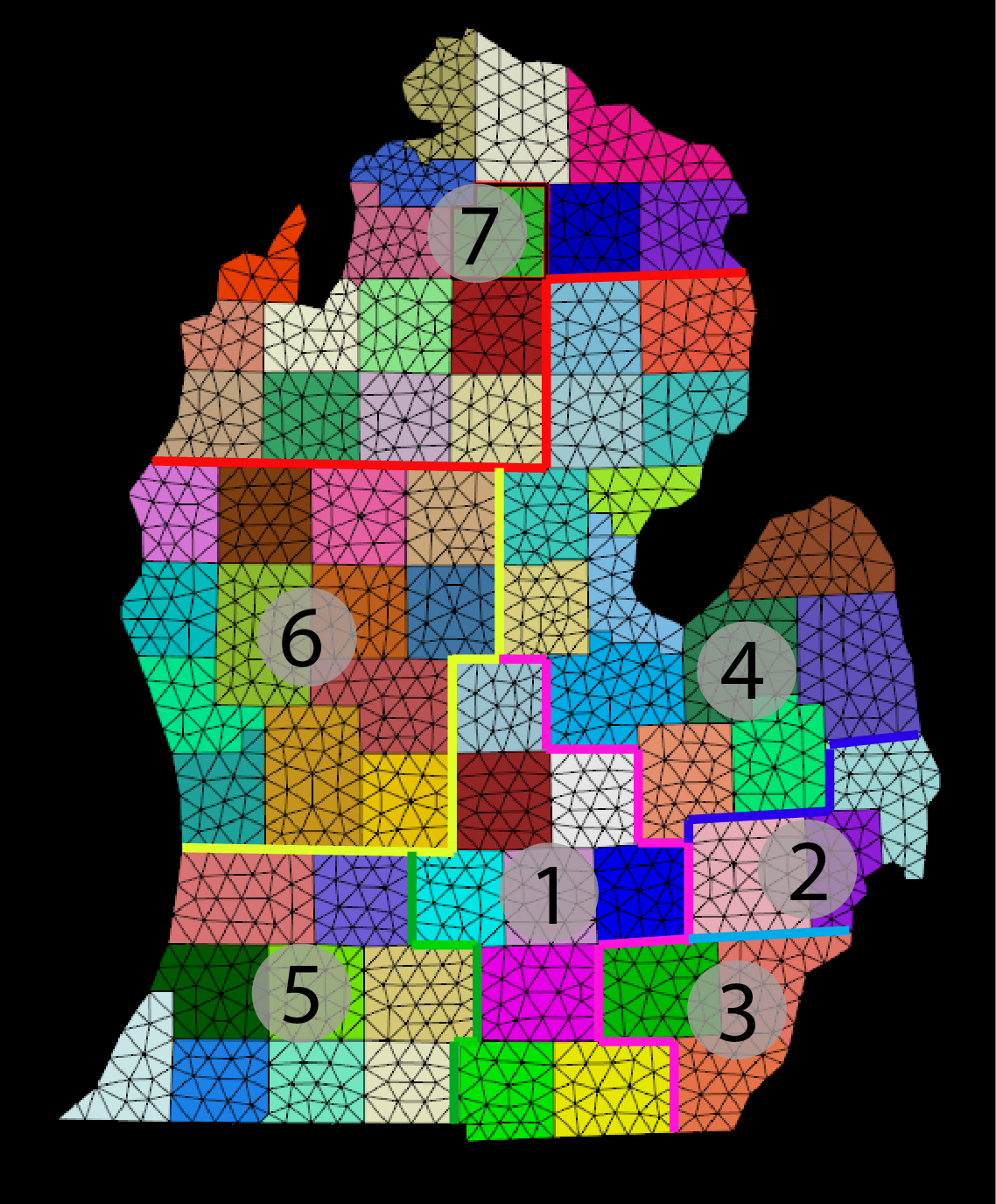}
    \caption{A finite element mesh of the map of Michigan delineating the counties. Only Regions 1-7 were used in the PDE inference problem.}
    \label{fig:2D-mesh}
\end{figure}

We first construct a two-dimensional mesh that fully resolves the counties as shown in Figure \ref{fig:2D-mesh}. Recall that only the lower peninsula, consisting of Regions 1-7 was included in the PDE inference problem. The data are available as cumulative sub-population numbers $I^\text{d}_m, R^\text{d}_m, D^\text{d}_m$ at the county level (Michigan's lower peninsula has 68 counties). We use a uniform density of each sub-population to compute $\widehat{I}^\text{d}_m, \widehat{R}^\text{d}_m, \widehat{D}^\text{d}_m$ within the county, and applied Gaussian filtering to smooth the discontinuities between counties. Note that the  discrete Gaussian filter can not be applied in a straightforward manner to unstructured meshes. Here we start with continuous Gaussian filtering over the infinite domain:  
\begin{align}
    u(\bx_0)&={\int_{-\infty}^\infty G(\bx_0,\bx)u_\text{raw}(\bx)\text{d}v}\\
    &=\int_\Omega G(\bx_0,\bx)u_\text{raw}(\bx)\text{d}v
\end{align}
where $u$ could be any of the four sub-population densities, and $G(\bx_0,\bx)=\frac{1}{2\pi\sigma^2}e^{-\frac{||\bx||^2}{2\sigma^2}}$ is the two dimensional Gaussian distribution function. The parameter $\sigma$ is the standard deviation of the Gaussian distribution which is related to the kernel size in the discrete Gaussian filter. Since $\int_\Omega G \text{d}A < 1$ we scale up the filtered displacement at each node: 
\begin{align}
    u(\bx_0)=\frac{\int_{-\infty}^\infty G(\bx_0,\bx)\text{d}v}{{\int_\Omega G(\bx_0,\bx)\text{d}v}}{\int_\Omega G(\bx_0,\bx)u_\text{raw}(\bx)\text{d}v}\\
    =\frac{1}{{\int_\Omega G(\bx_0,\bx)\text{d}v}}{\int_\Omega G(\bx_0,\bx)u_\text{raw}(\bx)\text{d}v}
\end{align}

\begin{figure}[h]
    \centering
    \includegraphics[width=0.65\textwidth]{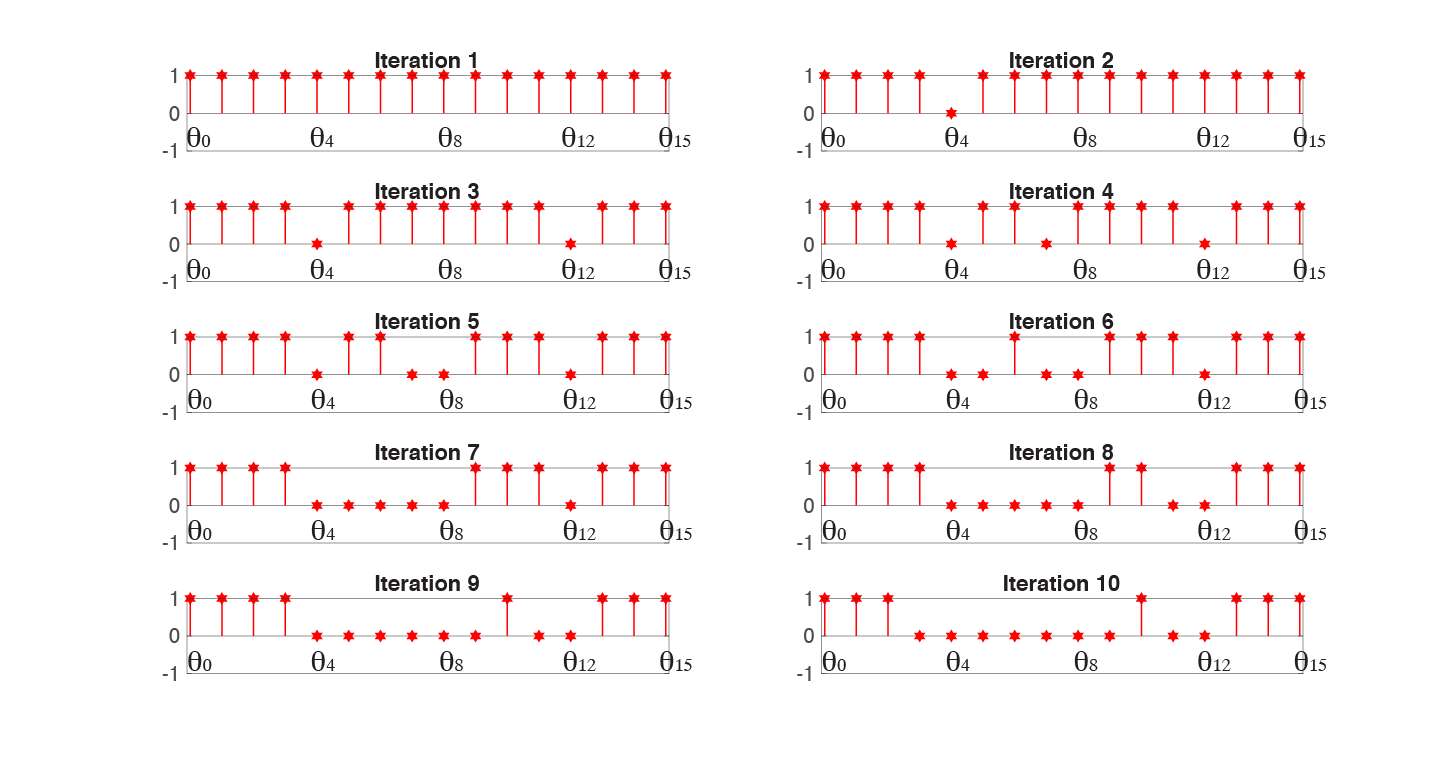}
    \includegraphics[width=0.3\textwidth]{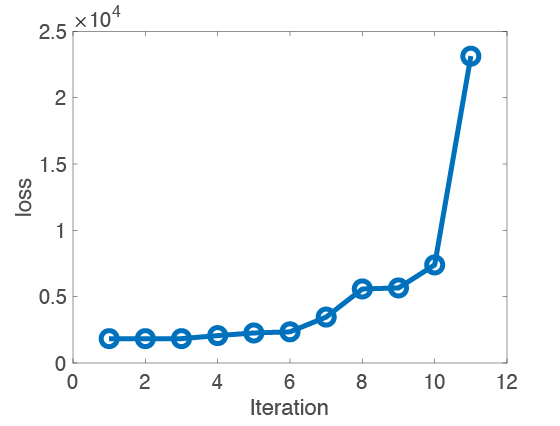}
    \caption{Left: stem-and-leaf plot illustrating system identification of active reaction parameters in the PDE SIRD model in Stage 1 of Variational System Identification. Each stem and leaf represents one term of $\theta_0,\dots \theta_{15}t^3$, scaled to 1 (active) or 0 (inactive).  Right: The changing loss as terms are eliminated from the set of time-dependent coefficients. System identification converges at Iteration 10 as the loss increases dramatically for further elimination of terms. }
    \label{fig:results-system-id-lp-2d-stem-loss-stage1}
\end{figure}

\begin{figure}[h]
    \centering
    \includegraphics[width=0.65\textwidth]{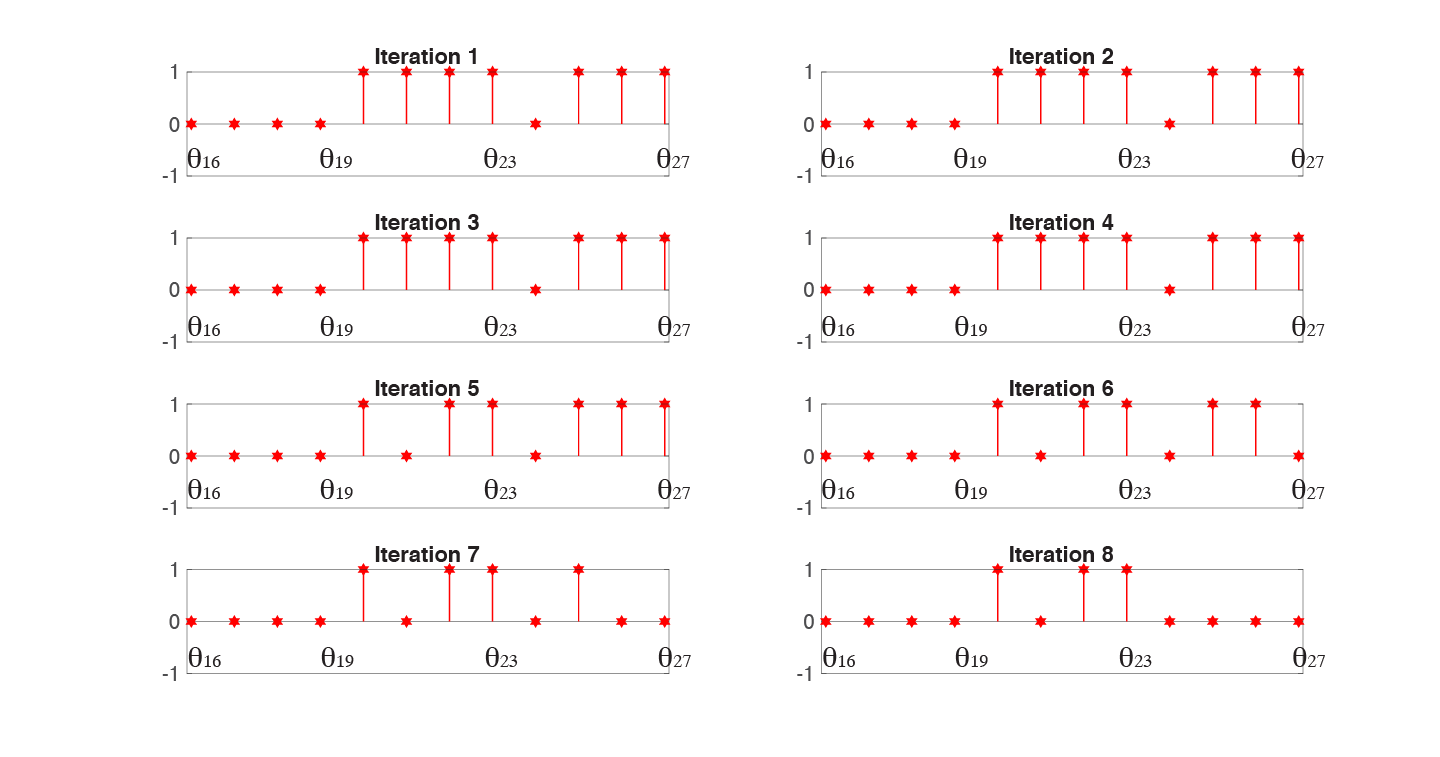}
    \includegraphics[width=0.3\textwidth]{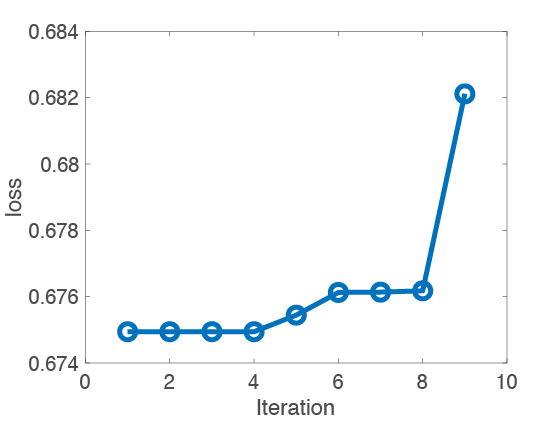}
    \caption{Left: stem-and-leaf plot illustrating system identification of active diffusion parameters in the PDE SIRD model in Stage 2 of Variational System Identification. Each stem and leaf represents one term of $\theta_{16},\dots \theta_{27}t^3$, scaled to 1 (active) or 0 (inactive). Right: The changing loss as terms are eliminated from the set of time-dependent coefficients. System identification converges at Iteration 8 as the loss increases dramatically for further elimination of terms.   }
    \label{fig:results-system-id-lp-2d-stem-loss-stage2}
\end{figure}

The spatio-temporal evolution of these fields was used in PDE inference via two-stage Variational System Identification as described in Section \ref{sec:pdeSIRD} followed by optimization constrained by the PDEs in (\ref{eq:S_2D_weak}-\ref{eq:D_2D_weak}) using adjoints. Stem-and-leaf plots and the losses for Stage 1 of Variational System Identification appear in Figure \ref{fig:results-system-id-lp-2d-stem-loss-stage1}. Recall that in this stage only the reaction terms $\beta(t),\gamma(t),\mu(t),\alpha(t)$ are identified. These inference results for active coefficients should be compared with the ODE SIRD model in Figure \ref{fig:results-system-id-lp-pred-para}.  This is followed by Stage 2 of Variational System Identification with stem-and-leaf plots and losses appearing in Figure \ref{fig:results-system-id-lp-2d-stem-loss-stage2}. Note that the diffusivities of the susceptible and recovered populations, $\mathcal{D}_\text{S} = 0$ and $\mathcal{D}_\text{R} = 0$. However, the infected population has a time-varying diffusivity $\mathcal{D}_\text{I}$ that declines.





\subsection{Results of system identification of two dimensional SIRD model with diffusion}

\begin{figure}[h]
    \centering
    \includegraphics[width=0.45\textwidth]{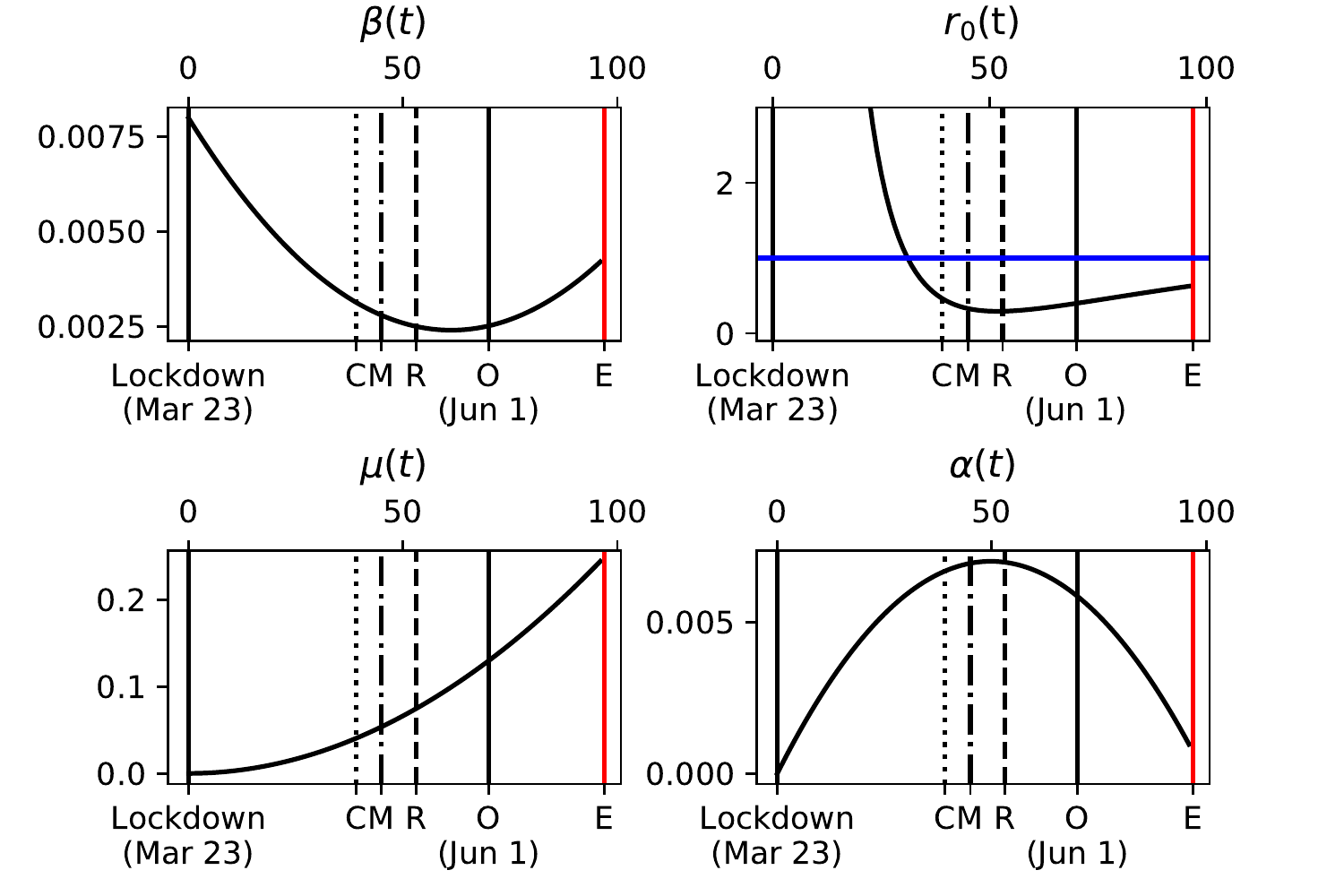}
    \includegraphics[width=0.45\textwidth]{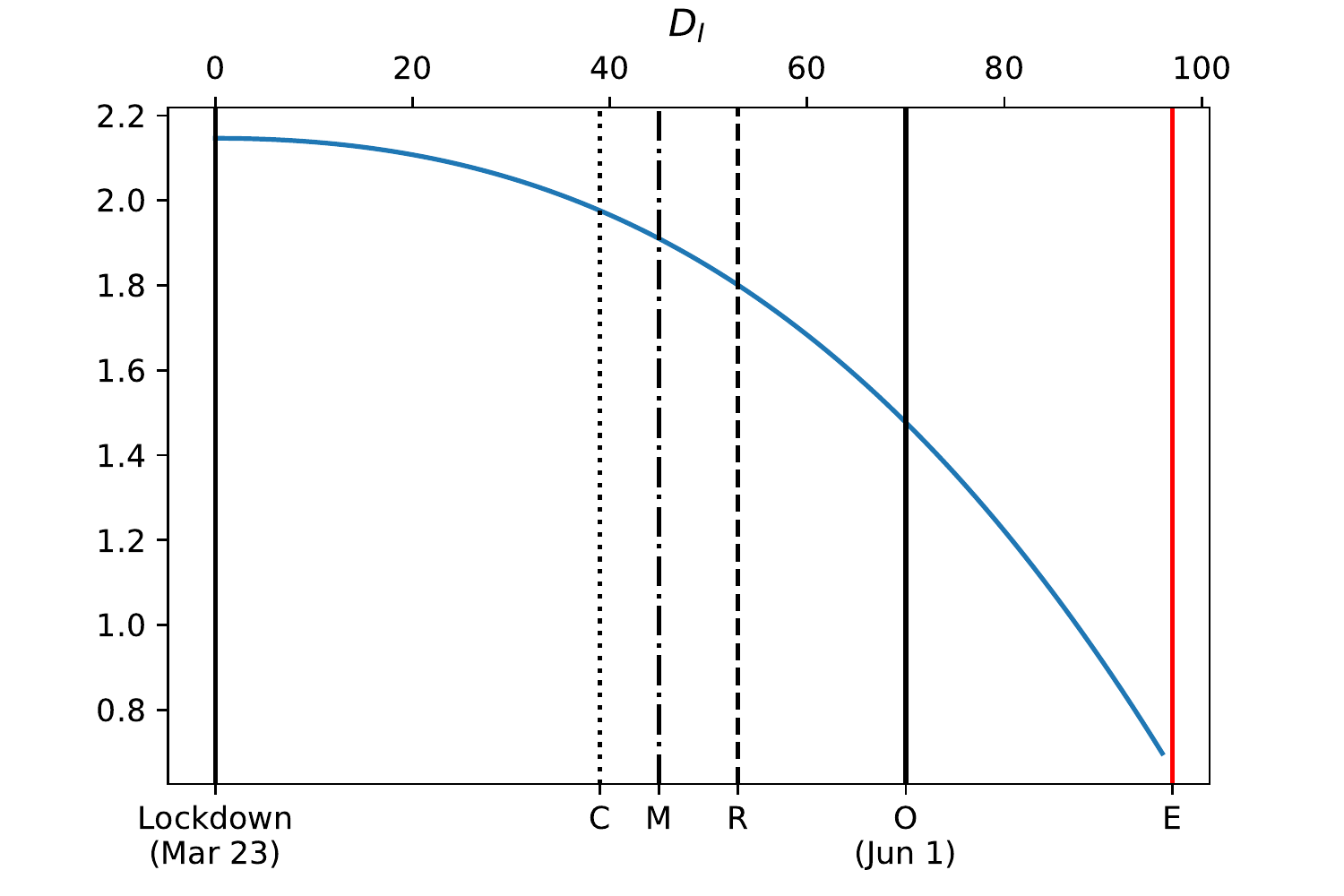}
    \caption{Left: The time-dependent reaction parameters in 2D SIRD model after tuning by PDE-constrained optiization: $\beta(t)=0.00798-1.82\times10^{-4}t+1.49\times10^{-6}t^2,\gamma(t)=0,\mu(t)=2.65\times10^{-5}t^2,\alpha(t)=2.82\times10^{-4}t-2.83\times10^{-6}t^2 $. Right: The sole time-dependent diffusivity is for the infected sub-population: $\mathcal{D}_\text{I}=2.146-8.12\times10^{-5}t^2-7.914\times10^{-7}t^3$.}
    \label{fig:results-system-id-lp-2d-par}
\end{figure}

\begin{figure}[h]
\subfigure[Infected
sub-population: data]{\includegraphics[width=0.6\textwidth]{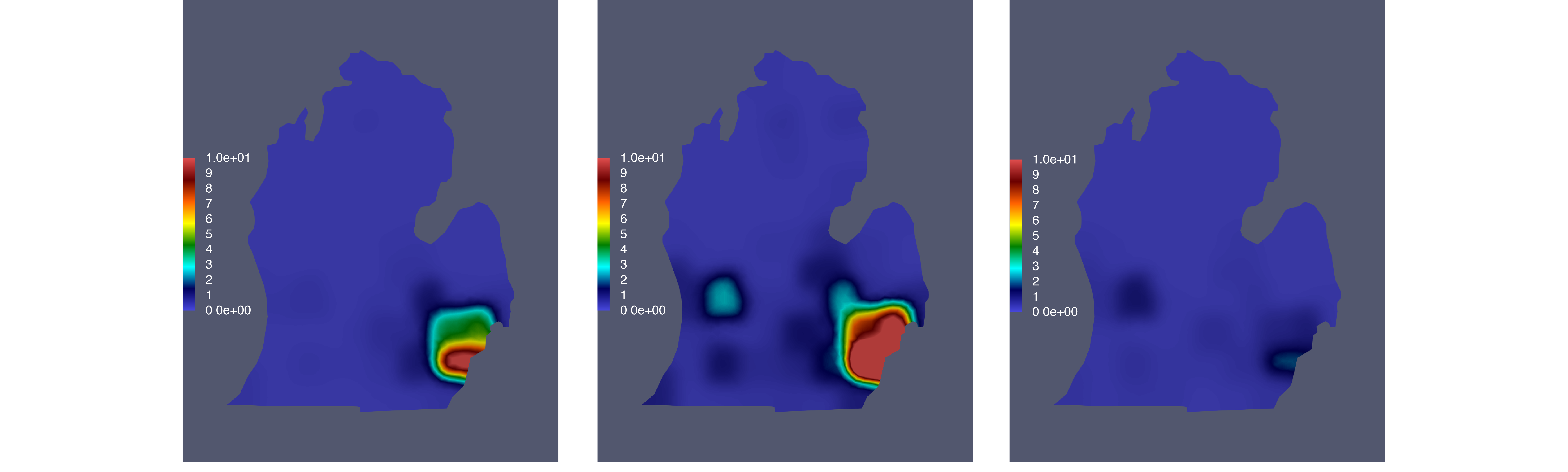}}\hspace{-1.5cm}
\subfigure[Infected
sub-population: PDE SIRD simulation]{\includegraphics[width=0.6\textwidth]{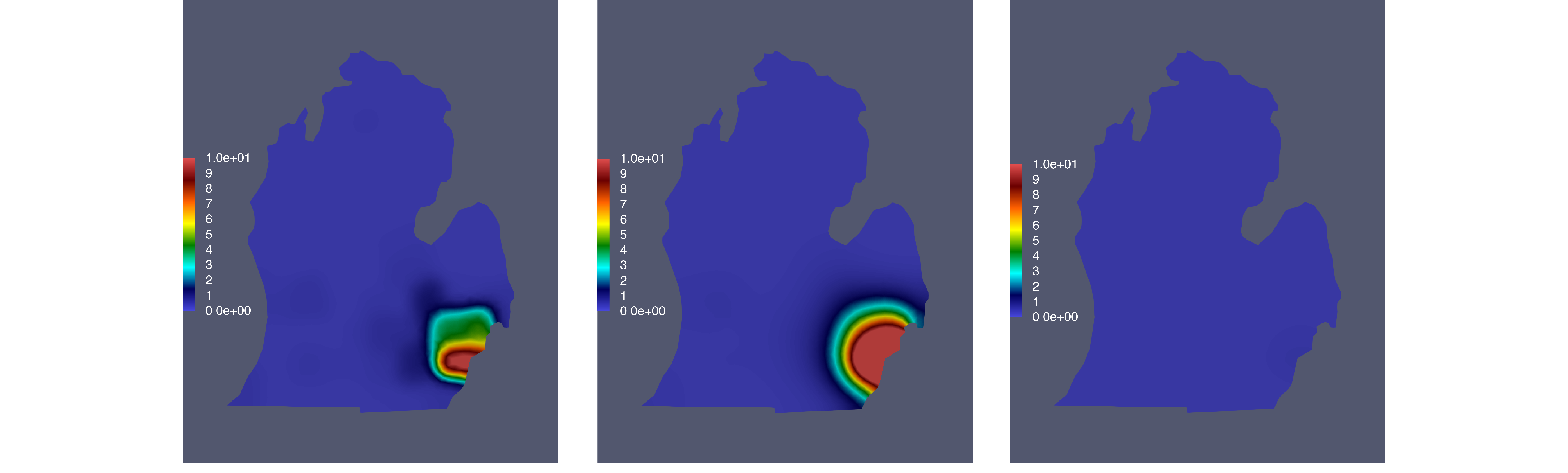}}
\hspace{-0.5cm}
\subfigure[Recovered
sub-population: data ]{\includegraphics[width=0.6\textwidth]{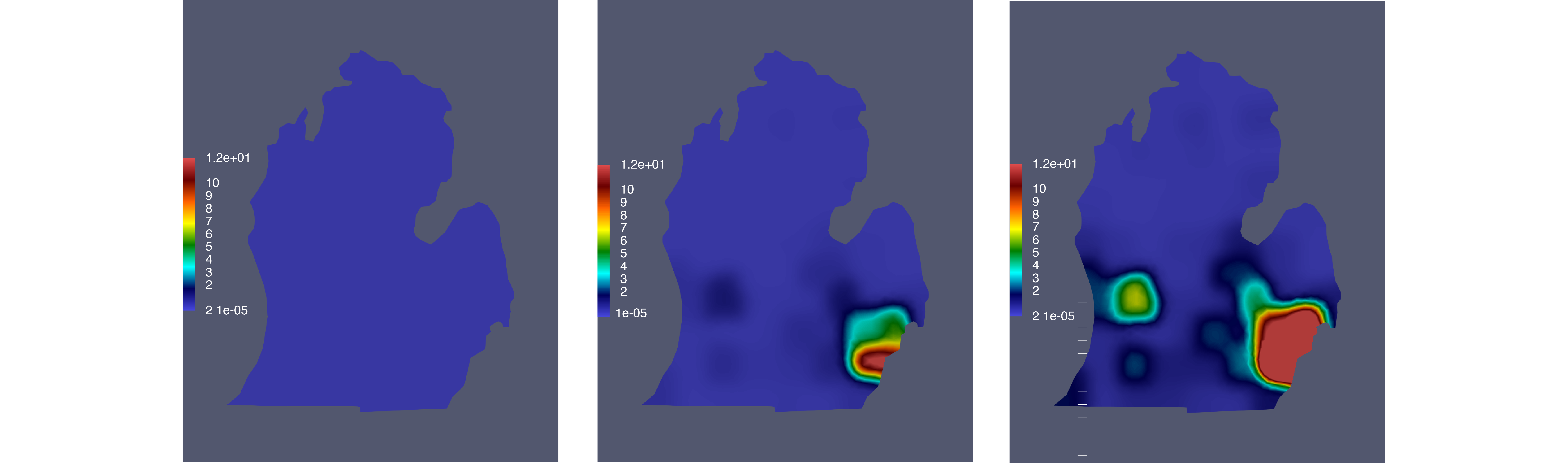}}\hspace{-1.5cm}
\subfigure[Recovered
sub-population: PDE SIRD simulation]{\includegraphics[width=0.6\textwidth]{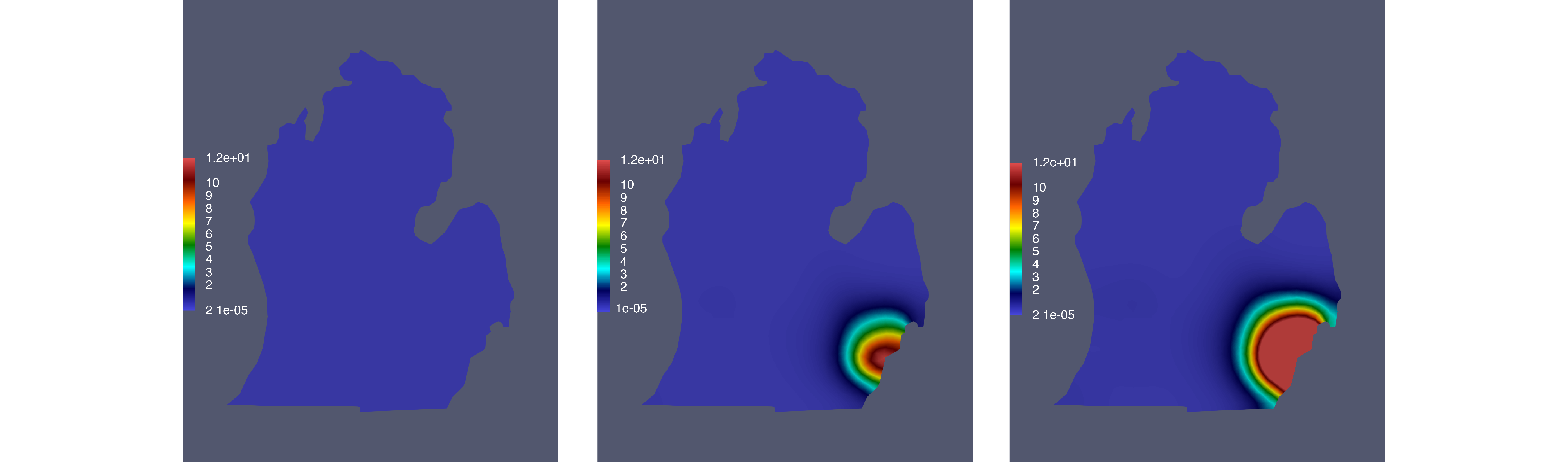}}
\caption{Comparison of the data on distributions of the infected (a) and recovered (c) sub-populations against forward PDE SIRD simulations with inferred quantities, (b) and (d), respectively. Data and simulation results are shown for Day 0 (lockdown, March 23), Day 40 (May 2, when the infected sub-population had its greatest spread across the state, but still restricted to Southeastern Michigan) and Day 96 (end of our data range, June 28, 2020).}
\label{tab:2D-sim}
\end{figure}

\begin{figure}[h]
    \centering
    \includegraphics[width=0.65\textwidth]{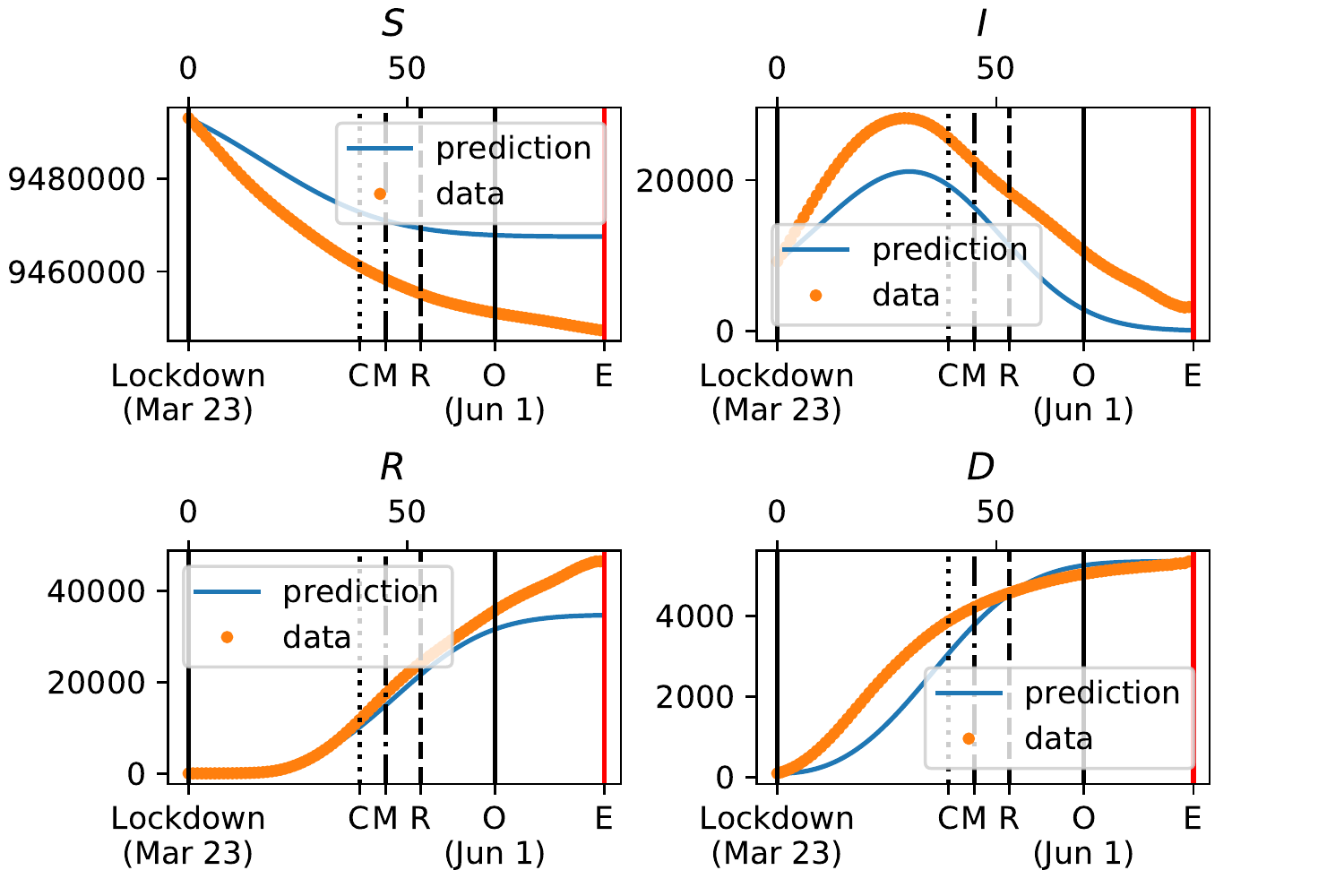}
    \caption{Simulation of the four compartments using the inferred PDE SIRD model, in comparison with the data. }
    \label{fig:results-system-id-lp-2d-prediction}
\end{figure}

Figure \ref{fig:results-system-id-lp-2d-par} shows the inference (two stage Variational System Identification followed by PDE-constrained optimization) for the coefficients $\beta(t),\mu(t),\alpha(t)$, the effective reproduction rate, $r_0(t)$ as well as the diffusivity $\mathcal{D}_\text{I}(t)$ in the PDE SIRD model. On comparing with Figure \ref{fig:results-system-id-lp-pred-para} some differences are revealed in the time dependence of $\beta(t), \mu(t), r_0(t), \alpha(t)$. This is to be expected in adopting the PDE SIRD model over the ODE form. The inference of time-dependent diffusion in the mobility of the infected sub-population, $\mathcal{D}_\text{I}$, naturally affects the other quantities. While the preliminary nature of these warrants caution, it is worth noting the inference of decreasing mobility of the infected sub-population in $\mathcal{D}_\text{I}$.   Figure \ref{tab:2D-sim} compares data and the forward simulation with inferred quantities for the distribution of the infected and recovered sub-populations on days corresponding to the initial lockdown, the maximum spread of the infected sub-population (May 2), and at the end of our data collection.  Notably, the restriction of the high density of the infected population, $\widehat{I}$ to Southeastern Michigan reflects the success, to date, of the state's public health response. While the correspondence is reasonable, the statewide sub-populations $S(t), I(t), R(t), D(t)$ obtained by integrating the corresponding densities over the lower peninsula, show a poorer match in Figure \ref{fig:results-system-id-lp-2d-prediction}. While the trends are reproduced, there are notable errors over time. A major improvement is possible in the PDE SIRD model by allowing the coefficients $\beta,\gamma,\mu,\alpha, \mathcal{D}_\text{S},\dots,\mathcal{D}_\text{R}$ to also vary over space. This would allow better representation of the system, in keeping with the inferred difference in $\beta(t),\mu(t),r_0(t),\alpha(t)$ over the eight Regions in Figure \ref{fig:paramregions}, which led to the excellent agreement between data and the forward ODE SIRD simulations in Figure \ref{fig:simregions}. From a purely data representation standpoint, the greater number of parameters will allow lower errors.

The code used for the inference, machine learning and forward simulations is available in the \texttt{mechanoChem} and \texttt{mechanoChemML} libraries at \url{https://github.com/mechanoChem/}.

\section{Conclusion}
\label{sec:concl}

We have brought machine learning inference techniques to bear upon the data on progression of COVID-19 across the state of Michigan by applying three distinct approaches: (a) Our methods of system identification to delineate the operational mechanisms, followed by (b) adjoint-based model-constrained optimization for refinement of the parameters, and (c) deep and Bayesian neural networks. Our interest in this study has been two-fold. 

The first has been to seek to infer the time-dependence of the coefficients in the classical ODE SIRD model, motivated by the evolving characteristics  of testing, quarantine and treatment protocols over the 97-day course of the pandemic as reflected in the data. As discussed in Section \ref{sec:results}, our inference methods reveal the course of rates of infection, recovery and death over the state and its eight regions, assuming uniform mixing in each case. Notably, our methods suggest that recovery confers immunity, but we hasten to add that this is a very preliminary conclusion. More detailed and fine-grained studies need to be undertaken to verify it, and of course, immunology will have the final say here. Also of note are our conclusions that while the infection rate has increased after an initial decline, as the state relaxed restrictions, the lower numbers of infectious individuals has meant a lower overall extent of transmission. This is also seen in the effective reproduction rate, which, while below one, has trended dangerously closer to that threshold of exponential growth. The uncertainty in our inference, given the data, is reflected in the results of the Bayesian neural networks in the same section. Of some interest here are the predictions made by BNNs for 30 days beyond the end of the data we have considered; that is until July 28, 2020.

The second facet of our interest is to try and infer spatial dependence by extending the SIRD models to PDEs by incorporating the population's mobility via diffusion. This is a different, and potentially intriguing, approach that complements the resolution of the problem down to the smaller Regions of the state as we did with the ODE SIRD model. On this front, we note that the inference needs to be extended to our methods of two-stage Variational System Identification followed by PDE-constrained optimization. Here, it is of note that the susceptible and recovered populations were found to have vanishing diffusivities (mobilities), while the infected population had a diffusivity that declined over the 97-day extent of the data that we used. This first extension to system inference of the PDE SIRD model returned reasonable comparisons with data on distributions of the sub-population densities, although the total numbers integrated over the state were not as well reproduced. As suggested by the notable differences in the ODE SIRD model coefficients for the eight regions of the state, the PDE SIRD model  with spatially varying coefficients may be a better representation. Building on these initial results, we see many possibilities for analysis and prediction of the future course and geographical spread of the COVID-19 Pandemic using the PDE SIRD model.

\section*{Acknowledgements}
We acknowledge the support of Defense Advanced Research Projects Agency (DARPA) under Agreement No. HR0011199002, ``Artificial Intelligence guided multi-scale multi-physics framework for discovering complex emergent materials phenomena''

\bibliographystyle{unsrtnat}
\bibliography{ref.bib}
\newpage

\appendix
\section{Additional regional results}
This appendix contains the inferred time-dependent coefficients and the NN prediction results for the eight regions of the Michigan state.
\subsection{DNN results for different regions}
\label{sec:DNN-results-regions}

\begin{figure}[p!]
    \centering
    \subfigure[region 1]{\includegraphics[height=38mm]{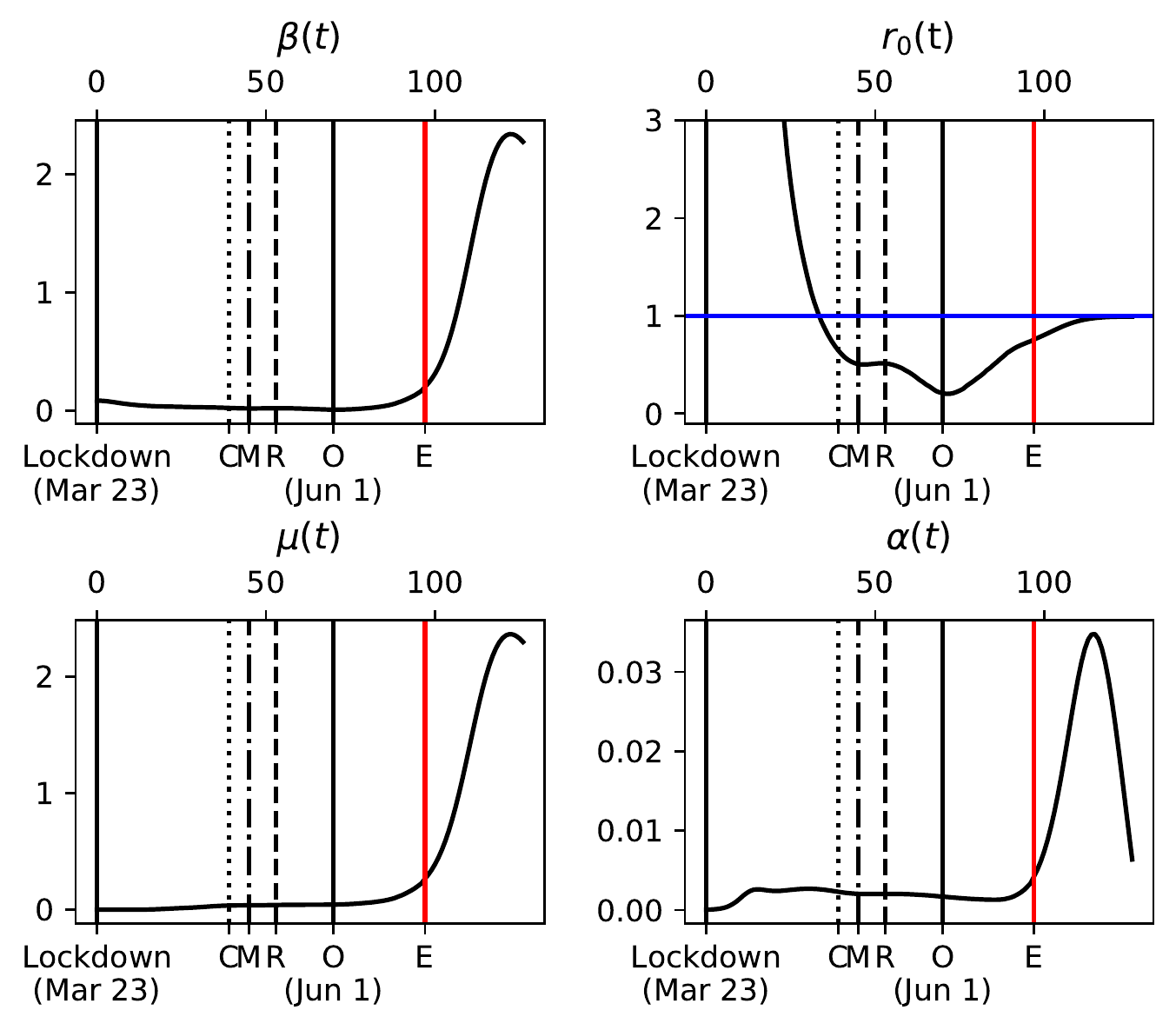}} \hspace{10mm}
    \subfigure[region 2]{\includegraphics[height=38mm]{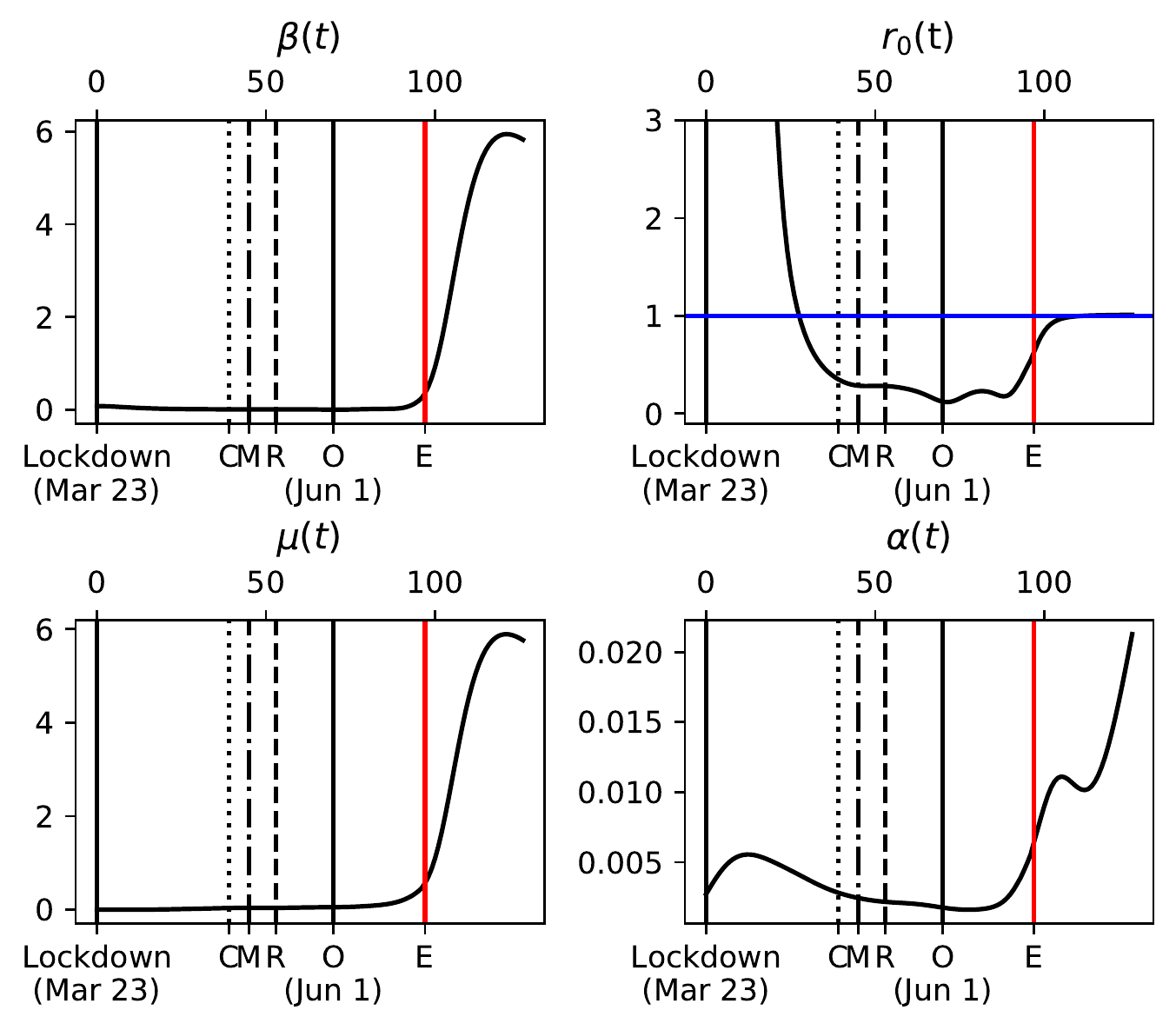}} \\
    \subfigure[region 3]{\includegraphics[height=38mm]{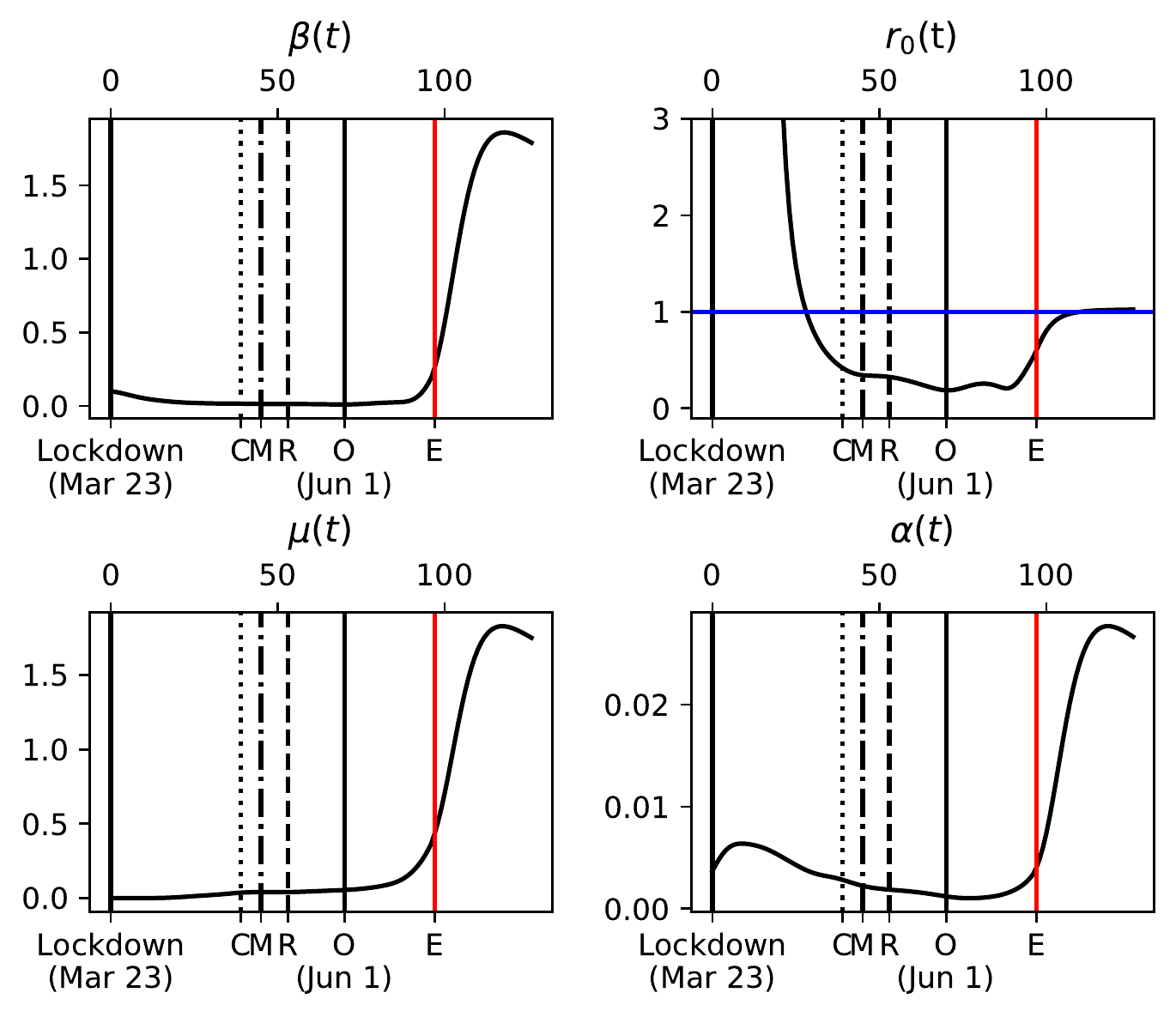}} \hspace{10mm}
    \subfigure[region 4]{\includegraphics[height=38mm]{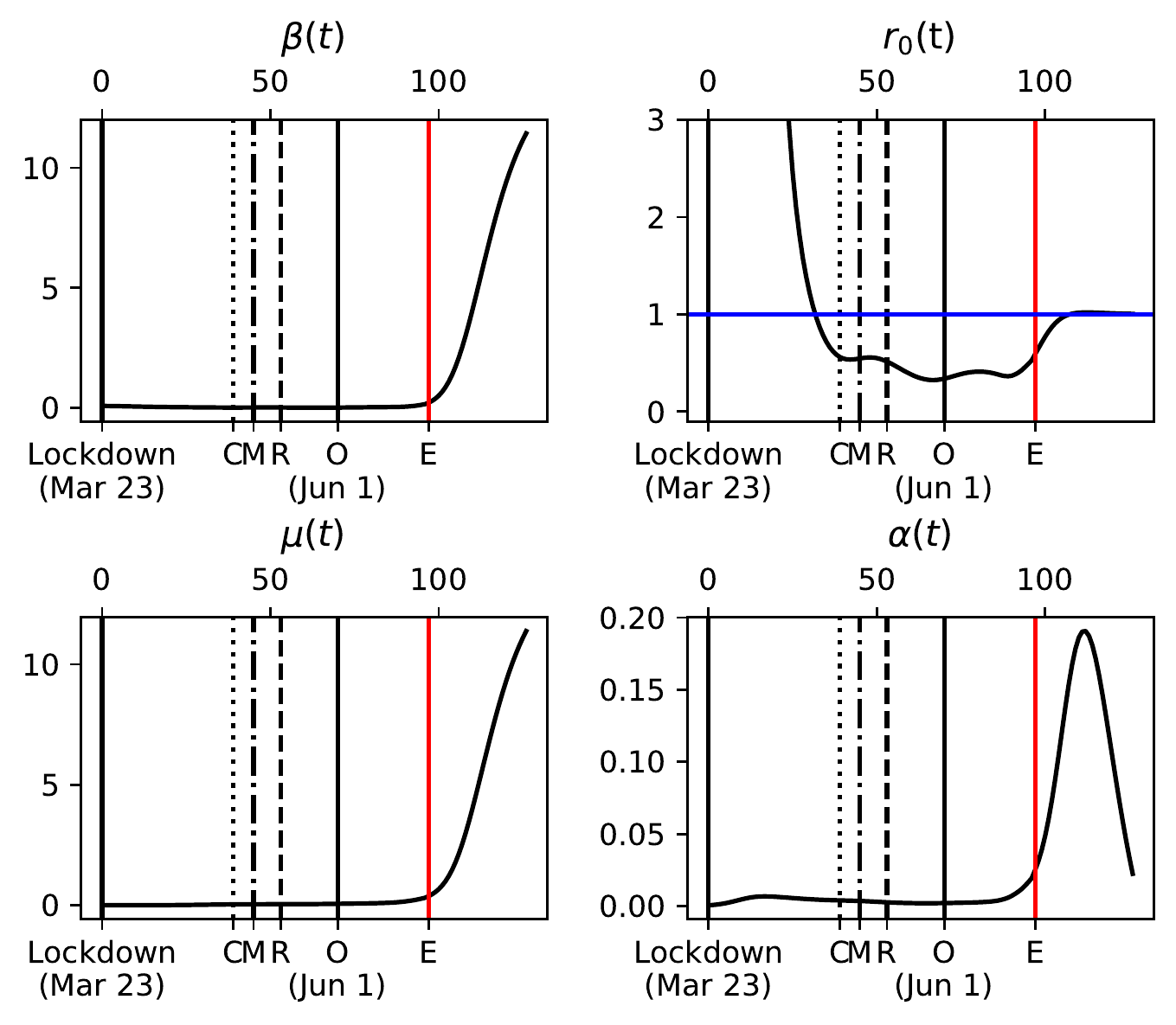}} \\
    \subfigure[region 5]{\includegraphics[height=38mm]{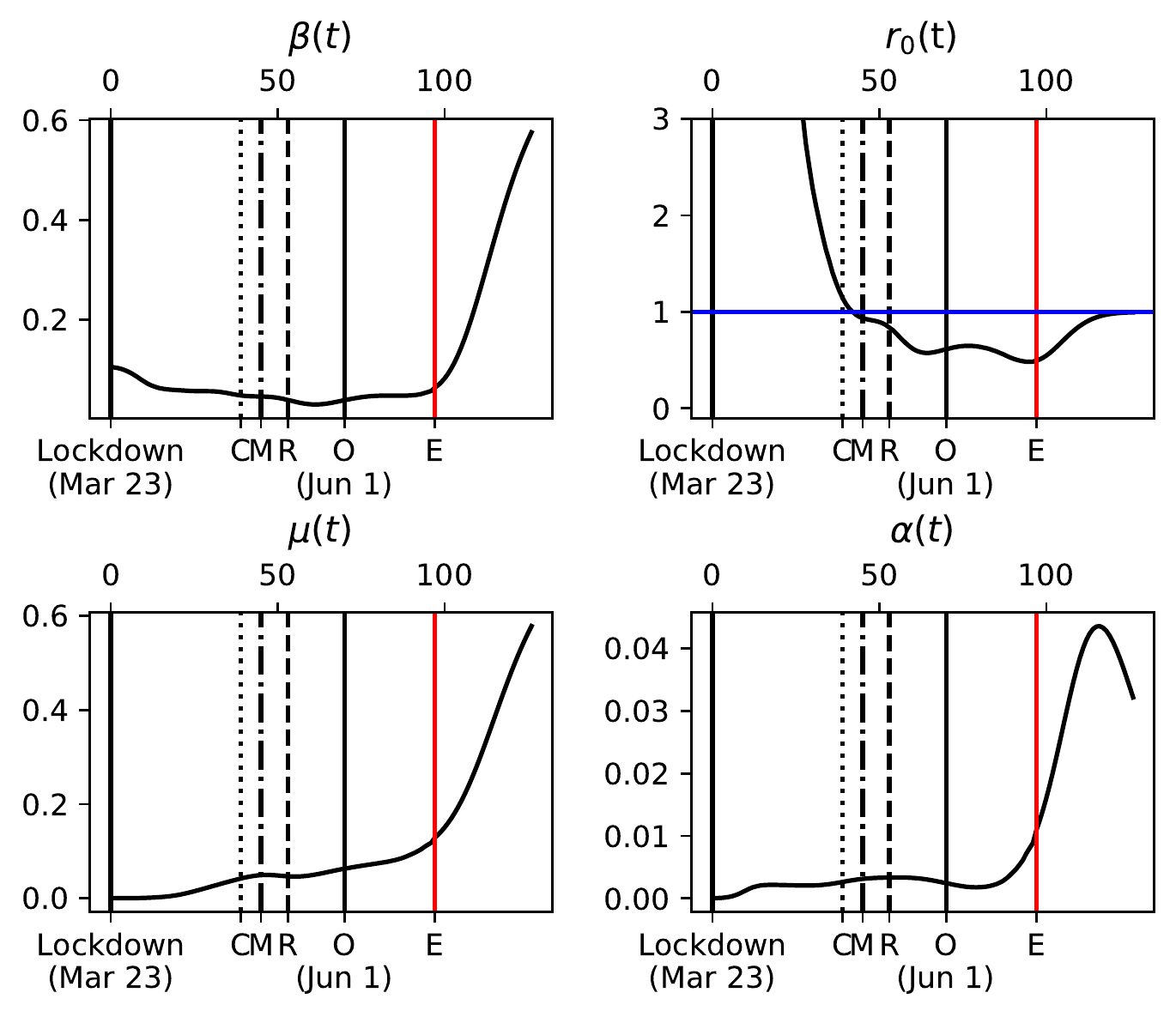}} \hspace{10mm}
    \subfigure[region 6]{\includegraphics[height=38mm]{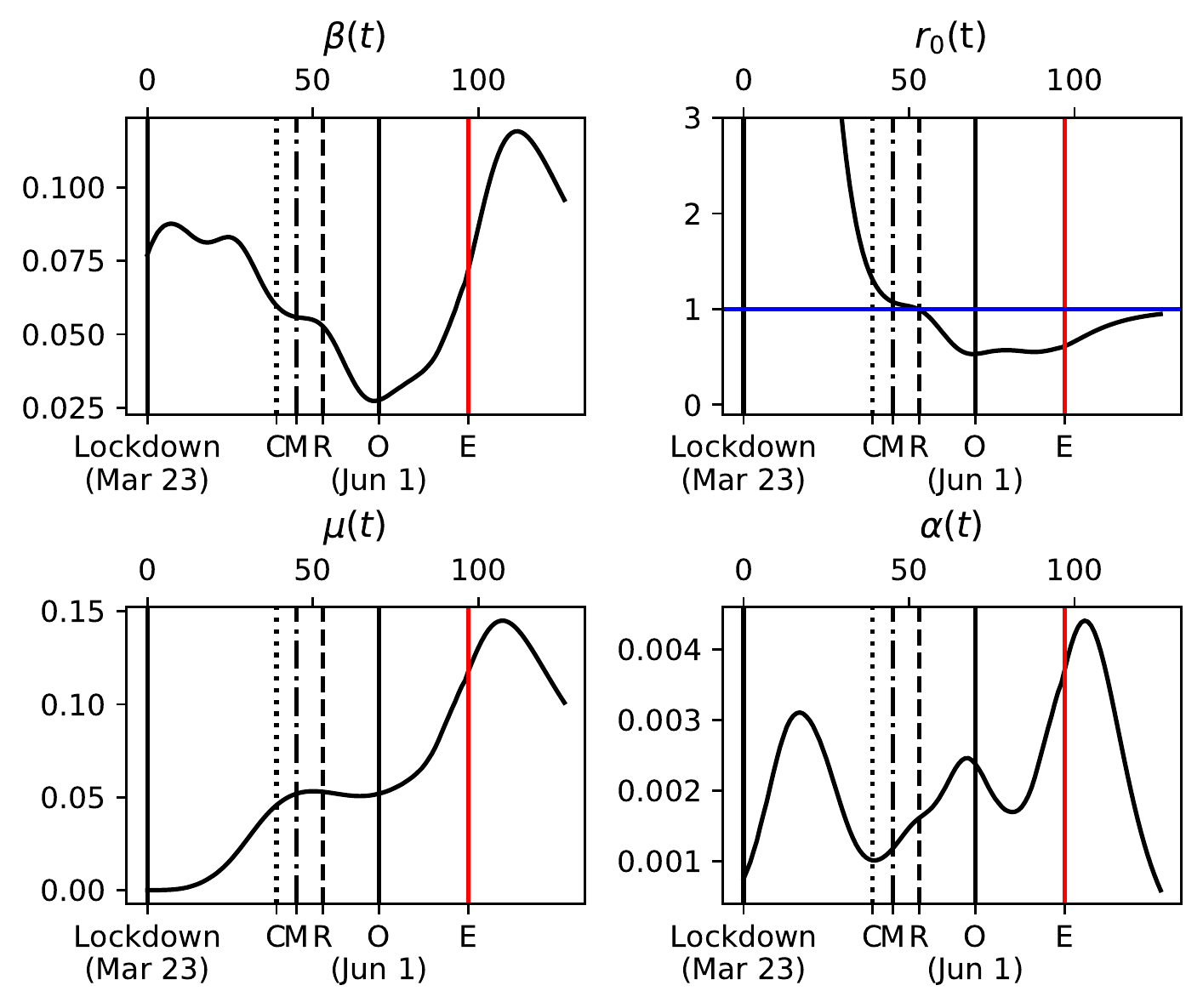}} \\
    \subfigure[region 7]{\includegraphics[height=38mm]{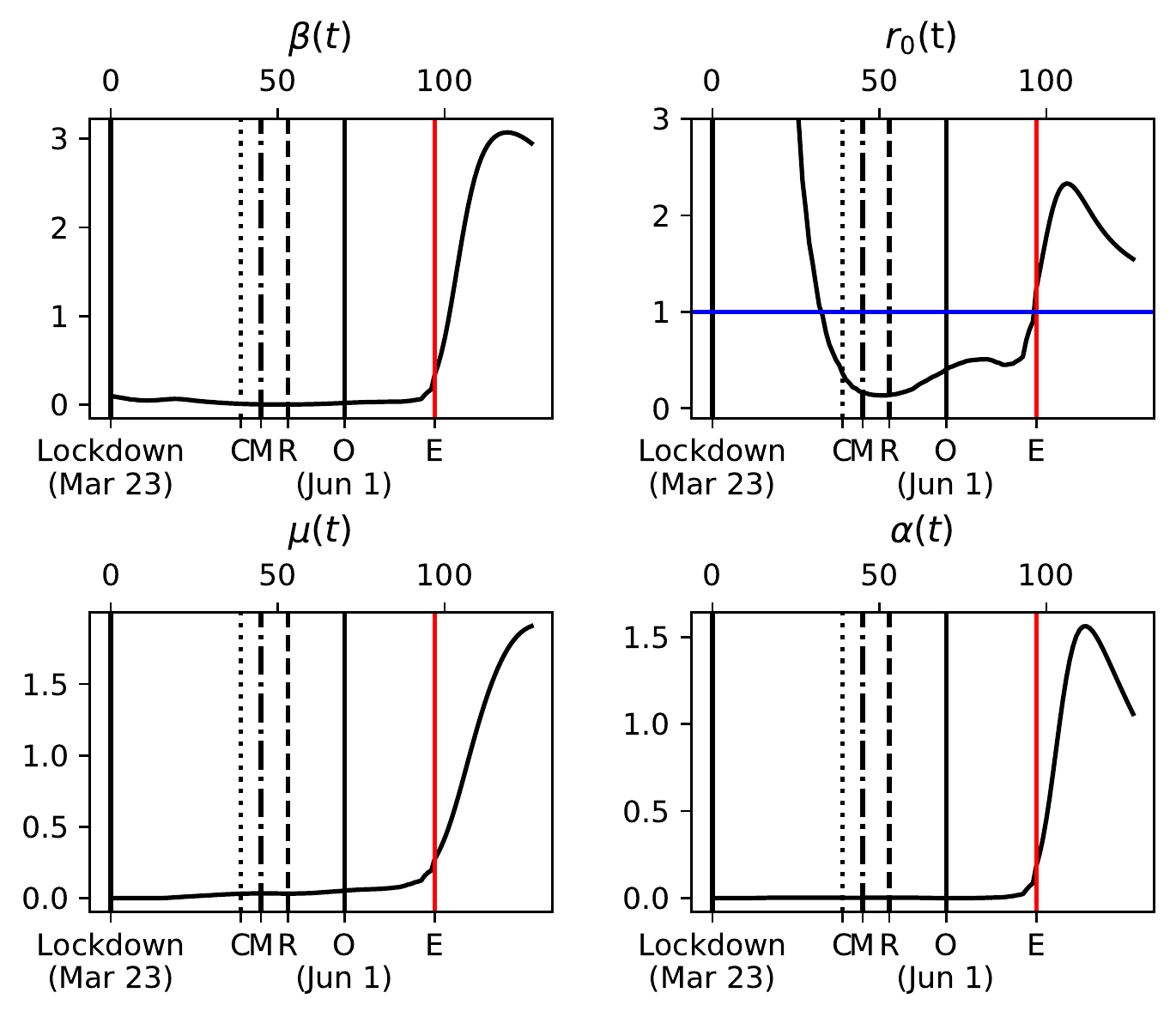}} \hspace{10mm}
    \subfigure[region 8]{\includegraphics[height=38mm]{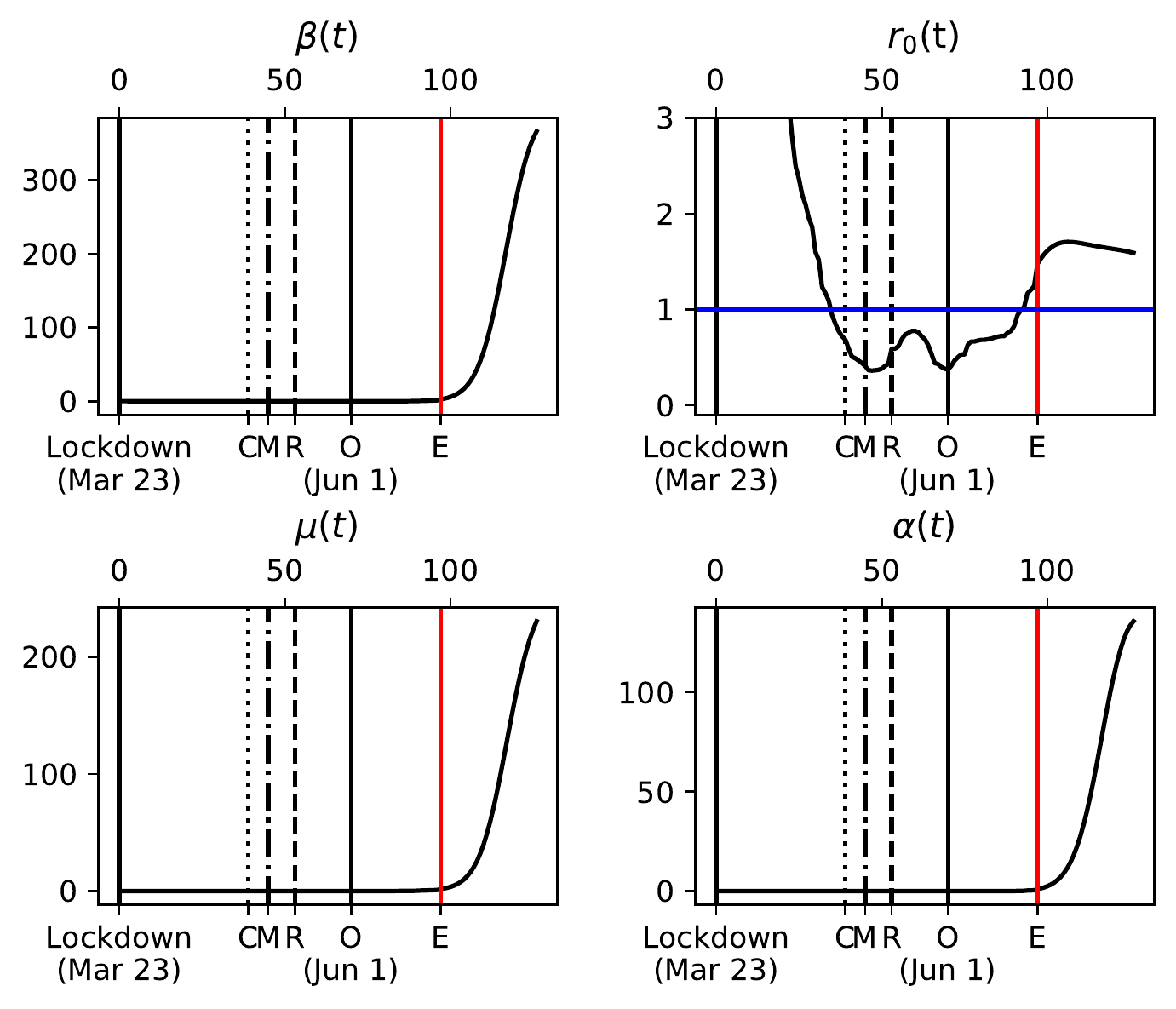}}
    \caption{Regions 1-8: Time-dependent coefficients identified by DNNs, where an increased infection rate after the open (O) of lockdown on June 1st is observed.}
\end{figure}
\begin{figure}[p!]
    \centering
    \subfigure[region 1]{\includegraphics[height=38mm]{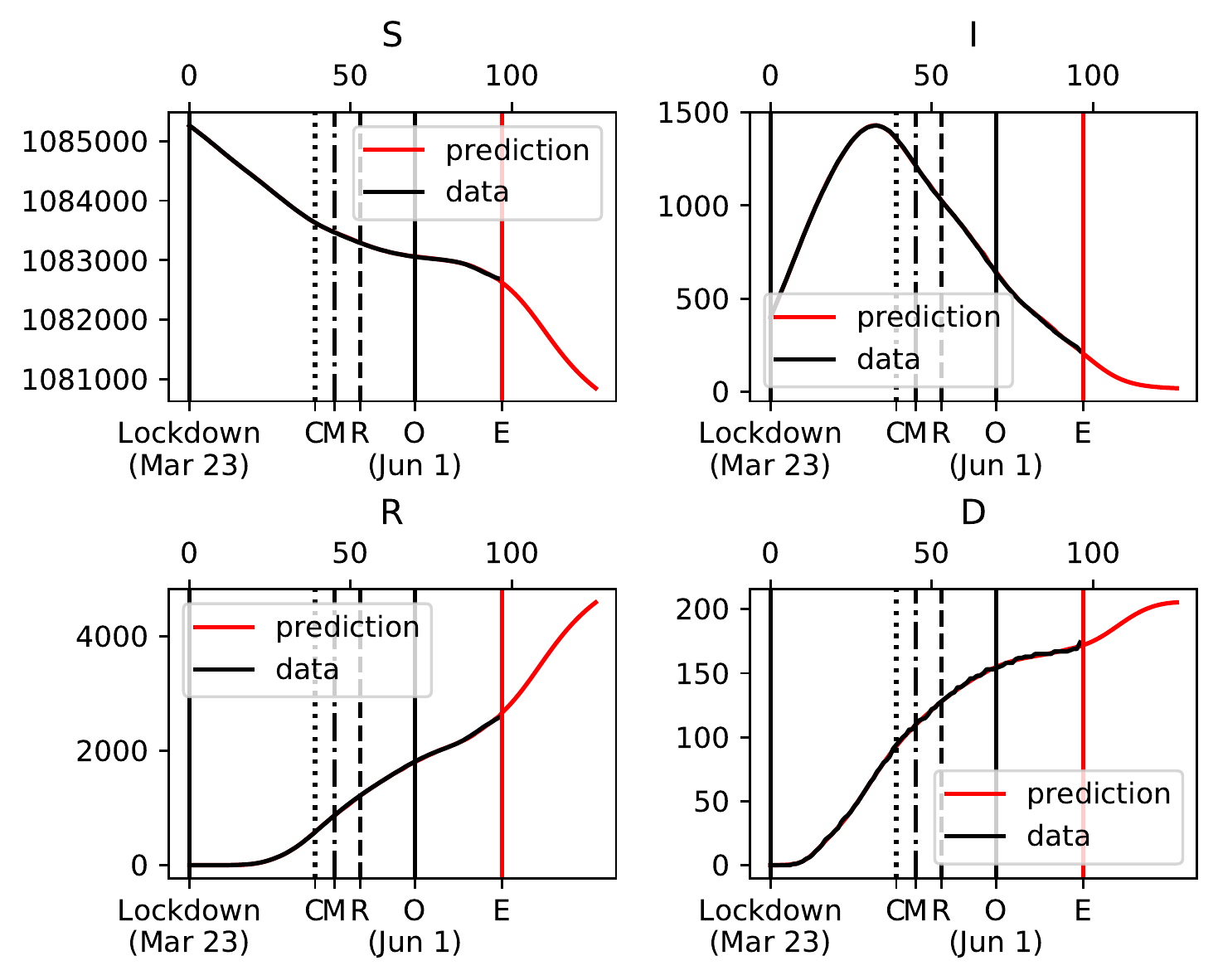}}\hspace{10mm}
    \subfigure[region 2]{\includegraphics[height=38mm]{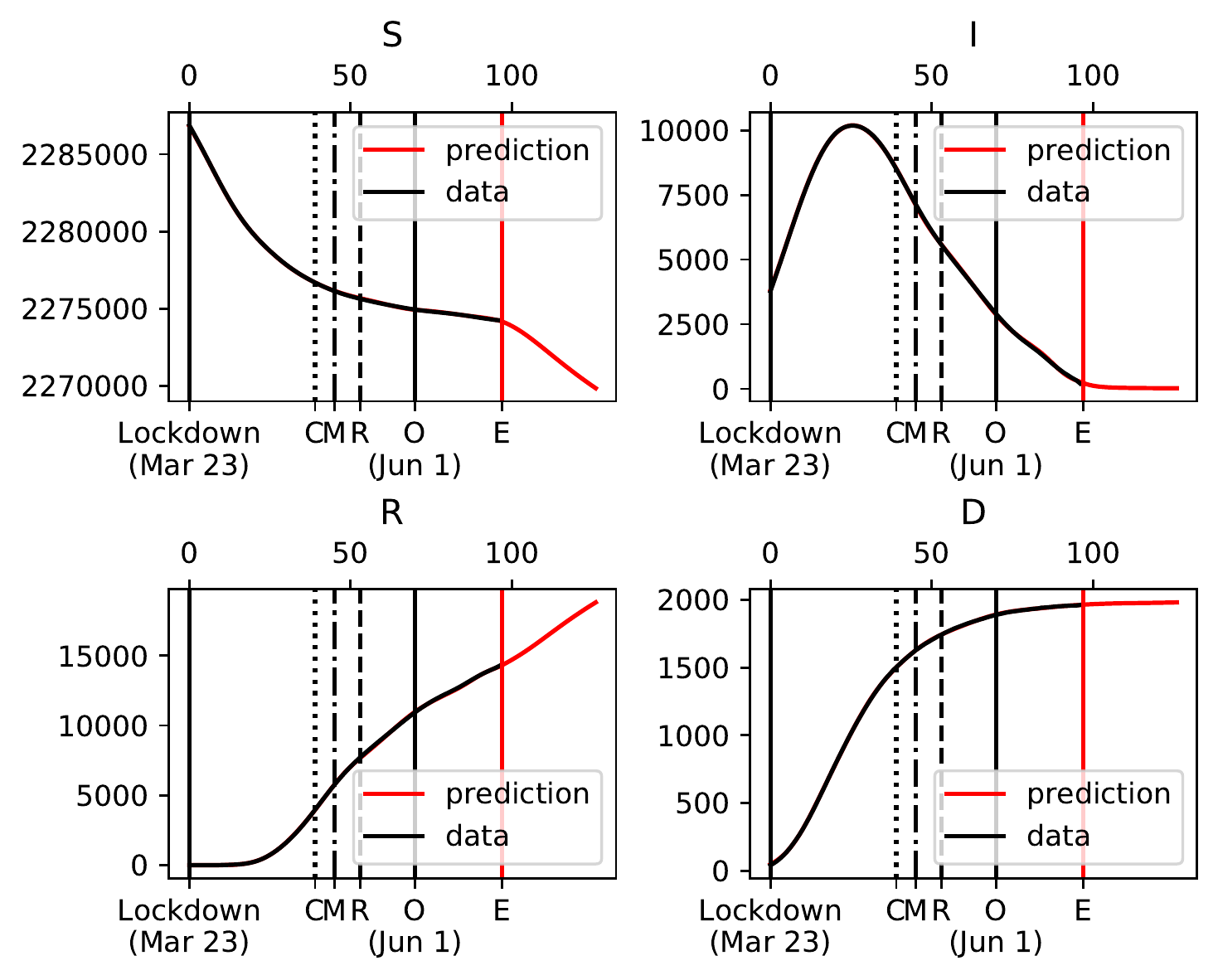}}\\
    \subfigure[region 3]{\includegraphics[height=38mm]{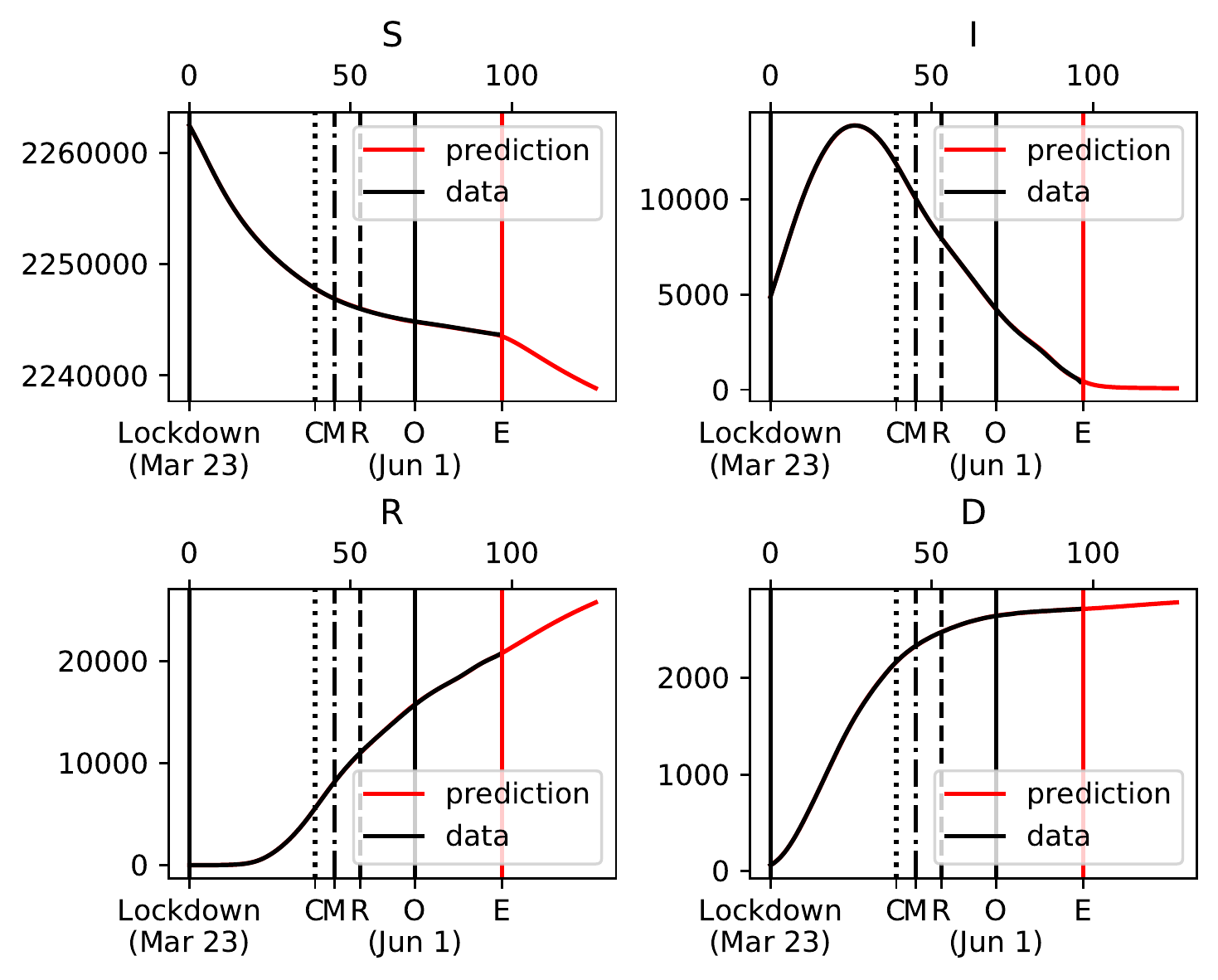}}\hspace{10mm}
    \subfigure[region 4]{\includegraphics[height=38mm]{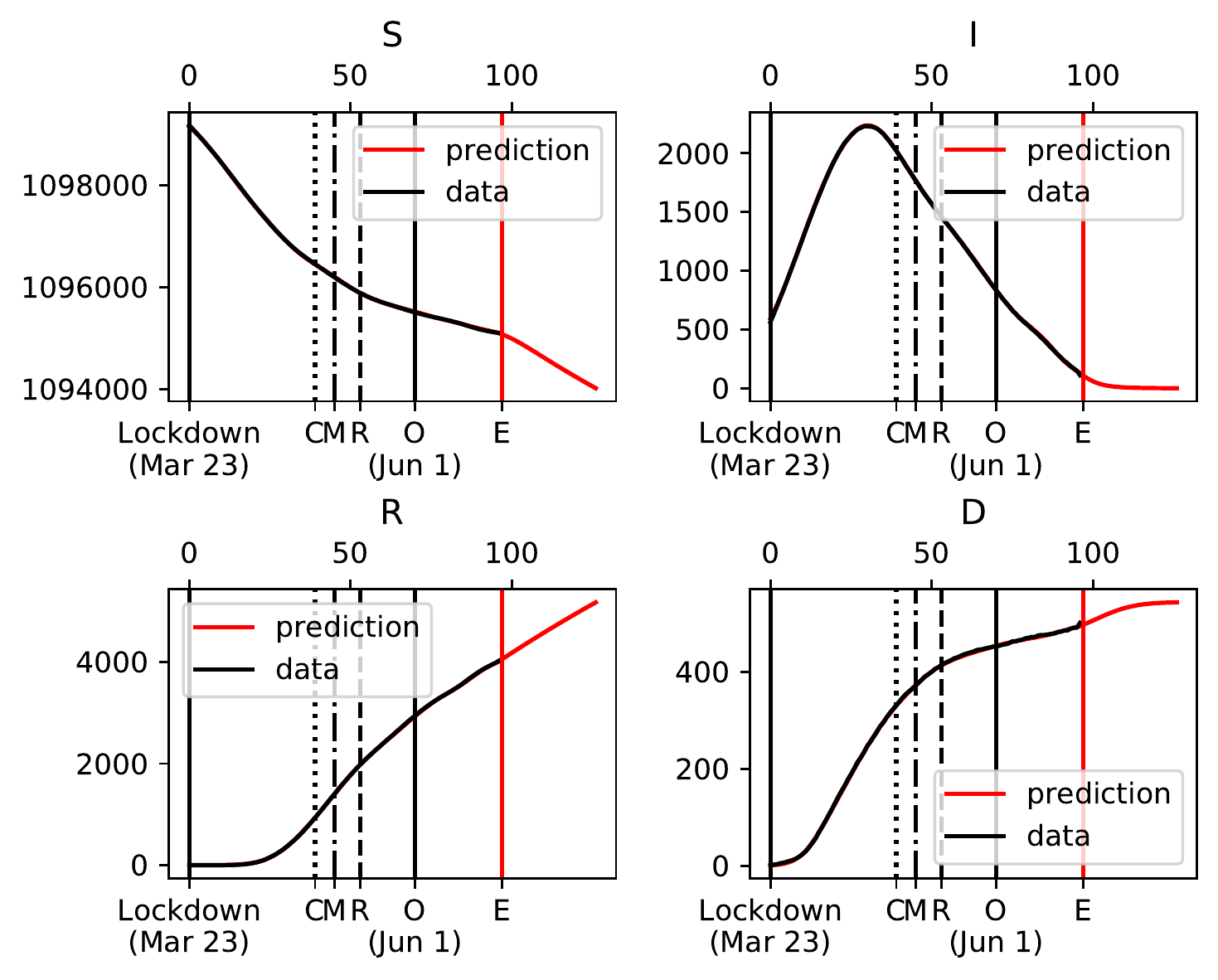}}\\
    \subfigure[region 5]{\includegraphics[height=38mm]{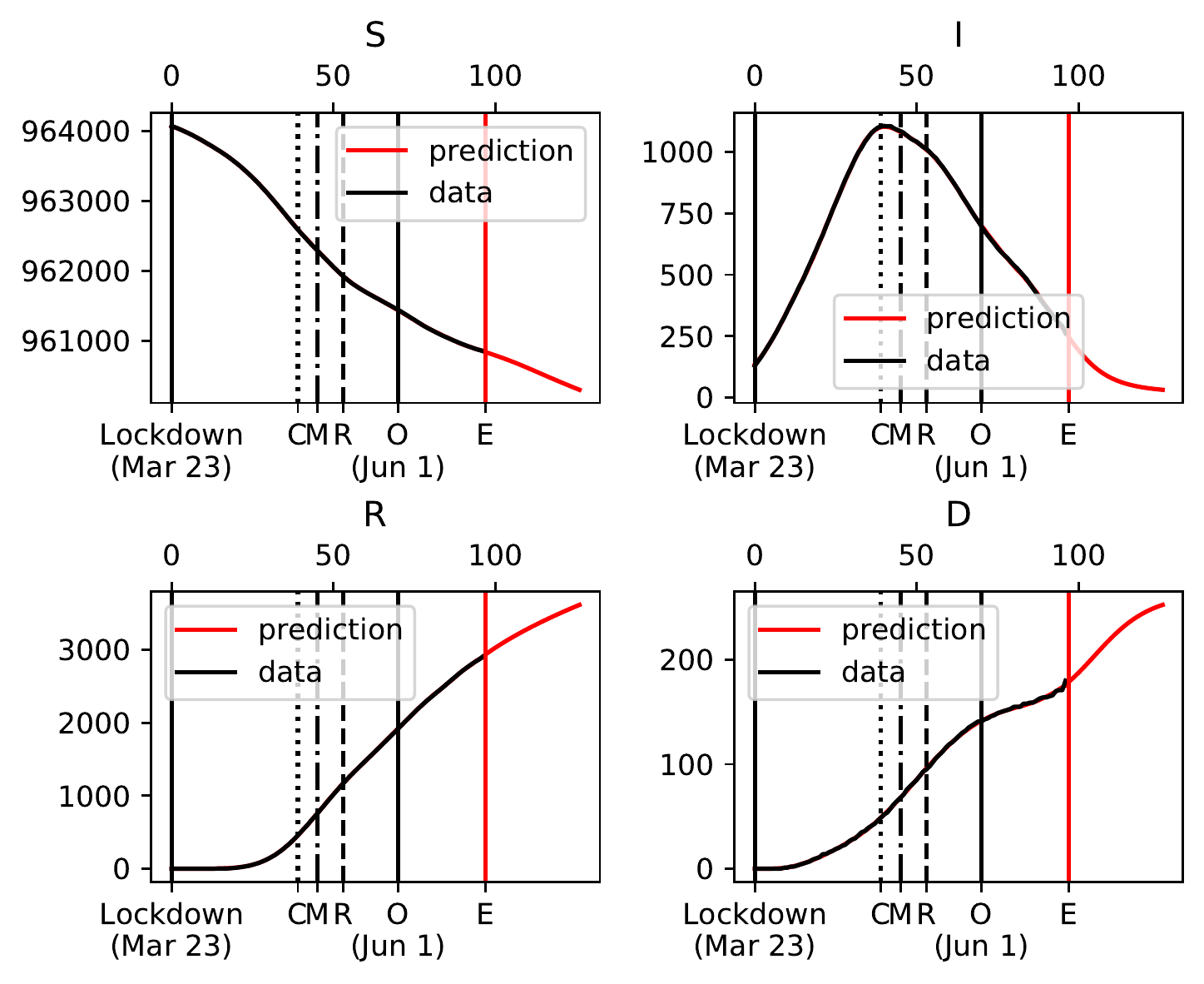}}\hspace{10mm}
    \subfigure[region 6]{\includegraphics[height=38mm]{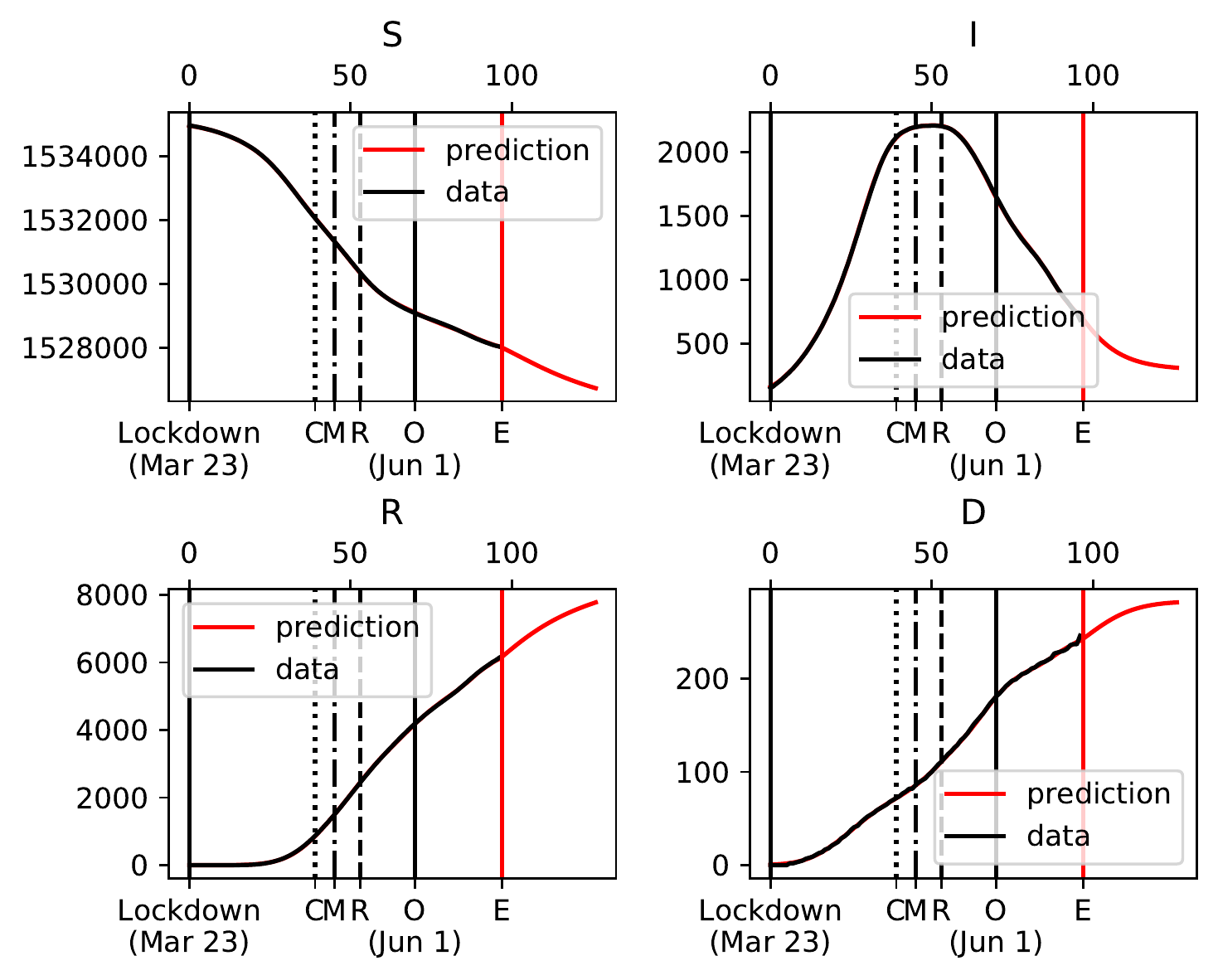}}\\
    \subfigure[region 7]{\includegraphics[height=38mm]{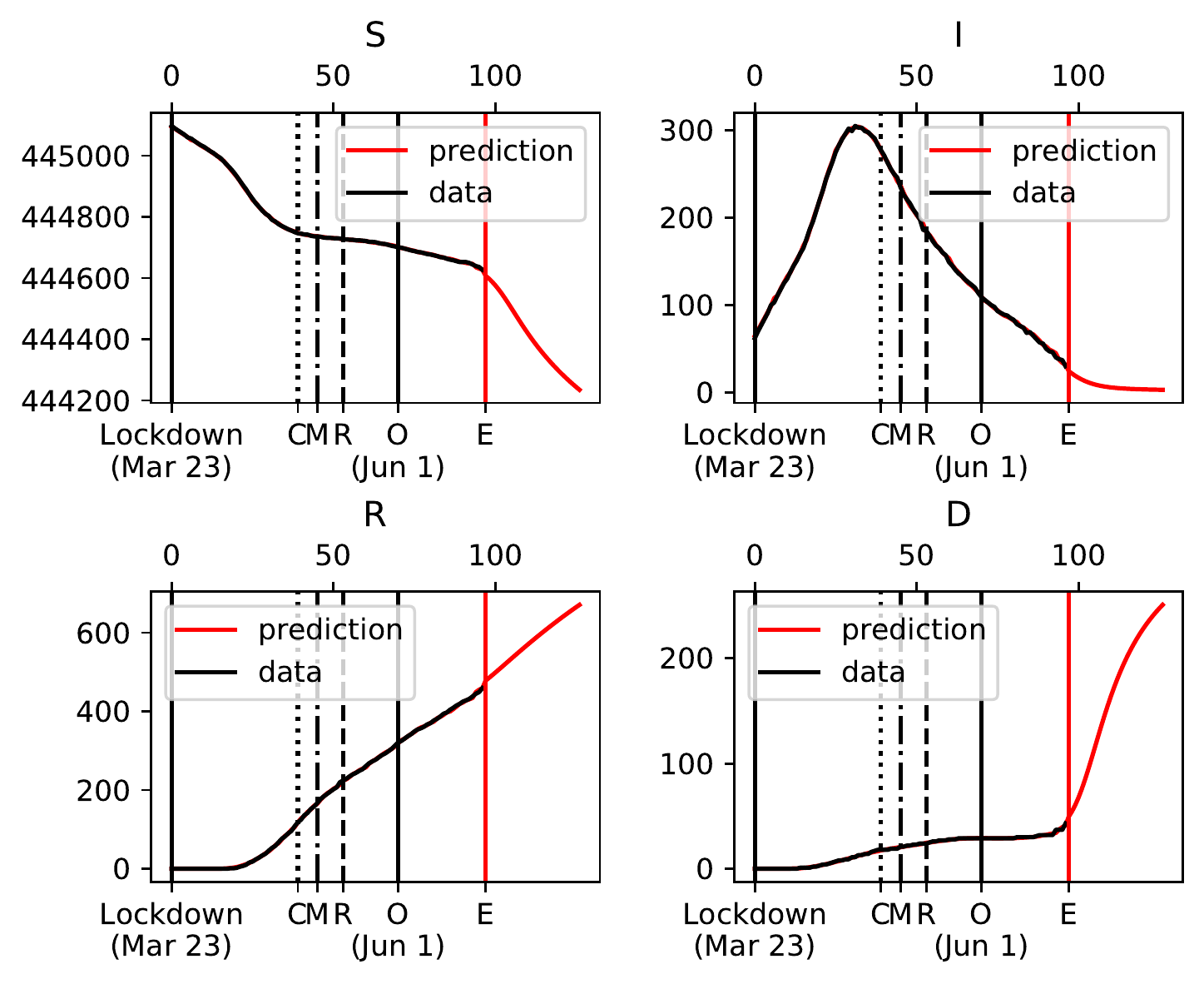}}\hspace{10mm}
    \subfigure[region 8]{\includegraphics[height=38mm]{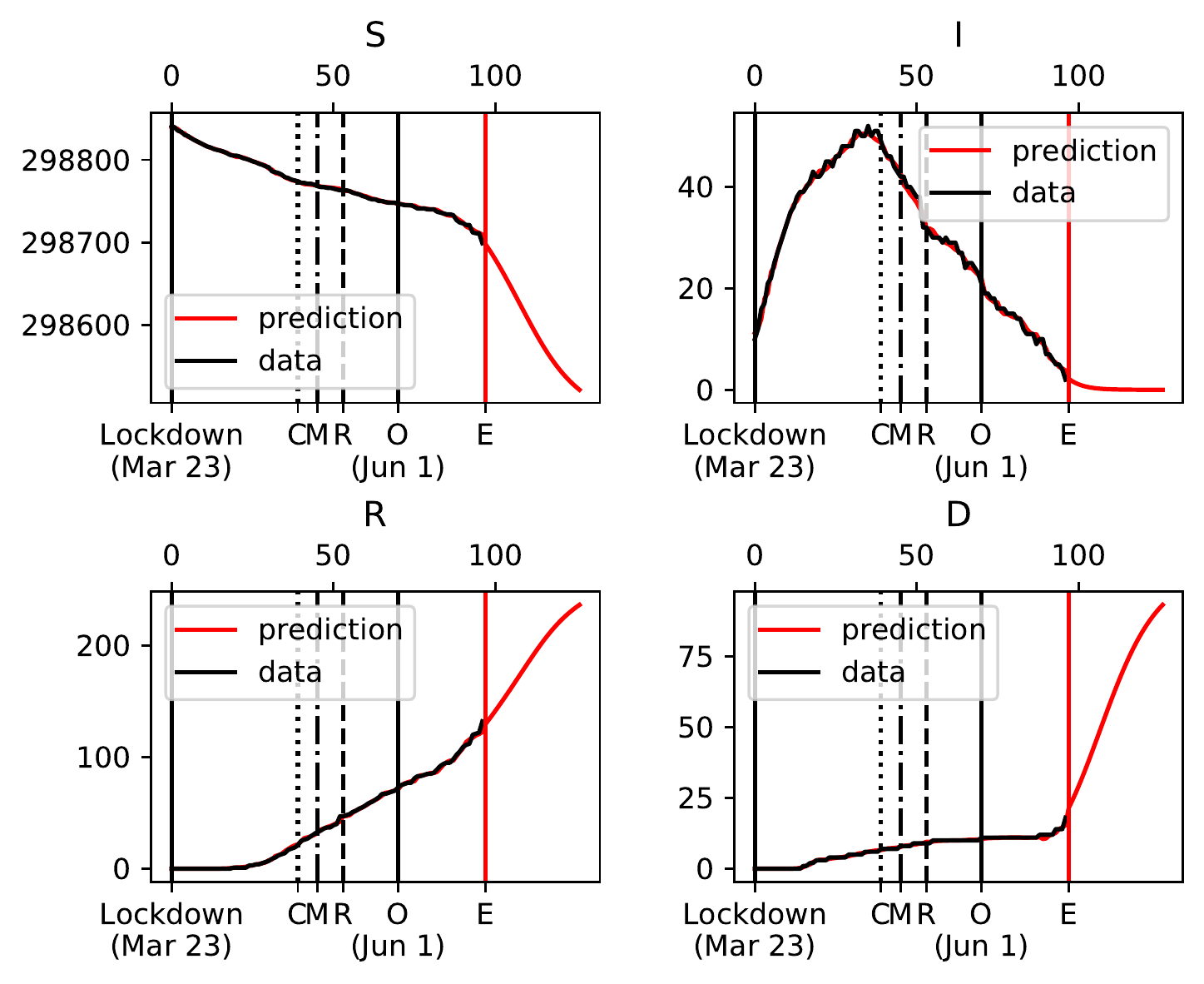}}
    \caption{Regions 1-8: DNNs learned $S(t), I(t), R(t), D(t)$ based on the existing discrete data point, where a 30-day prediction is made by DNNs.}
\end{figure}

\subsection{BNN results for different regions}
\label{sec:BNN-results-regions}

\begin{figure}[p!]
    \centering
    \subfigure[region 1]{\includegraphics[height=38mm]{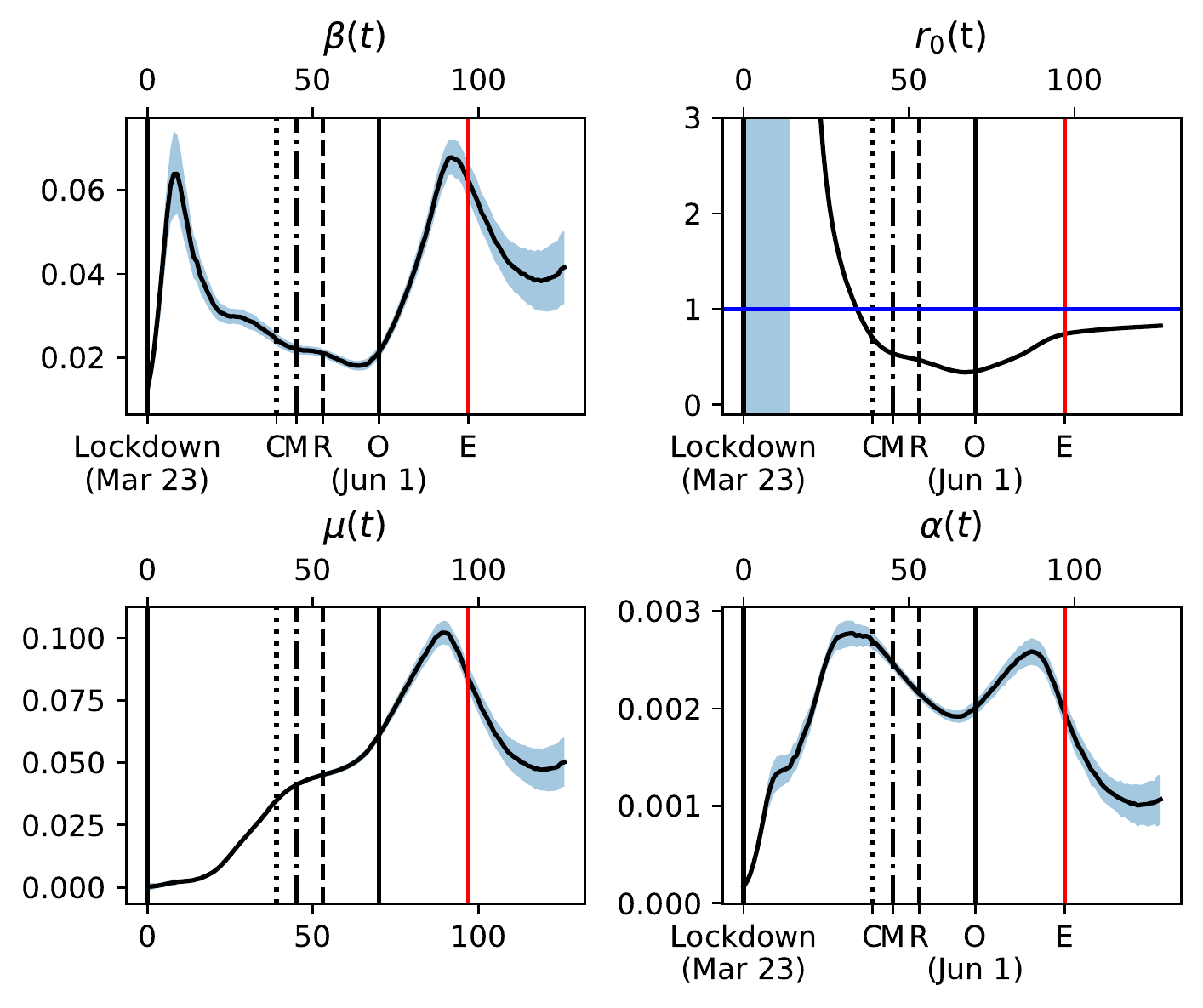}}\hspace{10mm}
    \subfigure[region 2]{\includegraphics[height=38mm]{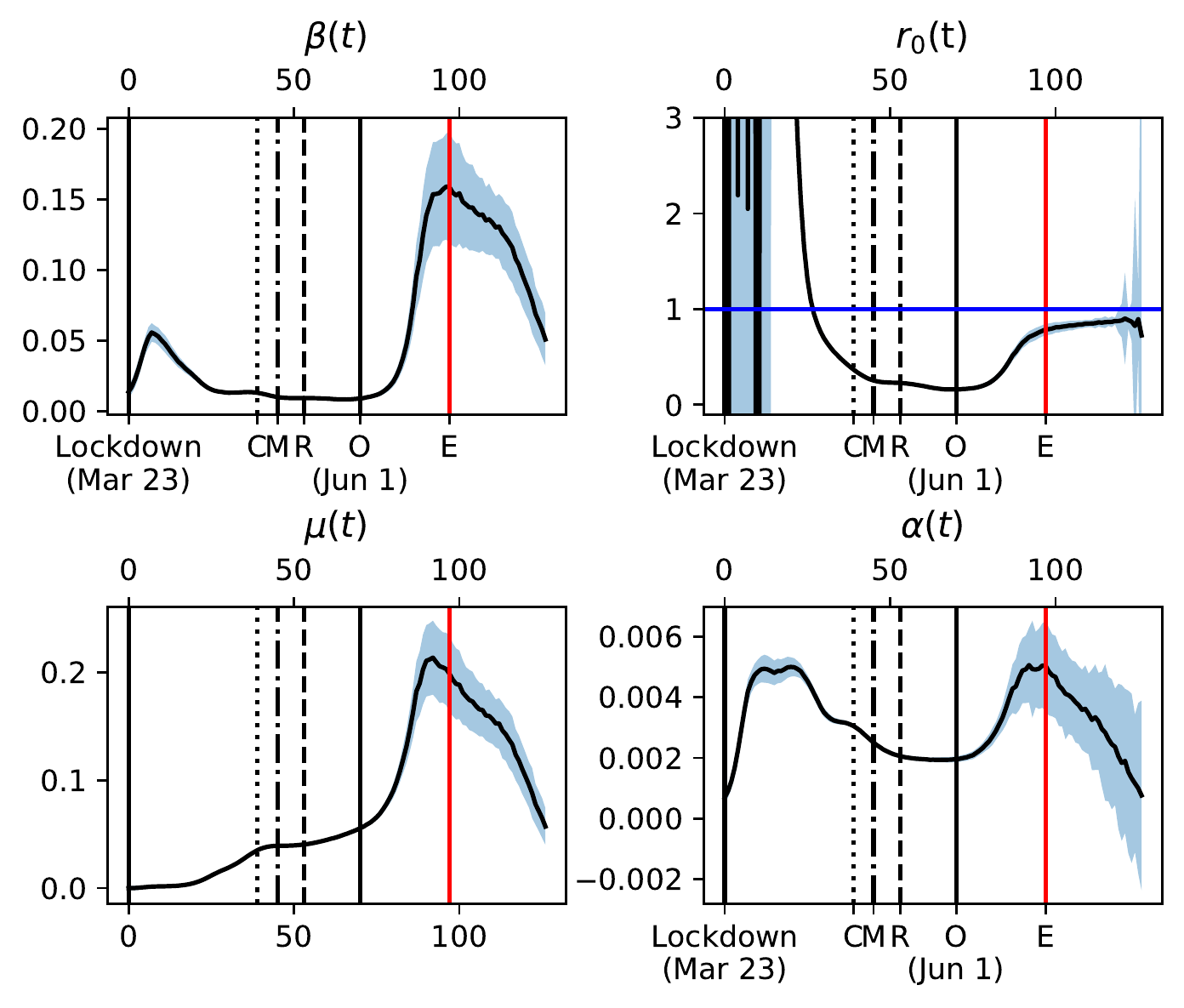}}\\
    \subfigure[region 3]{\includegraphics[height=38mm]{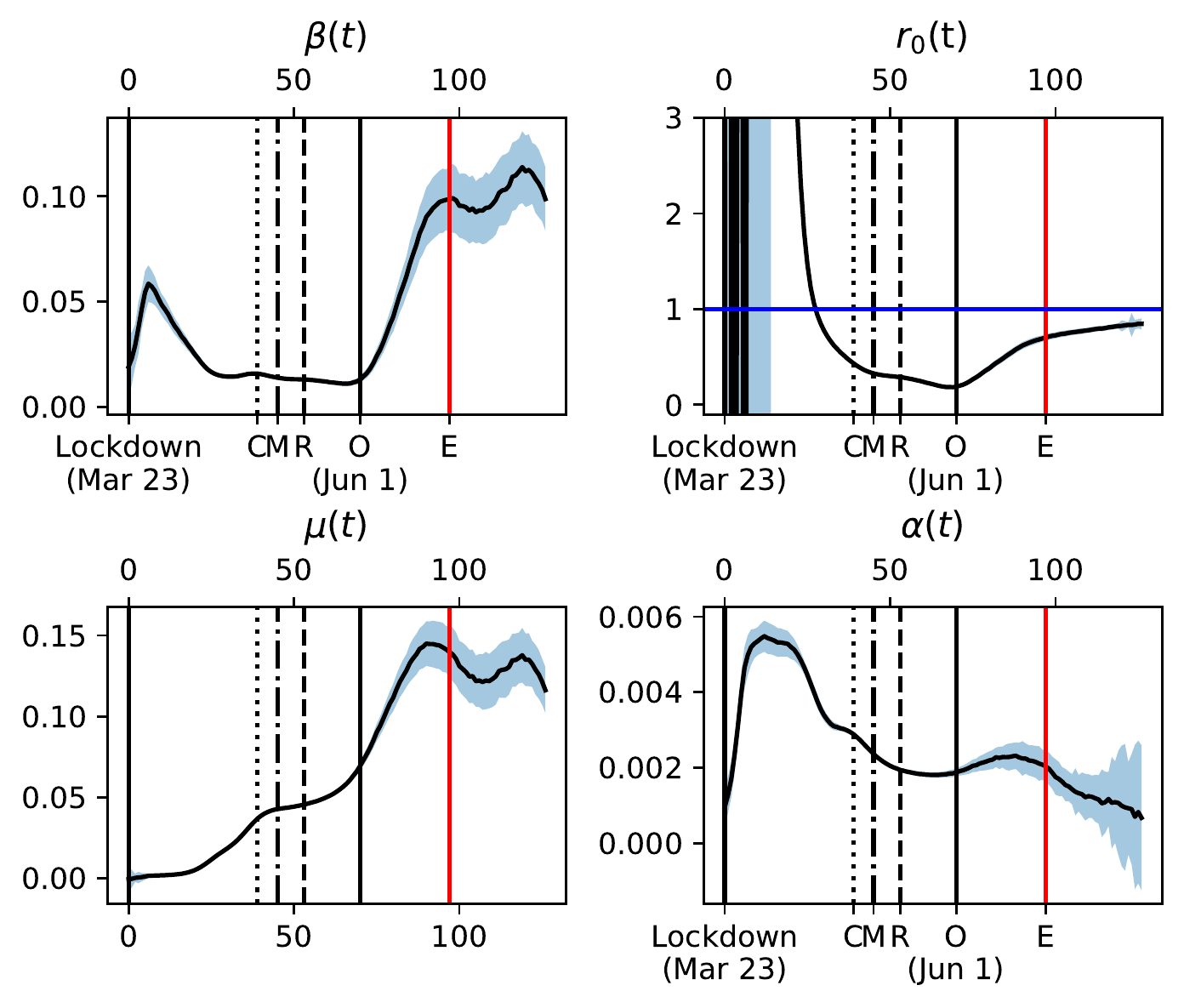}}\hspace{10mm}
    \subfigure[region 4]{\includegraphics[height=38mm]{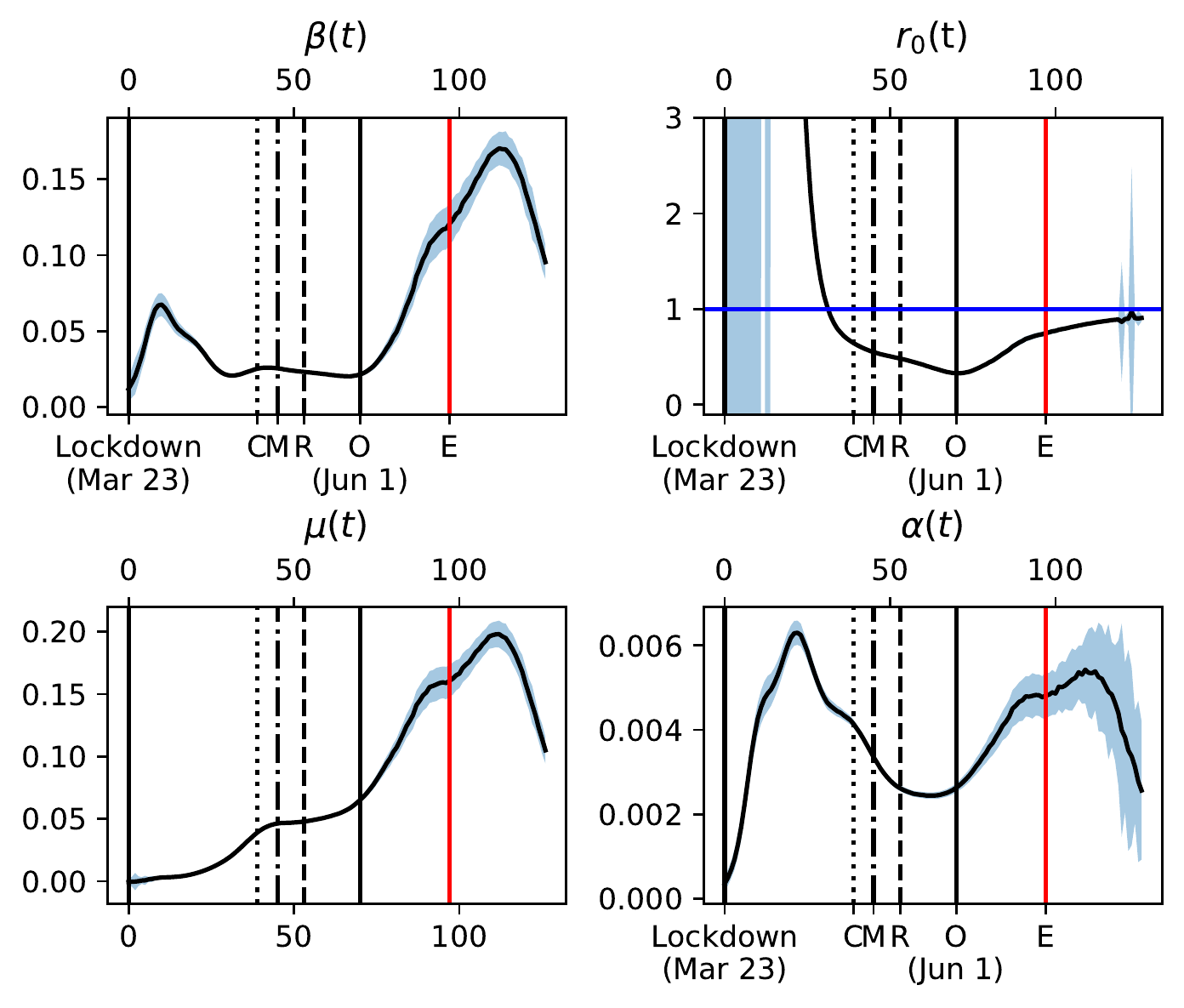}}\\
    \subfigure[region 5]{\includegraphics[height=38mm]{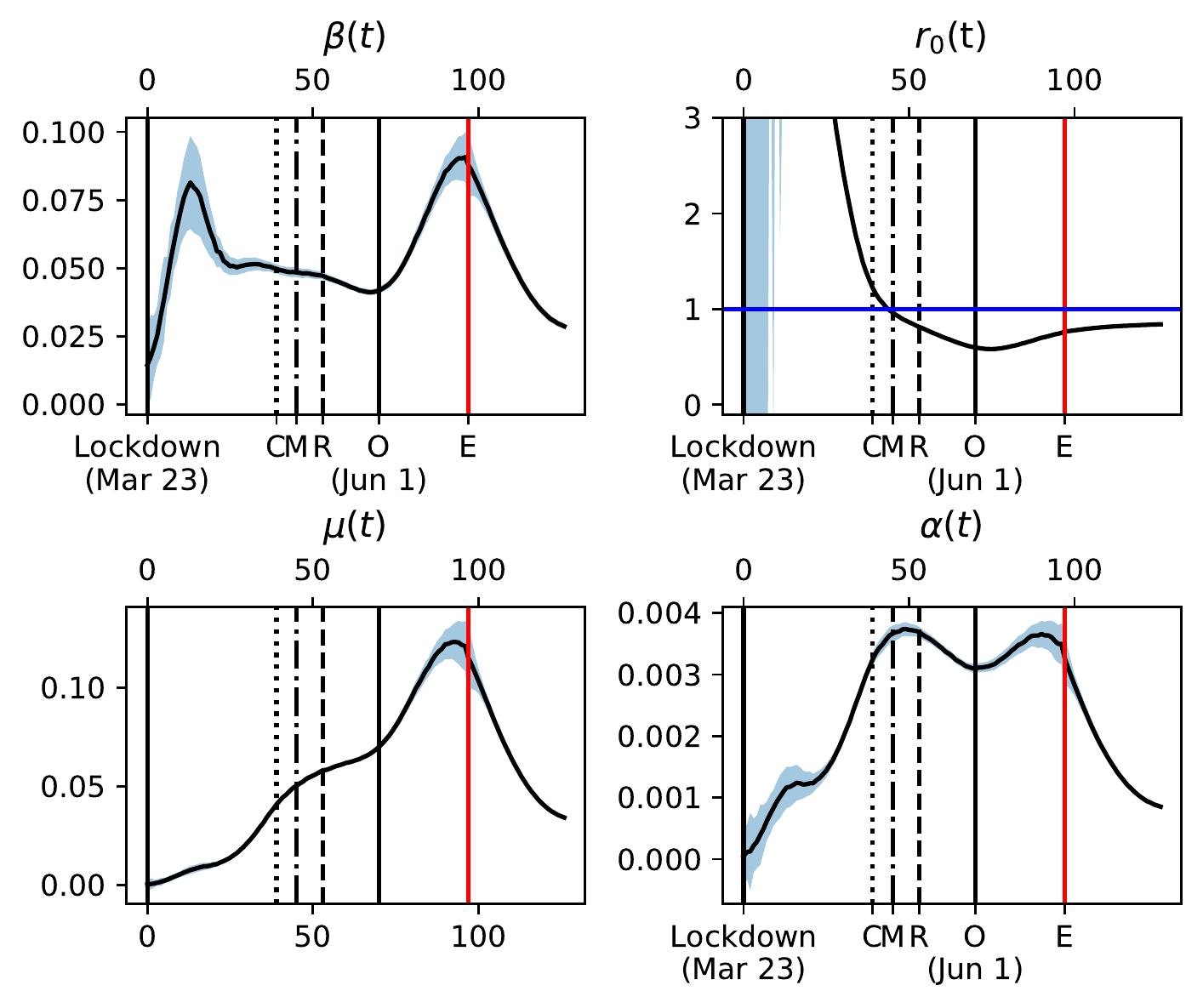}}\hspace{10mm}
    \subfigure[region 6]{\includegraphics[height=38mm]{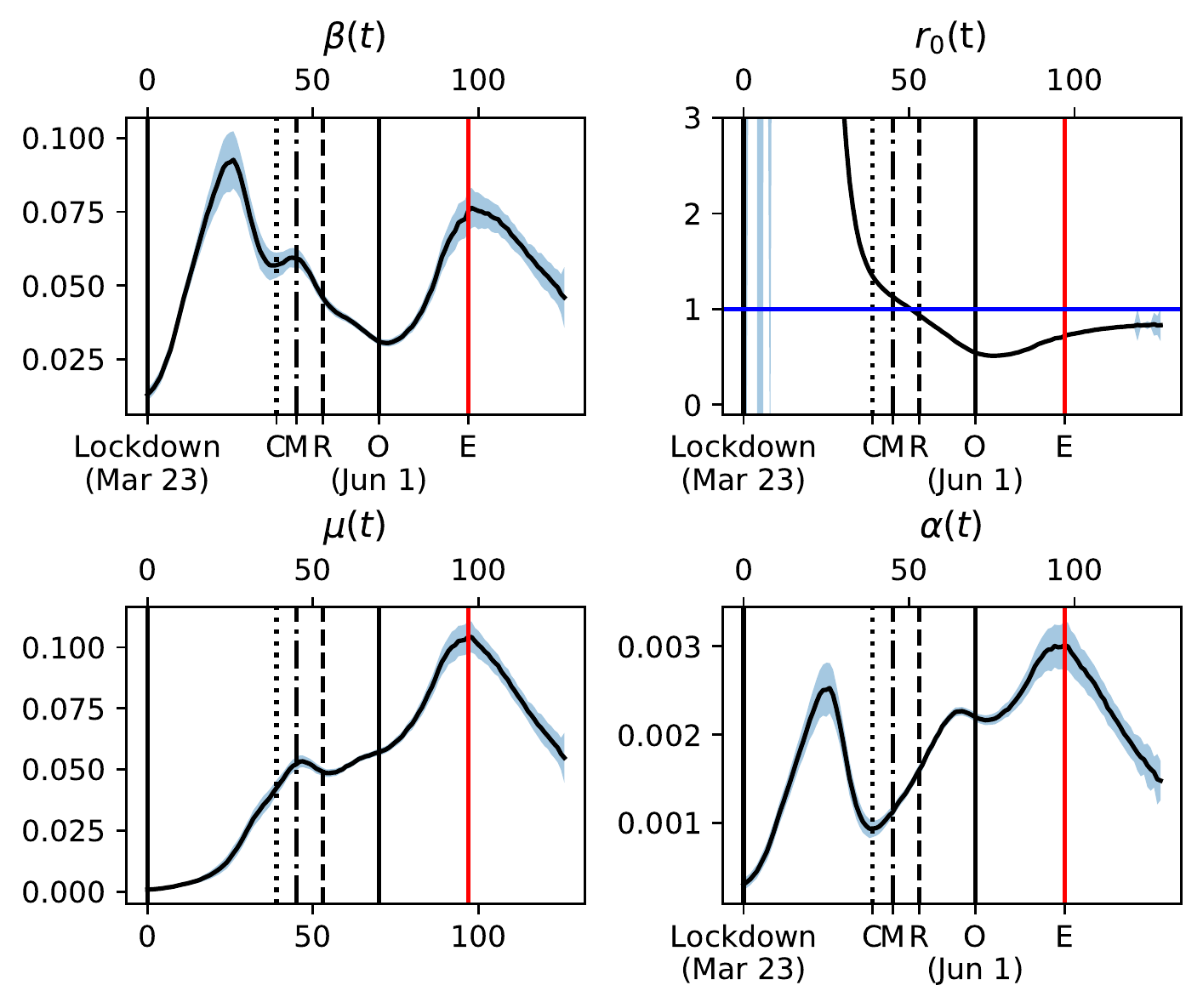}}\\
    \subfigure[region 7]{\includegraphics[height=38mm]{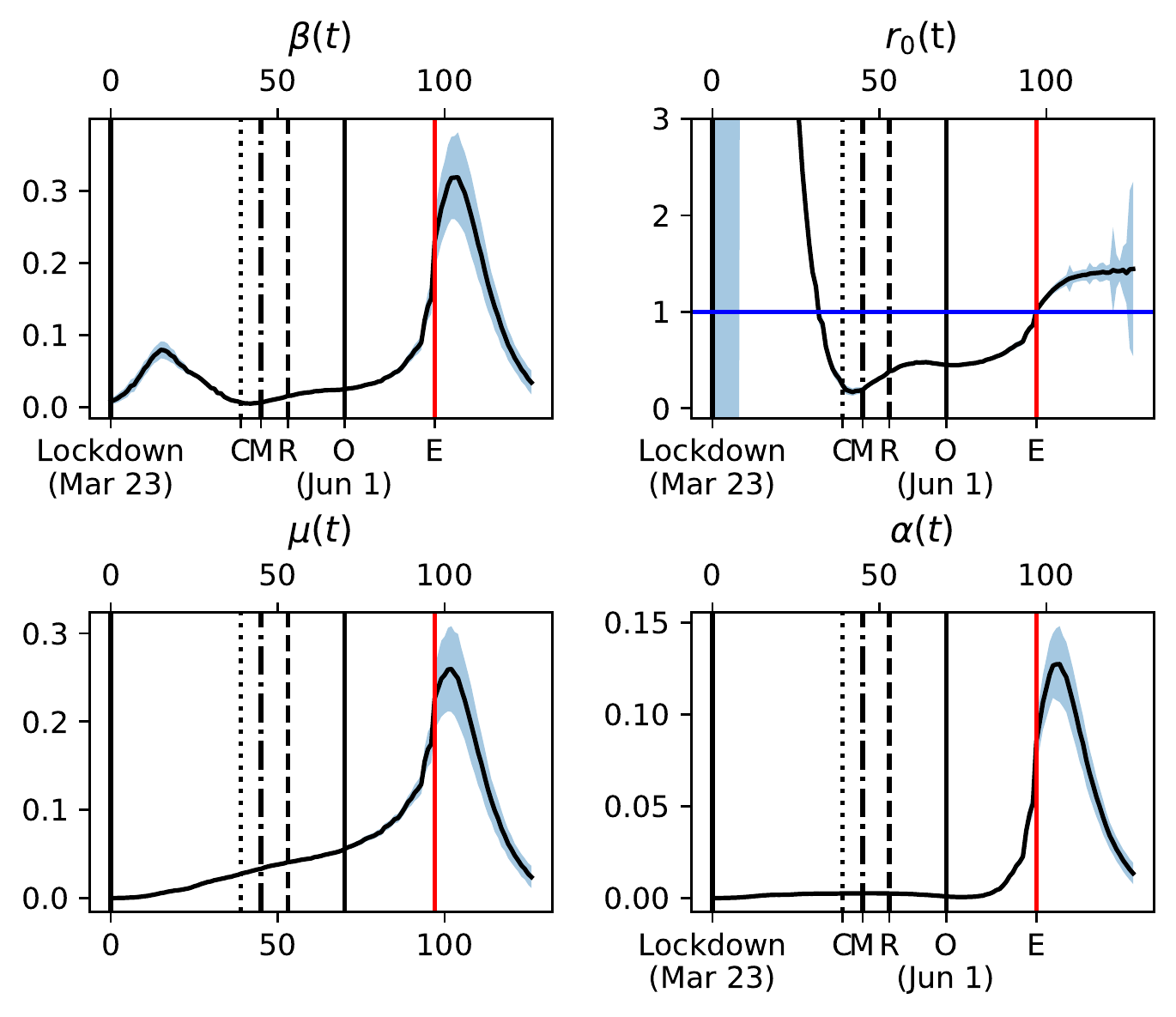}}\hspace{10mm}
    \subfigure[region 8]{\includegraphics[height=38mm]{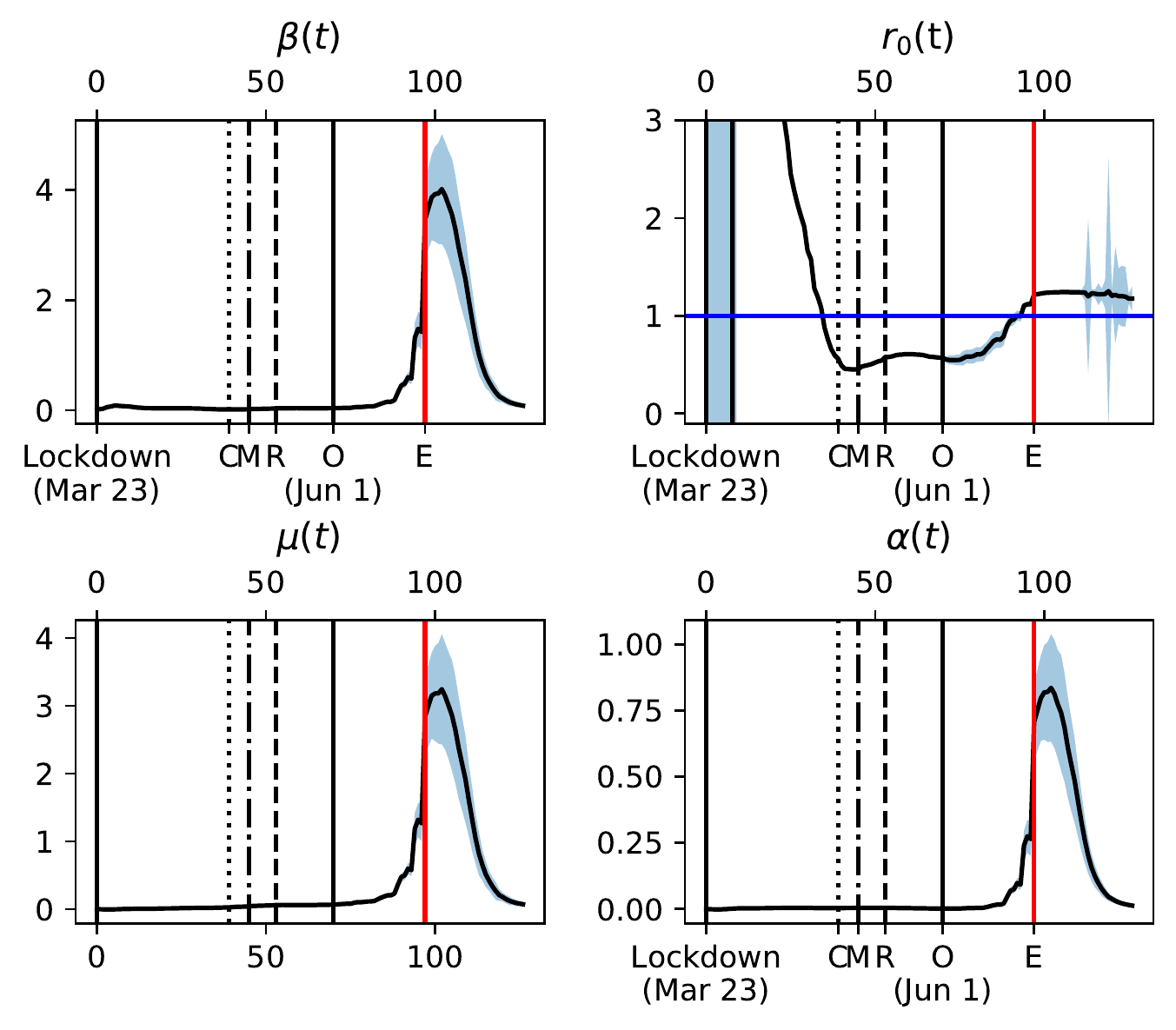}}
    \caption{Regions 1-8: Time-dependent coefficients identified by BNNs, where an increased infection rate after the open (O) of lockdown on June 1st is observed. Bands correspond to $\pm$ standard deviation over the mean.}
\end{figure}
\begin{figure}[p!]
    \centering
    \subfigure[region 1]{\includegraphics[height=38mm]{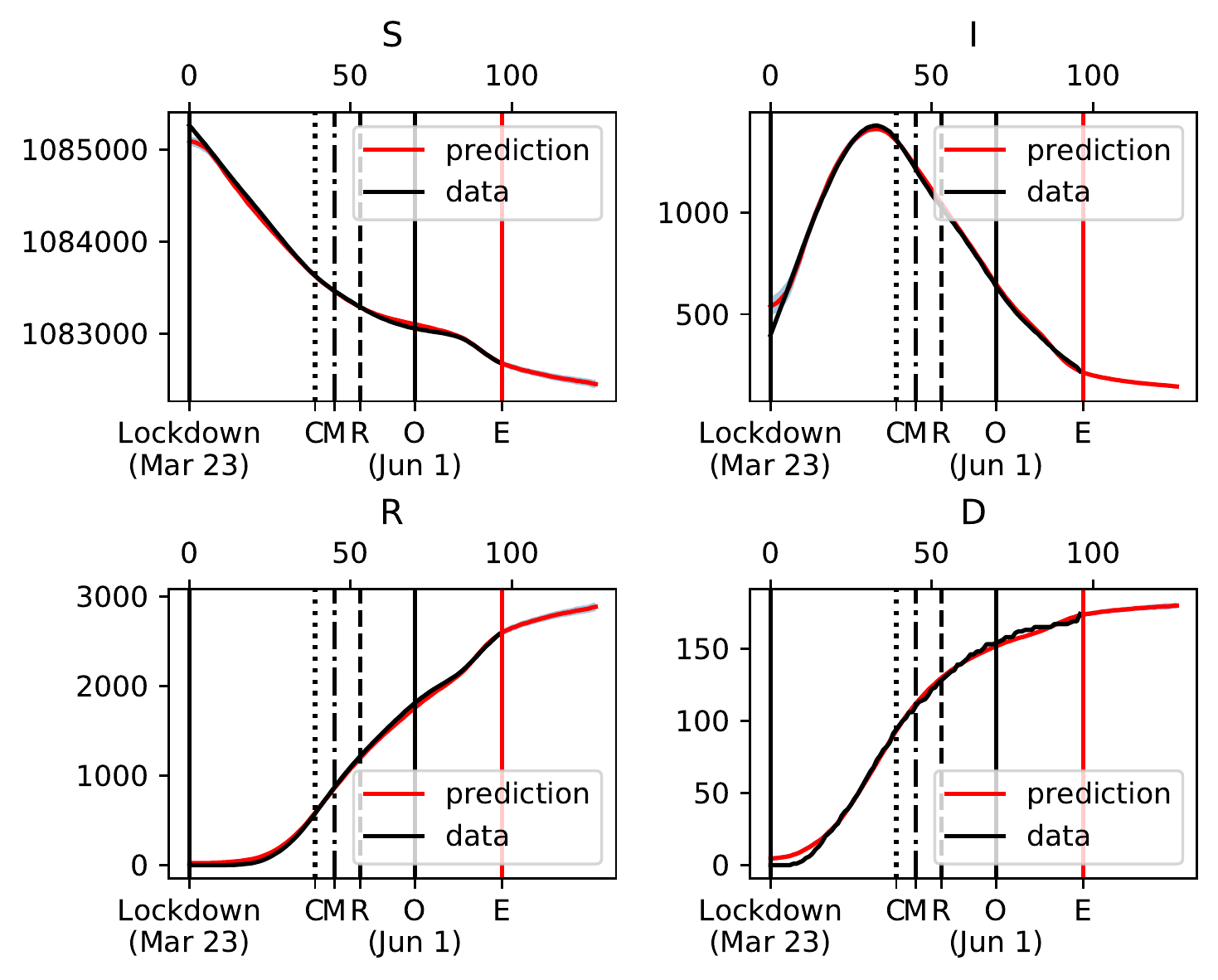}}\hspace{10mm}
    \subfigure[region 2]{\includegraphics[height=38mm]{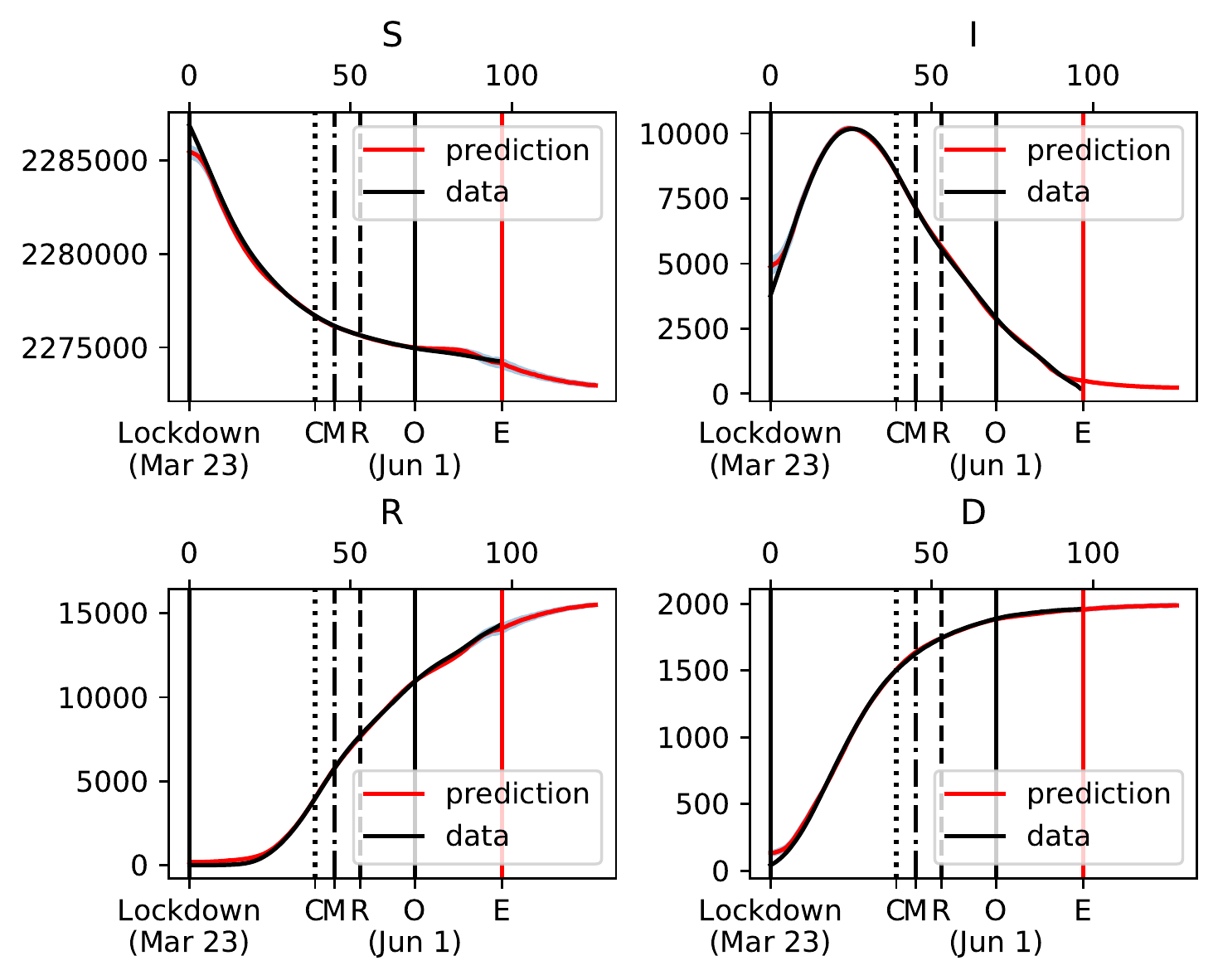}}\\
    \subfigure[region 3]{\includegraphics[height=38mm]{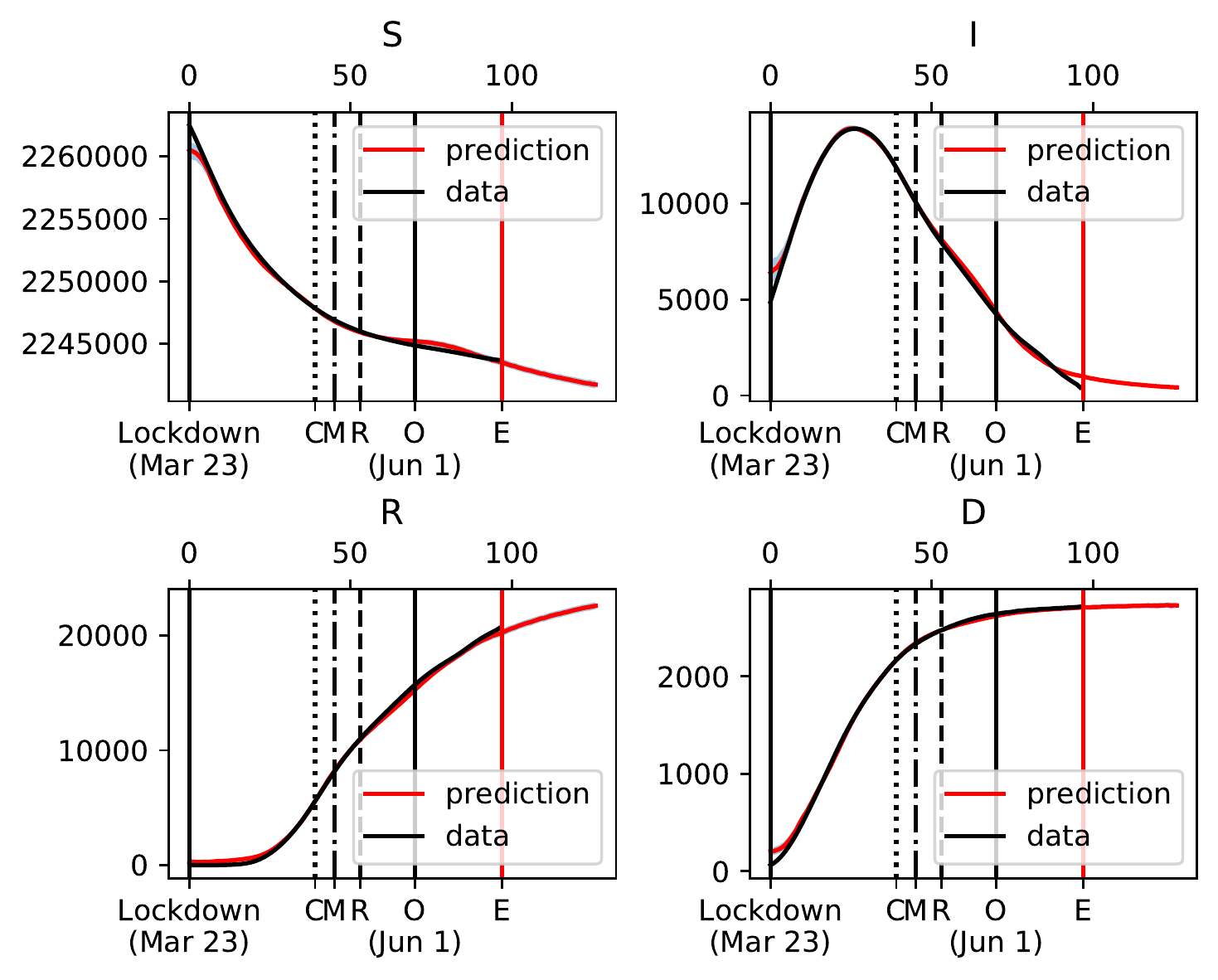}}\hspace{10mm}
    \subfigure[region 4]{\includegraphics[height=38mm]{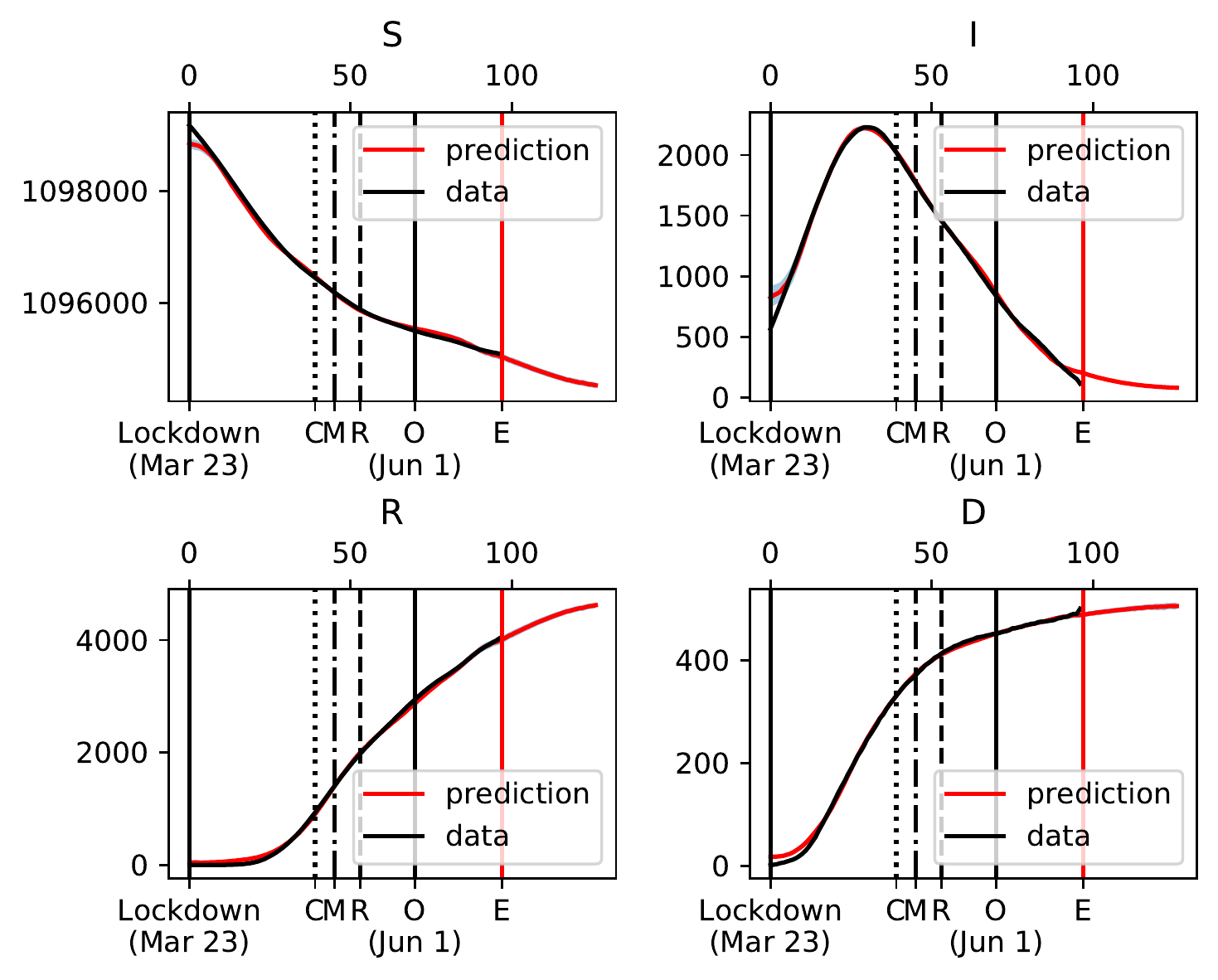}}\\
    \subfigure[region 5]{\includegraphics[height=38mm]{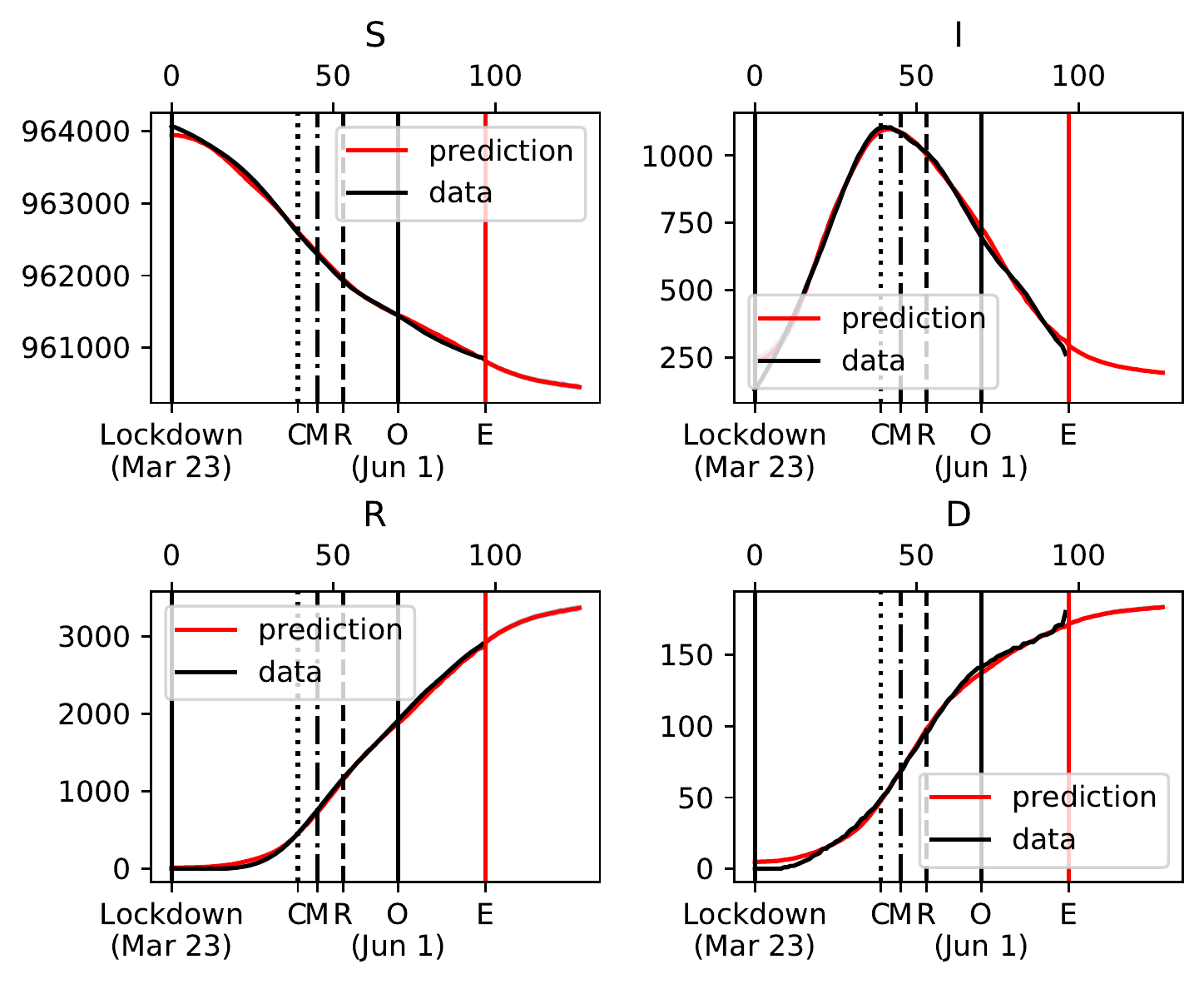}}\hspace{10mm}
    \subfigure[region 6]{\includegraphics[height=38mm]{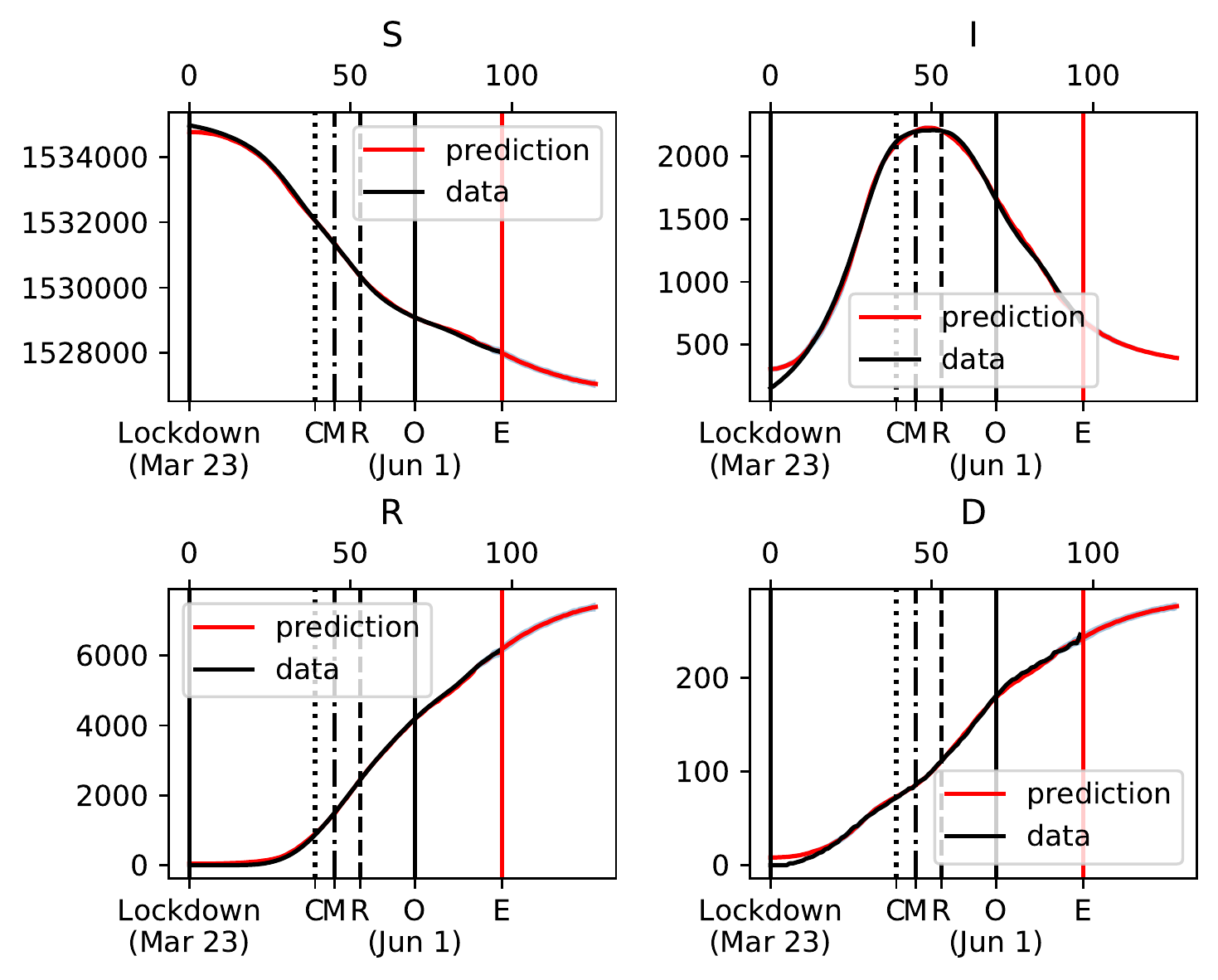}}\\
    \subfigure[region 7]{\includegraphics[height=38mm]{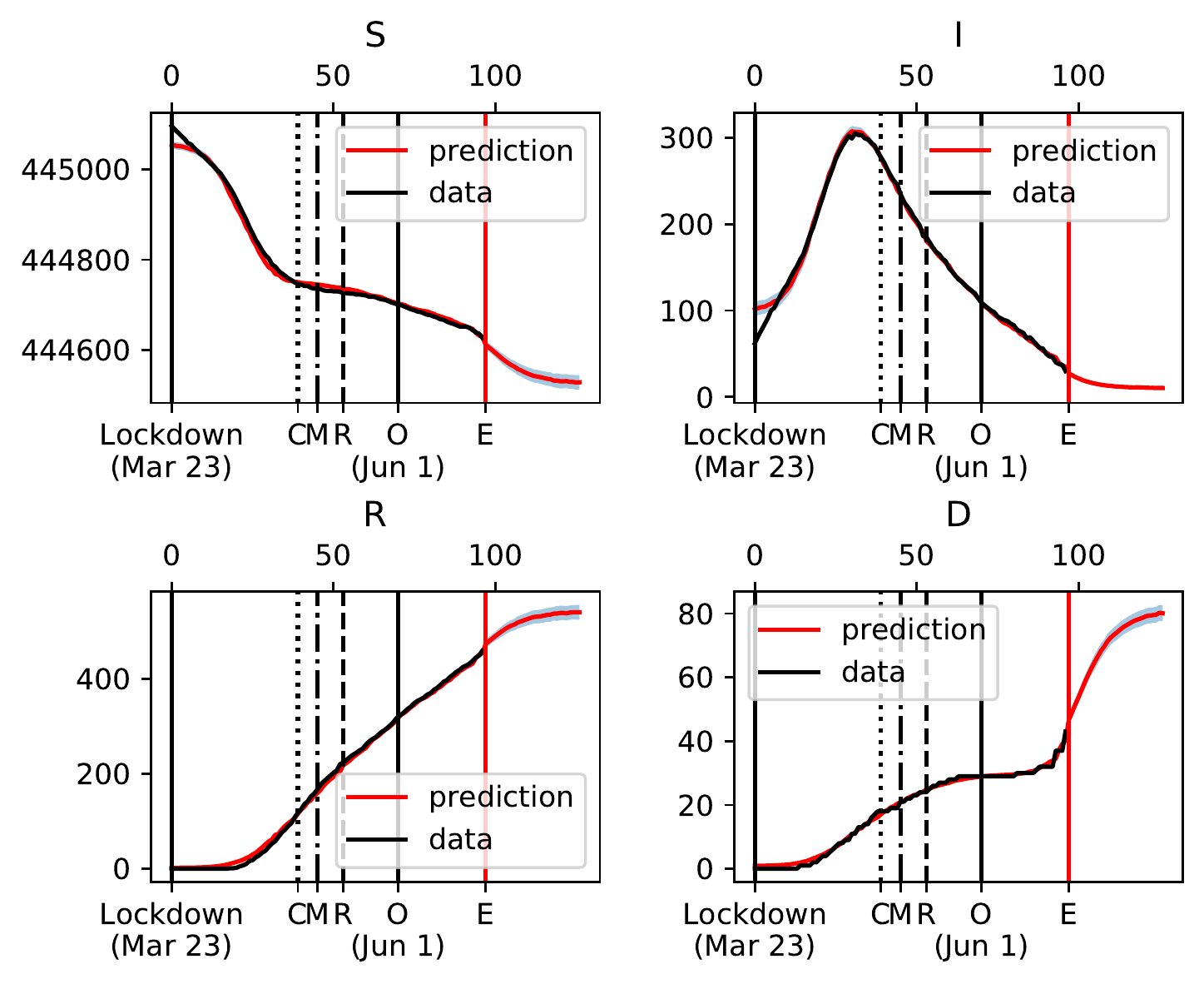}}\hspace{10mm}
    \subfigure[region 8]{\includegraphics[height=38mm]{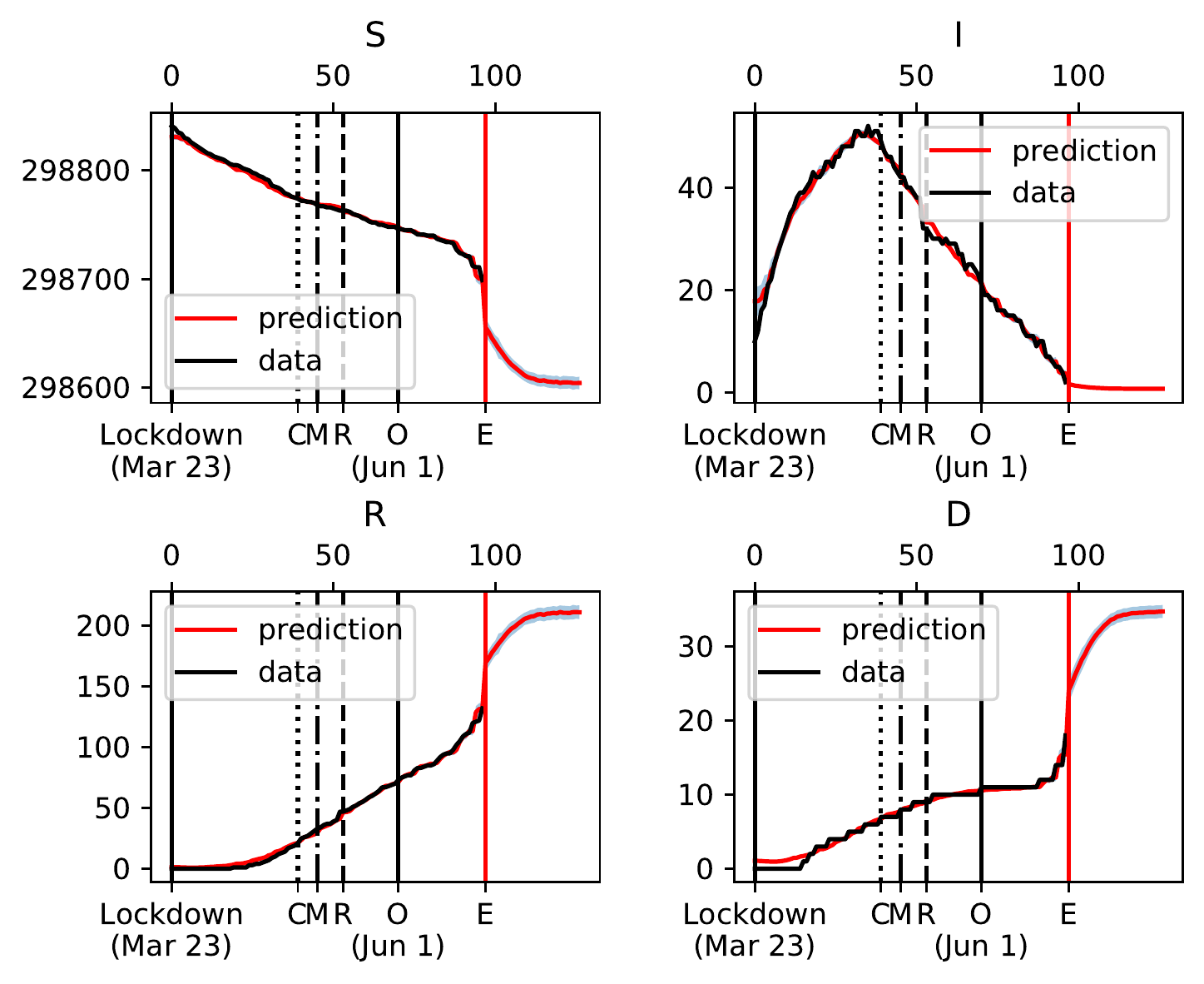}}
    \caption{Region 1-8: BNNs learned $S(t), I(t), R(t), D(t)$ based on the existing discrete data point, where a 30-day prediction is made by BNNs. Bands correspond to $\pm$ standard deviation over the mean.}
\end{figure}

\end{document}